\documentclass{article}

\usepackage[dblblindworkshop, final]{neurips_2025}

\usepackage[utf8]{inputenc} % allow utf-8 input
\usepackage[T1]{fontenc}    % use 8-bit T1 fonts
\usepackage{hyperref}       % hyperlinks
\usepackage{url}            % simple URL typesetting
\usepackage{booktabs}       % professional-quality tables
\usepackage{amsfonts}       % blackboard math symbols
\usepackage{nicefrac}       % compact symbols for 1/2, etc.
\usepackage{microtype}      % microtypography
\usepackage{xcolor}         % colors

\usepackage{graphicx}       
\usepackage{amsmath, amssymb} 
\usepackage{multirow}        
\usepackage[export]{adjustbox} 
\usepackage{subcaption}
\usepackage{tcolorbox}
\usepackage{makecell}
\tcbuselibrary{breakable}

\title{What AI Speaks for Your Community: Polling AI Agents for Public Opinion on Data Center Projects}

\workshoptitle{ResponsibleFM}

\author{%
  Zhifeng Wu \\
  UC Riverside \\
  \texttt{zwu178@ucr.edu} \\
  \And
  Yuelin Han \\
  UC Riverside \\
  \texttt{yhan116@ucr.edu} \\
  \And
  Shaolei Ren\thanks{Corresponding author}\\
  UC Riverside\\
  \texttt{shaolei@ucr.edu} \\
}

\begin{document}

\maketitle

\begin{abstract}
The intense computational demands of AI, especially large foundation models, are driving a global boom in data centers. These facilities bring both tangible benefits and potential  environmental burdens to local communities. However, the planning processes for data centers often fail to proactively integrate local public opinion in advance, largely because traditional polling is expensive or is conducted too late to influence decisions. To address this gap, we introduce an AI agent polling framework, leveraging large language models to assess community opinion on data centers and guide responsible development of AI. Our experiments reveal water consumption and utility costs as primary concerns, while tax revenue is a key perceived benefit. Furthermore, our cross-model and cross-regional analyses show opinions vary significantly by LLM and regional context. Finally, agent responses show strong topical alignment with real-world survey data. Our framework can serve as a scalable screening tool, enabling developers to integrate community sentiment into early-stage planning for a more informed and socially responsible AI infrastructure deployment.

\end{abstract}

\section{Introduction}
\label{main:intro}

The rapid expansion of artificial intelligence (AI), especially foundation models~\cite{foundation_models}, has fueled an unprecedented demand for computing resources, leading to a surge in the construction of large-scale data centers~\cite{EPRIreport}. These facilities are increasingly woven into the fabric of local communities, bringing both potential benefits and risks. On one hand, data centers can create jobs, generate tax revenues, and position regions at the forefront of the digital economy~\cite{datacenters_job}. On the other, they consume vast amounts of electricity, potentially stress limited water infrastructures if evaporative cooling is used, contribute to carbon emissions and local air pollution, and may strain local infrastructure or alter land use ~\cite{han2024unpaid, Wade2025electricity_grid, Shaolei_Water_AI_Thirsty_arXiv_2023_li2023making}.

Public perceptions of these trade-offs are sometimes framed in polarized terms: AI and data centers are either viewed as inherently ``bad'' or unequivocally ``good.'' Yet, real community sentiment is typically more nuanced and mixed. Many residents may simultaneously acknowledge the economic benefits while expressing concerns over environmental sustainability, health impacts, or long-term resilience. 

From the perspective of responsible data center deployment, it is often hard to take local voice into consideration in advance. The primary difficulty lies in capturing the complexity of public sentiment. Traditional polling requires significant financial and human resources~\cite{examine_survey_costs}. Most community feedback is only collected during public hearings, often after major commitments to a project have already been made. In some cases, the hearing process itself may be shaped by limited participation, sample bias, or incomplete information, resulting in feedback that does not fully represent the broader community.

A critical gap exists in mechanisms for obtaining early, scalable and diverse community input for socially responsible AI data center deployment. In this paper, we explore the use of foundational models---in particular, large language models (LLMs)---as AI agents to poll public opinion on data center projects, the physical homes of AI and foundational models. By leveraging the reasoning and world knowledge capabilities of LLMs~\cite{survey_llm_agents}, our approach introduces a scalable and cost-effective framework to approximate the breadth of perspectives that might emerge in community engagement. This framework enables stakeholders to rapidly screen community sentiment across diverse regions, identifying key concerns to inform socially responsible planning.

\begin{figure}[htbp!]
    \centering
    \includegraphics[width=0.9\linewidth]{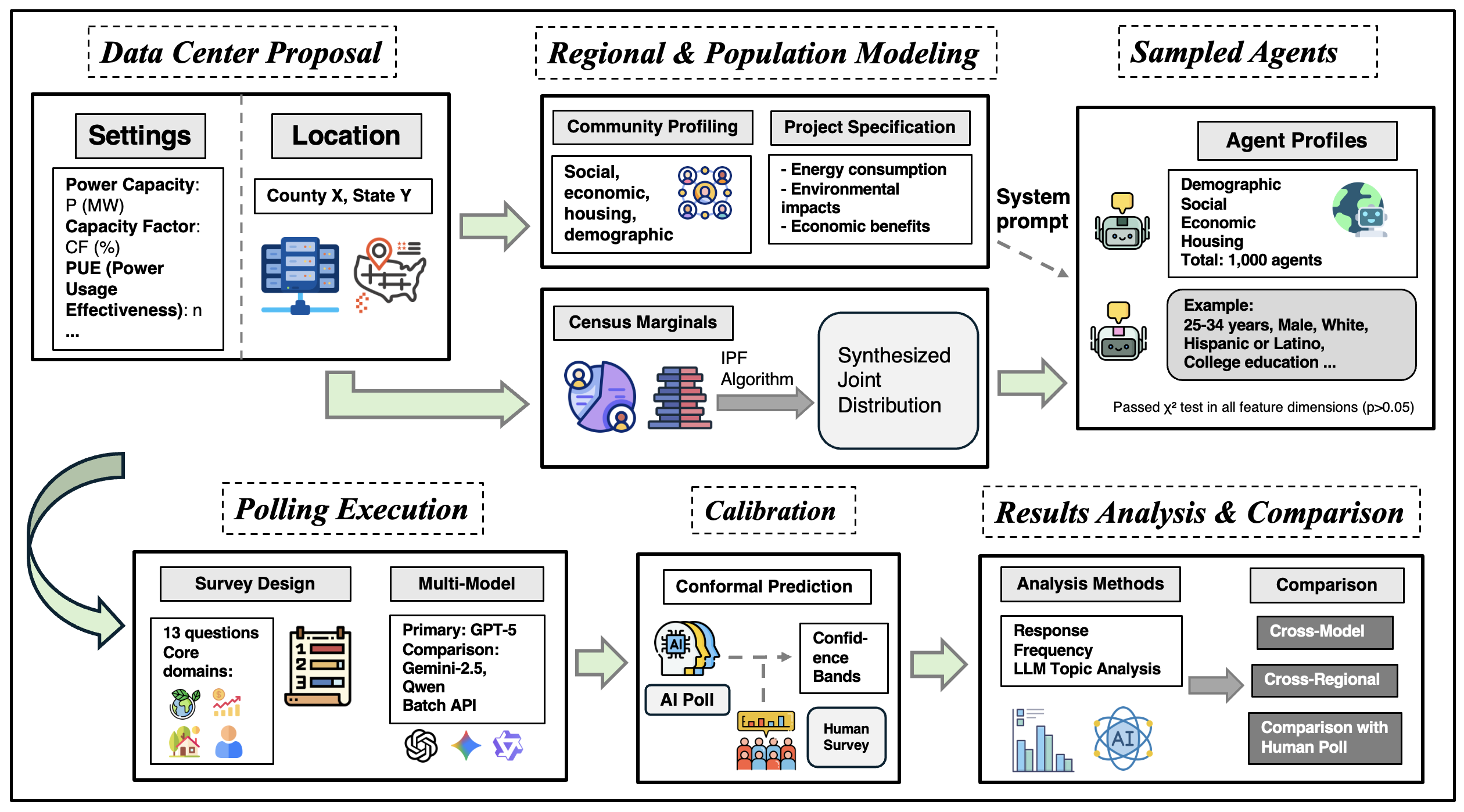}
    \caption{\textbf{AI agent polling framework for data center public opinion assessment}. The framework synthesizes county-level demographics with project specifications to generate representative virtual agents, validated through chi-square tests. Multi-model polling across GPT-5, Gemini-2.5-Pro, and Qwen-Max captures responses to 13 questions spanning 5 core domains, enabling cross-model, cross-regional, and human poll comparative analysis.}
    \label{fig:community-informed_AI}
\end{figure}

The AI agent polling framework for public opinion is shown in Figure ~\ref{fig:community-informed_AI}. Our methodology consists of six key stages. First, we establish a data center proposal. Then we generate representative AI agent samples using Iterative Proportional Fitting (IPF) from county-level demographics. Additionally, we model community profiling and project specifications, providing them as system prompts to the sampled agents. After that, we conduct detailed polling by using modern LLMs (e.g., GPT-5, Gemini-2.5-Pro, and Qwen-Max), covering multiple questions across diverse core domains. Results can then be calibrated with a small set of real-world polling data using conformal prediction to provide statistical guarantees. Finally, we perform multi-level analyses including cross-model, cross-regional, and human poll comparisons.

Our experiments in two distinct U.S. counties reveal several key findings. First, overall attitudes vary by region: Taylor County agents show higher support, 
while Loudoun County agents remain largely neutral. Second, there are some commonalities in specific items: water consumption, tax revenue, and utility bills emerge as top concerns or priorities for AI agents across regions. Third, agent opinions exhibit model-specific patterns. For example, Qwen agents stress economic factors and exhibit higher trust in government than their GPT-5 and Gemini-2.5 counterparts. Finally, while direct quantitative comparison is not feasible, AI agent responses show strong topical alignment with recent national polls on primary concerns and perceived benefits~\cite{heatmap_poll}.

\textbf{Disclosure.} \emph{We recognize that using AI agents to gauge public opinion may introduce significant challenges including, but not limited to, legitimacy, bias, and representativeness. Synthetic opinions generated by AI agents may not substitute for actual voices from the involved communities, and there is a potential risk that biased training data could skew results. Our goal is not to claim definitive measurements of public sentiment but to use a transparent approach that encourages earlier and more informed public discussions. }

\section{Related Works}

The prior research has extensively evaluated the environmental effects of AI data centers, including water consumption, health impact, grid impacts, and life cycle assessments~\cite{han2024unpaid, Wade2025electricity_grid, lifecycle_datacenter,Carbon_SustainbleAI_CaroleWu_MLSys_2022_wu2022sustainable}. Studies also examine the economic impacts of data centers~\cite{NVTC2024datacenter_impact, datacenters_job}. Beyond these macro-level analyses, recent research investigates local impacts of AI data centers %, including health effects, financial burdens, and water consumption~
\cite{Ngata2025cloud_next_door}. These research efforts provide valuable insights into the diverse consequences of AI infrastructure and establish foundations for understanding the impacts that communities may experience when hosting data center projects. 

{Public opinion polling} traditionally serves as the primary method for gauging local sentiment, involving surveys from representative samples~\cite{measuring_public_opinion}. Currently, the combination of fixed-line and mobile phone surveys remains one of the most widely used methodologies in the field~\cite{survey_methods_traditional}. With the rise of big data, data-augmented approaches like social media analysis have emerged~\cite{social_media_public_opinion_analyses, Shaik_2023}. Currently, data center constructors often rely on traditional polling methods or public hearing to gauge community feedback, typically implementing these efforts only during the late phase of construction.

The widespread adoption of foundation models enables new approaches to opinion research. Researchers can now conduct opinion polling through AI agents that leverage the world knowledge and reasoning capabilities of LLMs~\cite{ai_town, cot}. In political research, ~\cite{election_simulation} develops an LLM-based AI agent framework for election simulation. Similarly, prior studies demonstrate that LLM-powered interactive systems can extract themes from public discussions that align with formal reports ~\cite{policy_pulse}.
Regarding public engagement, \cite{Noveck2025governing_ai} suggests that AI could be utilized for evaluating public engagement processes due to its time and cost advantages.

While this research demonstrates the potential of AI agents for public opinion, the focus has seldom turned to the physical infrastructure enabling foundation models and artificial intelligence: the data centers themselves. The traditional engagement used by data center builders faces scalability and timeliness challenges, often failing to provide the early community feedback required for responsible planning. The maturation of LLM-based foundational models for AI agents~\cite{surveyllmbasedsystem} offers a cost-effective and scalable methodology to fill this specific gap. To our knowledge, this study is the first to apply the AI agent polling framework to assess public opinion on the boom of data centers.

\section{Methods}

The section presents the methodology for our community-based AI agent polling framework. The overall pipeline comprises four core components: regional context modeling, virtual agent construction, AI agent polling, and conformal prediction calibration. A ``community'' refers to a U.S. county, matching the level at which census data are provided. More details are provided in Appendix ~\ref{appendix}.

\subsection{Regional Context Modeling}
To ensure that AI agents make decisions that closely approximate real-world scenarios, we establish background information for both the data center project and the target community. It contains three components: state-level context, county-level profiling, and proposed data center project description.

For the state in which the target community is located, we incorporate state-level data center electricity consumption data from publicly available reports~\cite{EPRIreport, DOEreport}. The state-level data center electricity consumption provides each agent with a macro-level understanding of the current scale of data center operations in the region.

Since AI agents represent county-level residents, they must possess community relevance comparable to typical residents to improve accuracy. To achieve this local knowledge baseline, we implement a data acquisition and processing module that programmatically interfaces with the U.S. Census Bureau's 2023 American Community Survey 5-year estimates API~\cite{census_acs_5year} to obtain detailed population statistics for the target county.

Finally, we construct a standardized project profile for the proposed data center, which is provided to the AI agents as part of the prompt to ensure they are informed about the project. The profile is grounded in publicly available reports and industry standards, encompassing energy consumption, environmental impacts and economic implications.

These three components above, collectively constitute the ``Global Context'' provided to each AI agent to inform their responses.

\subsection{Virtual Agent Construction}
To facilitate representative results, we create AI agents that statistically mirror the demographic composition of target communities. This process involves two components: the acquisition of multi-dimensional demographic data and the sampling of the agent population. 

Firstly, we extract demographic distributions from the U.S. Census Bureau's 2023 American Community Survey (ACS)~\cite{census_acs_5year}. These encompass four categories~\cite{us_census_acs_2024}: social, economic, housing, and demographic characteristics, as detailed in Table~\ref{tab:census_variables}.

\begin{table}[!htbp]
  \scriptsize  \captionsetup{width=\linewidth}
    \caption{Census variables used in agent demographics. All characteristics are based on the U.S. Census Bureau's official categories~\cite{us_census_category}.}
    \renewcommand{\arraystretch}{1.5}
    \centering
    \begin{tabular}{lll}
    \toprule
    \textbf{Census Category} & \textbf{Agent Attribute} & \textbf{Classification} \\
    \hline
    \multirow{4}{*}{\begin{tabular}[c]{@{}l@{}}Social\\Characteristics\end{tabular}} 
    & Education Level & No degree, Associate's degree, Bachelor's degree, etc. \\
    \cline{2-3}
    & Marital Status & Never Married, Married, Divorced, etc. \\
    \cline{2-3}
    & Language at Home & English, Spanish, Other Indo-European languages, etc. \\
    \cline{2-3}
    & Citizenship & Native - born in state of residence, Native - born in different state, etc. \\
    \hline
    \multirow{2}{*}{\begin{tabular}[c]{@{}l@{}}Economic\\Characteristics\end{tabular}} 
    & Employment Status & Construction, Manufacturing, Wholesale trade, etc\\
    \cline{2-3}
    & Household Income & Less than \$10K, \$10K-\$15K, \$15K-\$25K, etc. \\
    \hline
    \multirow{2}{*}{\begin{tabular}[c]{@{}l@{}}Housing\\Characteristics\end{tabular}} 
    & Housing & Owner-occupied: Less than \$50K, Rent: Less than \$500, etc\\
    \cline{2-3}
    & Vehicles & No Vehicle, 1 Vehicle, 2 Vehicles, etc \\
    \hline
    \multirow{4}{*}{\begin{tabular}[c]{@{}l@{}} Demographic\\Characteristics\end{tabular}} 
    & Age Group & Under 5 years, 5 to 9 years, 10 to 14 years, etc \\
    \cline{2-3}
    & Sex & Male, Female \\
    \cline{2-3}
    & Race & Black or African American, Asian Indian, Asian, Japanese, etc. \\
    \cline{2-3}
    & Ethnicity & Hispanic, Non-Hispanic \\
    \bottomrule
    \end{tabular}
    \label{tab:census_variables}
\end{table}

The population statistics acquired previously provide only the marginal distributions for individual attributes. To sample realistic agents with correlated characteristics, we must first synthesize a joint probability distribution. We employ the Iterative Proportional Fitting (IPF) algorithm, a standard method in population synthesis~\cite{IPFreview}. The resulting agent population's demographic distribution is then verified against the Census data using chi-square goodness-of-fit tests (see Appendix~\ref{appendix:chi_square}).

\subsection{AI Agent Polling}
This process involves designing a structured questionnaire to capture public sentiment toward the proposed data center and implementing a scalable pipeline to execute surveys and analyze results.

Firstly, to ensure the survey effectively captures community sentiment, the questionnaire is designed to address key aspects of public opinion regarding data center development. 

The instrument contains 13 questions, comprising 12 single- and multiple-select items and one open-text question. These questions are structured around five core domains: (1) economic impacts, (2) environmental concerns, (3) community engagement, (4) anticipated personal impacts, and (5) overall project support. The open-text question solicits residents' primary concerns or messages regarding the proposed data center, enabling capture of nuanced perspectives that may not emerge through multi-choice questions. Full questionnaire is available in Appendix ~\ref{appendix:questionnaire}.

The subsequent step is to implement the pipeline for executing the survey. Our polling pipeline is an automated workflow that surveys AI agents and processes their responses. It operates in three stages.

First, to optimize for large-scale execution, the prompt is bifurcated. The static regional context, which includes state-level data, the county-level profile, and project specifications, is designated as the system message. The dynamic components, consisting of each agent's unique demographic profile and the survey questionnaire, form the user prompt. This separation enables the API provider to cache the constant system message~\cite{openai2024promptcaching}. Second, we leverage the batch APIs of the selected providers. We choose the latest LLM versions from OpenAI, Google, and Alibaba available as of September 30th, 2025. GPT-5 from OpenAI serves as the primary model throughout all experiments, while Gemini-2.5-Pro from Google and Qwen-Max from Alibaba are used for cross-model comparison to identify model-specific patterns. This asynchronous method facilitates the parallel processing of thousands of agent responses on the provider's infrastructure. Finally, the resulting structured dataset undergoes a two-pronged analysis. For the 12 multiple-choice questions, we perform a quantitative frequency analysis to compute response distributions. For the final open-ended question, we employ LLM-driven topic analysis to identify and quantify the emergent themes in the community's feedback.
    
\subsection{Conformal Prediction Calibration}
To obtain results with statistical guarantees and alignment with the real world, we leverage conformal prediction calibration~\cite{tutorialconformalprediction}. The workflow proceeds in two phases. First, a calibration phase uses a small set of real-world survey results ($y$) to compute nonconformity scores ($s = |y - \hat{y}|$) against the corresponding AI agent polling results ($\hat{y}$). This process determines a score threshold $\hat{q}$ based on a target confidence level $\alpha$. Second, a deployment phase uses this threshold $\hat{q}$ to construct a prediction interval (e.g., $[\hat{y}_{\text{new}} - \hat{q}, \hat{y}_{\text{new}} + \hat{q}]$) for new agent poll results, which is guaranteed to contain the true population probability with at least $1-\alpha$ probability. This component provides a principled pathway for establishing formal statistical guarantees on agent polling predictions. We note that while included for methodological completeness, this calibration step is not implemented in the current study due to the scope of this preliminary research and the lack of financial resources. However, we will compare our results with an existing national pool (Section~\ref{sec:comparison_human_polls}).

\section{Experiments}\label{main:exp}
Our experimental analysis comprises two components. First, we conduct a baseline case analyzing AI agent responses to a proposed data center in Taylor County, Texas, with cross-model comparison across three LLMs to identify model-specific variations. Second, we perform cross-regional comparison between Taylor County, Texas and Loudoun County, Virginia to assess how distinct demographic and economic contexts influence agent responses. Finally, we compare our AI agent results with recent human polling data to contextualize our findings. 

\textbf{Implementation scope.} This study implements five of the six methodological stages. The calibration component requires resources 
(e.g., funding, time, personnel) beyond this preliminary study. Validation relies on comparison with human polls (Section~\ref{sec:comparison_human_polls}).

\textbf{Basic settings} \quad We select Taylor County, Texas, as our baseline case, which is the home to one of the world's largest data centers. For cross-regional comparison, we contrast this with Loudoun County, Virginia, which hosts the largest concentration of data centers in the world \cite{Virginia_AirPermitsDataCenter_Report_JLARC_2024}. Each experiment polls 1,000 virtual agents per model or region responding to a hypothetical 100 MW data center proposal through a 13-question survey. In practice, some data centers built in these regions are significantly larger \cite{openai_stargate_oracle_2025}.

\textbf{Model} \quad We employ three popular LLMs representing diverse institutional and cultural contexts: OpenAI GPT-5, Google Gemini-2.5-Pro, and Alibaba Qwen-Max\footnote{Shortened for readability: GPT-5, Gemini-2.5, and Qwen.}. This selection enables cross-model validation and identifying model-specific biases stemming from different training data, institutional backgrounds, and cultural perspectives.

\textbf{API costs} \quad Our large-scale agent polling incurs total API costs of \$36.2 per run\,. Cost-optimization strategies reduce expenses by at least 50\% compared to standard usage. Model-specific costs are \$23.3 for GPT-5 (2,000 agents across baseline and regional comparisons), \$11.2 for Gemini-2.5 (1,000 agents), and \$1.7 for Qwen (1,000 agents). It usually takes more than 24 hours per run.

Detailed specifications, census demographics, data center configurations, and model parameters are provided in Appendix~\ref{appendix}.

\subsection{Baseline Case Analysis}
\subsubsection{Baseline Results}
\begin{figure}[htbp!]
    \centering
    \subfloat[Community attitudes]{ 
        %\raisebox{4mm}
        {\includegraphics[height=0.13\textwidth, valign=b]{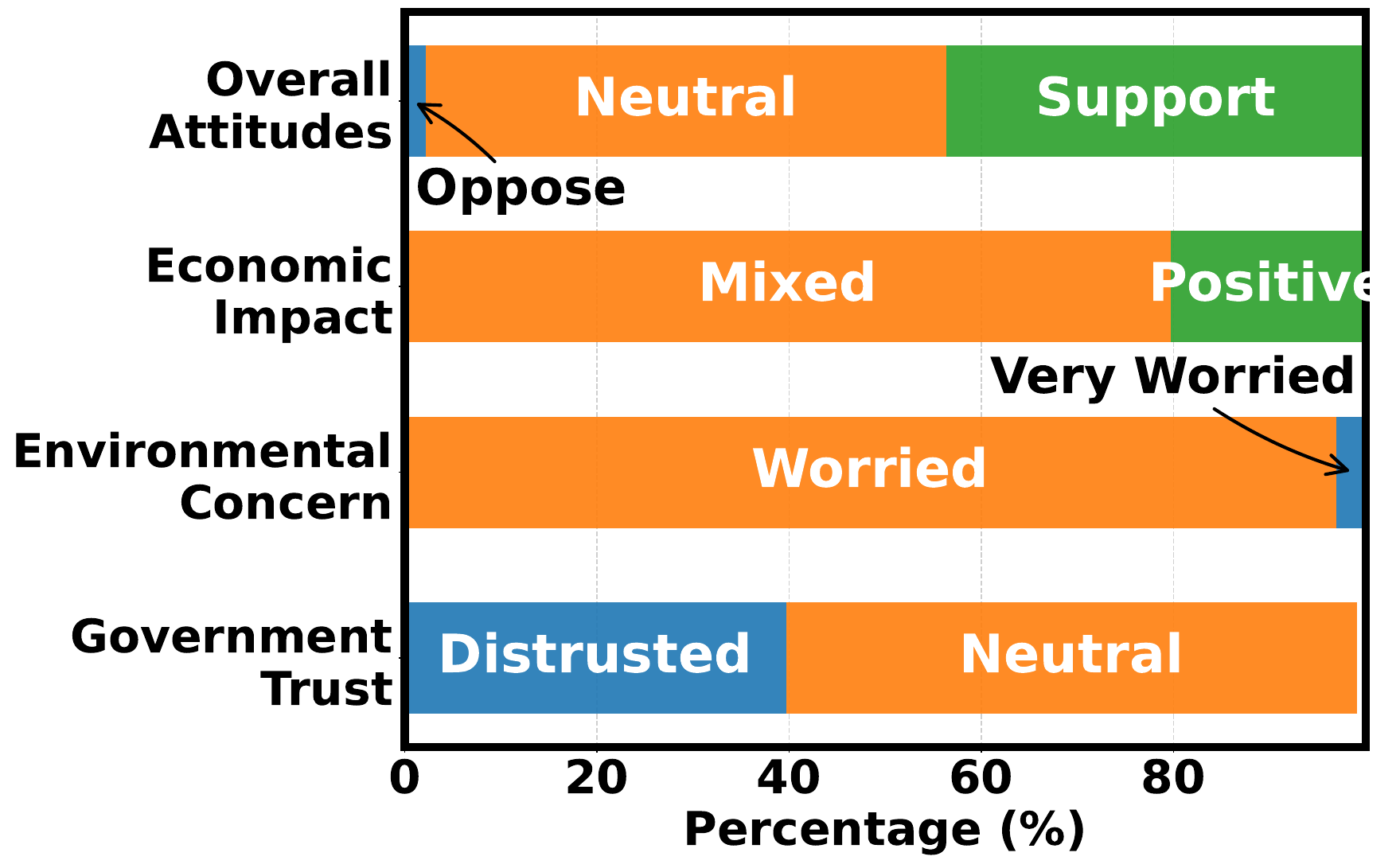}\label{fig:community_attitudes}
        }
    }
    \subfloat[LLM topic analysis]{
        %\raisebox{4mm}
        {\includegraphics[height=0.13\textwidth, valign=b]{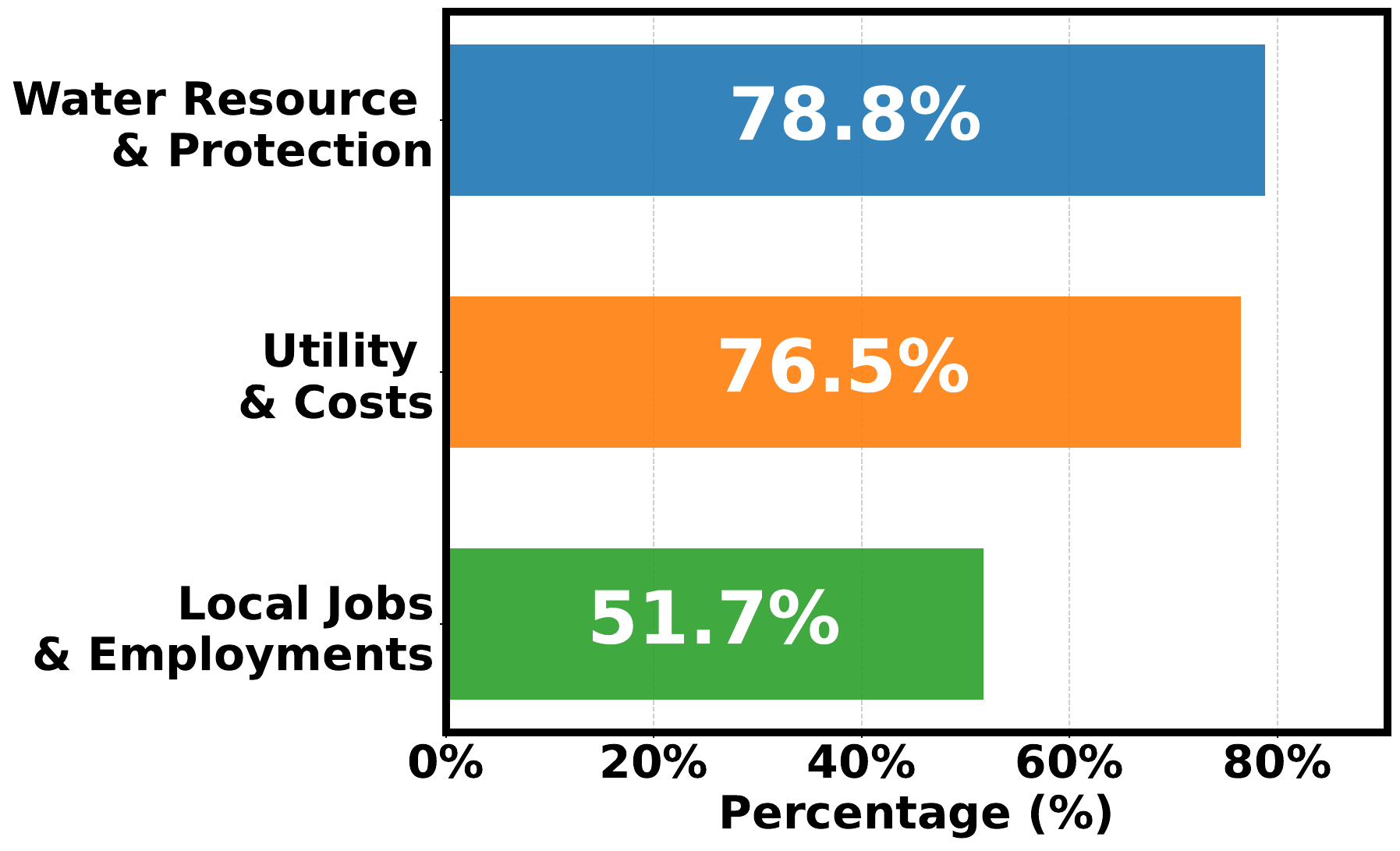}\label{fig:topic_analysis}
        }
    }
    \subfloat[Detailed topic analysis breakdown]
    {\includegraphics[height=0.13\textwidth, valign=b]{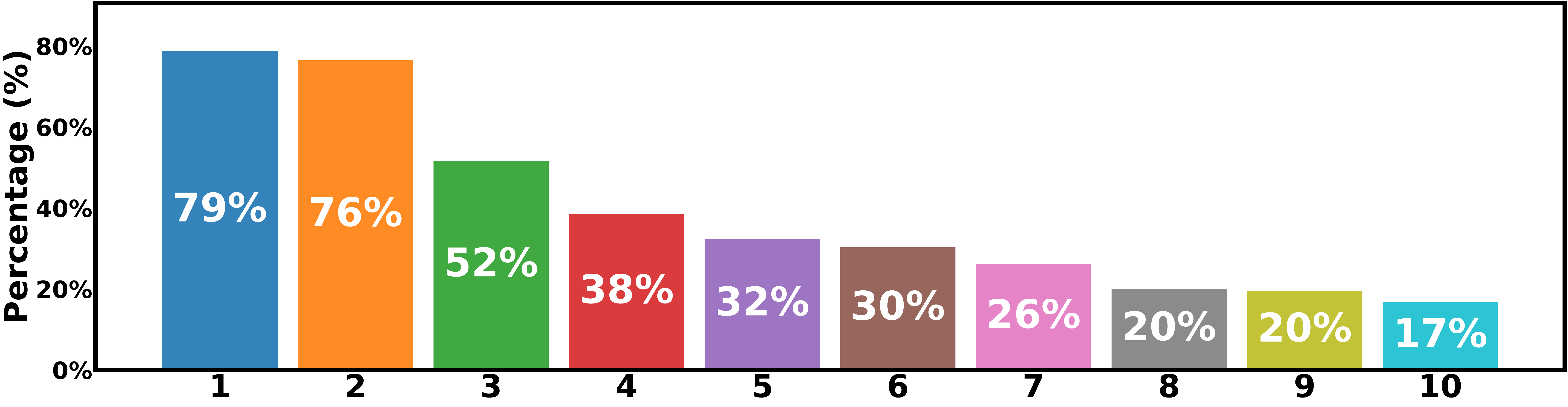}\label{fig:top_analysis_full}
    }

    \caption{\textbf{Taylor County results using GPT-5 ($n$=1000)}. (a) Community attitude distribution showing neutral-leaning support attitudes, mixed economic impact perception, widespread environmental concerns, and neutral government trust. (b) LLM topic analysis of open-ended community feedback revealing top-three themes. (c) Full topic breakdown (1. Water Resource Protection, 2. Utility Costs, 3. Local Jobs \& Employments, 4. Clean Energy, 5. Economic Benefits, 6. Transparency \& Public Reporting, 7. Accountability \& Enforcement, 8. Grid Impact \& Reliability, 9. Housing Costs, 10. Taxes \& Public Finance).  Charts display only selected response categories, and complete survey options are in Appendix ~\ref{appendix}.}
    
    \label{fig:taylor_county_results}
\end{figure}

Our baseline analysis presents an overview of general attitudes from our AI agent polling in Taylor County as shown in Figure~\ref{fig:taylor_county_results}.

In terms of overall position, slightly over half (54.2\%) agents express neutral attitudes toward the data center, while nearly half (43.6\%) show positive attitudes. Regarding economic impacts, 80\% of agents view the economic effects as mixed, while 20\% perceive them as positive. A high level of environmental concern is also evident, with 97\% of agents expressing they are worried. On the topic of government oversight, most agents (60\%) maintain neutral attitudes, while a 40\% express distrust. An analysis of open-ended responses further shows that the top three topics of concern are water resource \& protection, utility costs, and local jobs \& employments. 

\subsubsection{Cross-Model Comparison}

\begin{figure}[htbp!]
    \centering
    \subfloat[Economic attitudes]{\includegraphics[height=0.13\textwidth, valign=b]{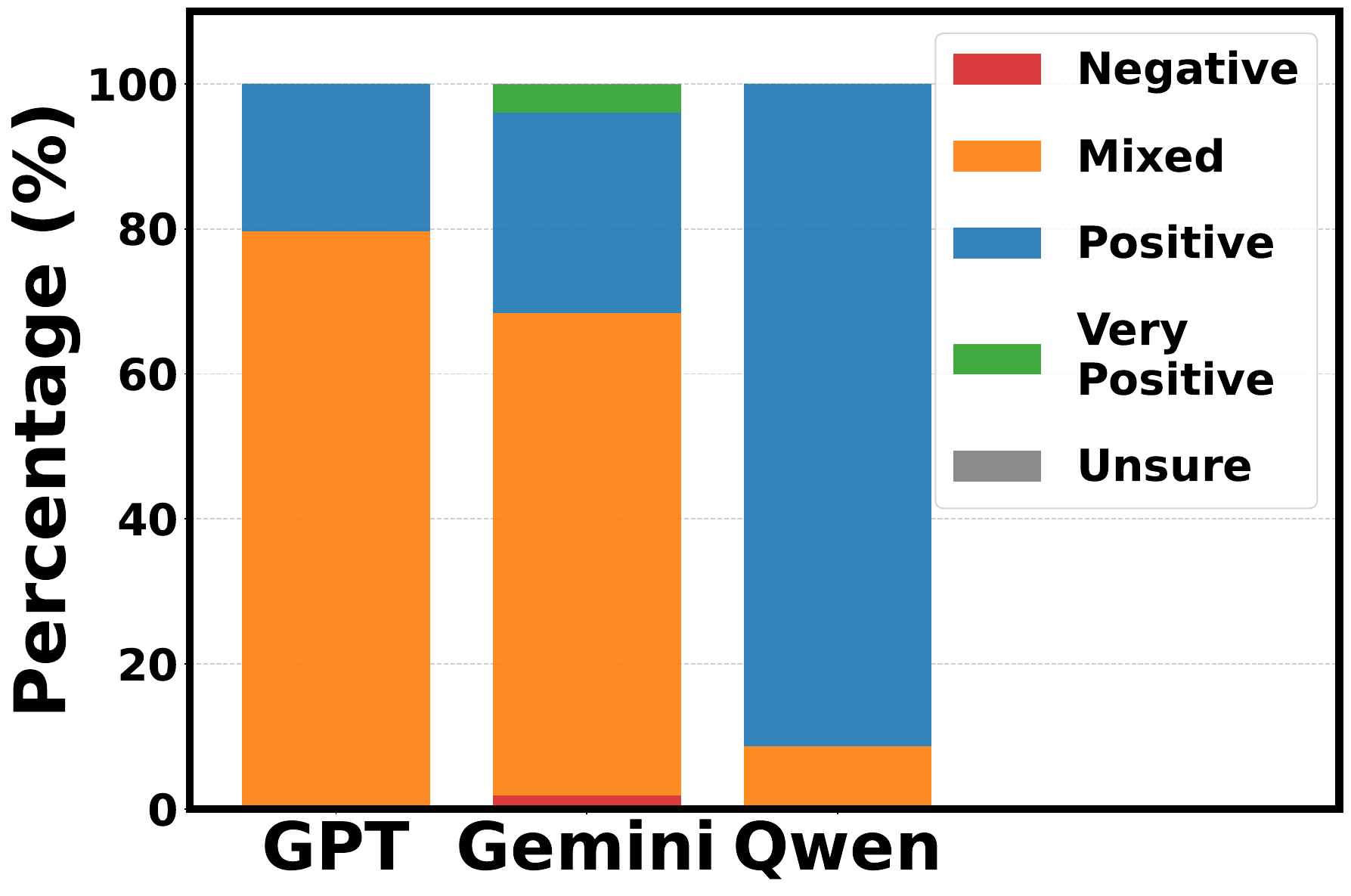}
    }
    \subfloat[Top-3 economic benefits]
    {\includegraphics[height=0.13\textwidth, valign=b]{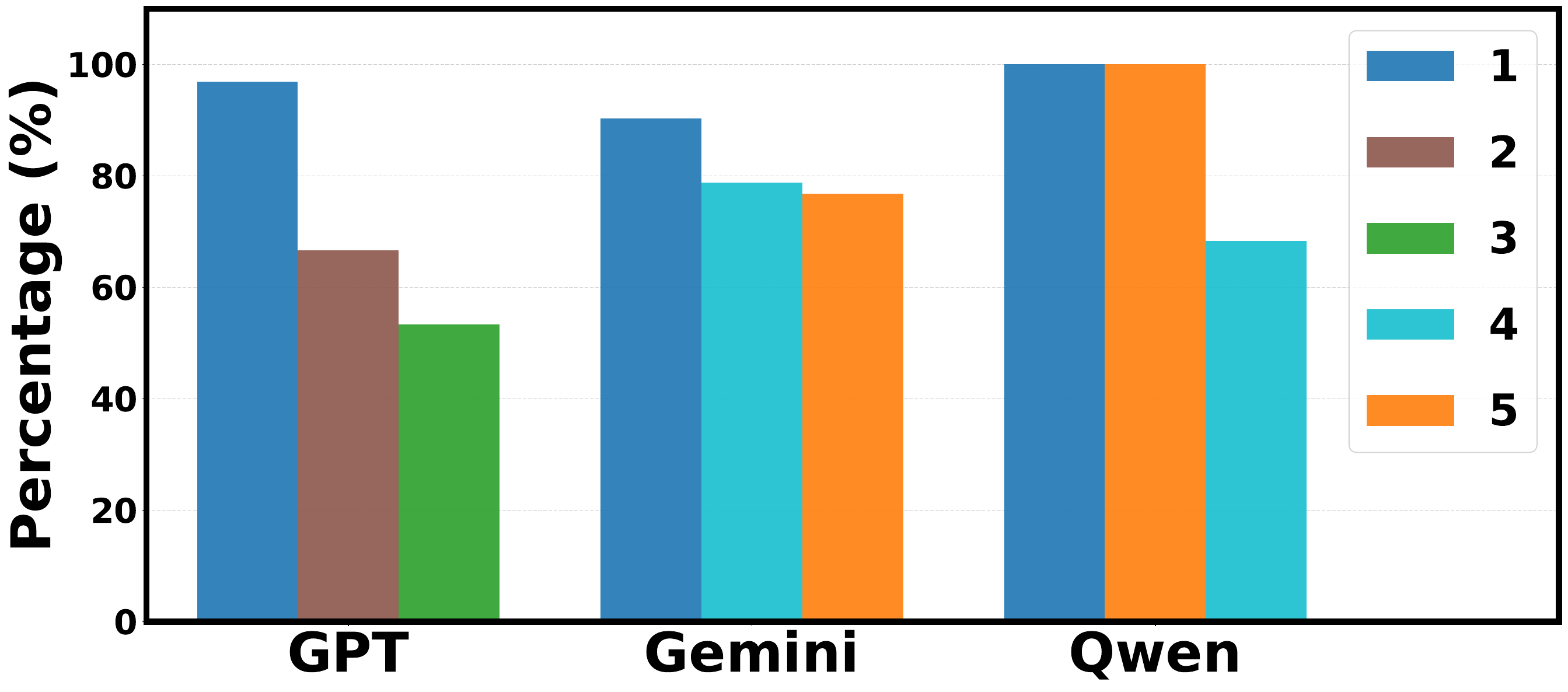}\label{fig:economic_priorities_models}
    }
    \subfloat[Government trust]{\includegraphics[height=0.13\textwidth, valign=b]{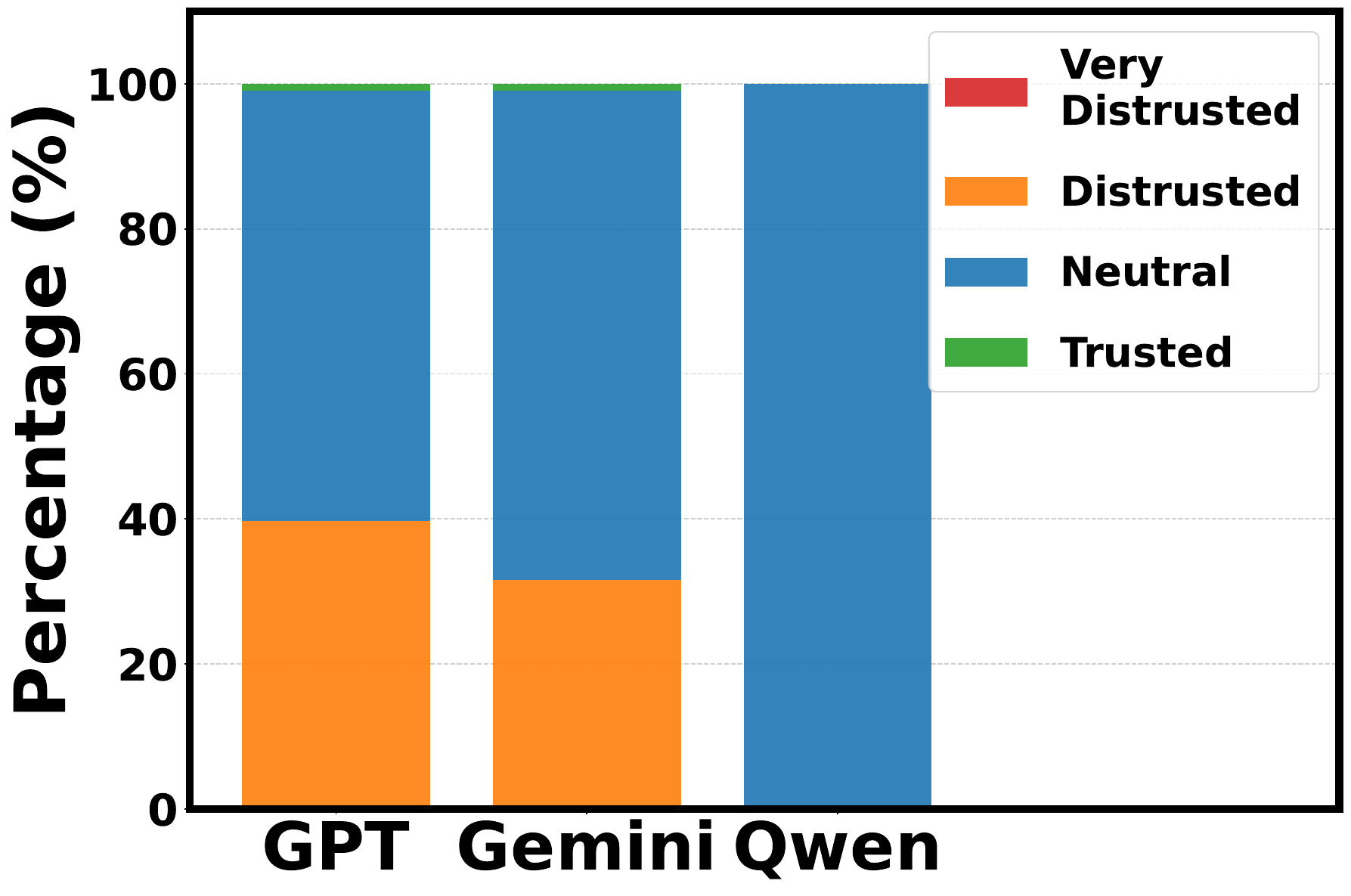}
    }
    \subfloat[Top-3 information source]
    {\includegraphics[height=0.13\textwidth, valign=b]{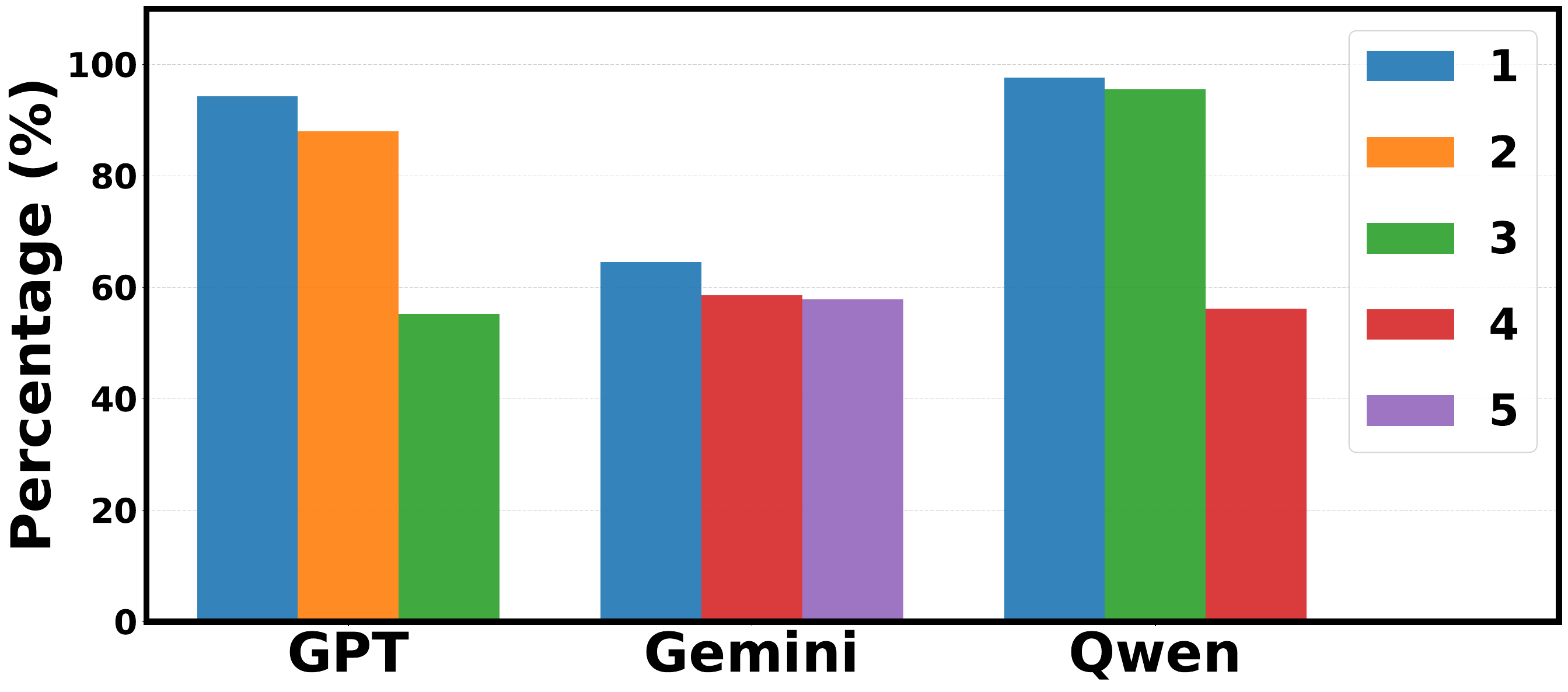}
    }
    \caption{\textbf{Key differences in cross-model polling results ($n$=1000)}. (a) Overall economic attitudes; (b) Top economic benefits (1: Tax Revenue, 2: Infrastructure Upgrades, 3: Business Growth, 4: Economic Diversity, 5: Job Creation); (c) Government trust; (d) Top trustworthy information sources (1: Academic Research, 2: Federal/State Agencies, 3: Local Government, 4: Community Organizations, 5: Local Media). Note: Selected categories; see Appendix ~\ref{appendix} for complete data.}
    \label{fig:cross_model_differences}
\end{figure}

We conduct identical experiments in Taylor County using Gemini and  Qwen to assess how polling results vary across different  LLMs. While all models identify similar underlying patterns regarding data center development, such as a high level of environmental concern, notable model-specific variations emerge in certain issues as shown in Figure~\ref{fig:cross_model_differences}:

\textbf{Economic issues}: Agents from GPT-5 and Gemini-2.5 primarily view the project's economic impacts as mixed, whereas Qwen's agents are notably more optimistic, with 91\% expressing positive attitudes. This difference is explained by their economic priorities; GPT-5 and Gemini-2.5 agents have relatively diverse preferences, while Qwen agents almost uniformly prioritize tax revenue and job creation. This divergence likely reflects different economic development philosophies embedded in the LLMs’ training data. For instance, Qwen's uniform focus on concrete economic outcomes may reflect training data influenced by development-oriented economic paradigms where large-scale infrastructure projects are viewed primarily through the lens of measurable economic advancement. The more varied responses from GPT-5 and Gemini-2.5 may represent training data incorporating diverse concerns.

\textbf{Governance and preferred information sources}: In terms of regulation, both GPT-5 and Gemini-2.5 exhibit substantial distrust in the government's regulatory capacity (40\% and 32\%, respectively), while nearly no agents from Qwen express distrust. This pattern extends to preferred information sources. Although academic research is a top choice for all models, Qwen's agents show a much stronger preference for local government as a trusted source compared to the other models. These differences may stem from the varying cultural and institutional contexts represented in each model's training data. For example, the high government trust shown by Qwen might be influenced by data from societies with strong state institutions, while the skepticism from GPT-5 and Gemini-2.5 could be more aligned with cultural contexts that emphasize independent oversight and academic validation.

\subsection{Cross-Regional Analysis}
To explore how the regional context influences public attitudes, we conduct a comparative analysis between Loudoun County, Virginia, known as ``Data Center Alley,'' and Taylor County, Texas. Our analysis focuses on overall attitudes and economic impacts. While some common patterns emerge, we find and explain several notable differences between the two regions. These findings demonstrate that our framework's outputs are location-specific, highlighting its potential utility in the site-selection process for data center projects.

Agents in Taylor County, Texas, show notably higher support for the proposed data center project than those in Loudoun County, Virginia (43.6\% vs 9.7\%). Regarding the specific conditions that would make agents more supportive, environmental protection emerges as a top consideration for nearly all agents in both Taylor and Loudoun counties. However, agents in Taylor County show a stronger preference for economic benefits, with 94\% selecting lower utility bills and about 51\% choosing local job guarantees. This emphasis may be attributable to two factors: Taylor County has a lower median household income, and its hot climate drives up air-conditioning usage, making residents particularly sensitive to the prospect of lower utility bills. Furthermore, as a smaller county with more limited job opportunities compared to Loudoun, the prospect of new employment and corresponding industry upgrades is more attractive to its residents. In contrast, Loudoun County agents prioritize governance, with approximately 90\% selecting stricter oversight as a condition for their support. This focus likely stems from the community's extensive experience with data center development, which may have led to a greater awareness of the importance of refined oversight rules and procedures. Additionally, the higher level of trust in government in Loudoun County may also contribute to this preference, which is a topic discussed in the following section.

Regarding the perceived economic impacts, agents in Taylor County express more positive attitudes, with 20\% viewing the project's effects as positive. This may be because agents representing Taylor County residents attach greater importance to job creation and related industry opportunities than their Loudoun County counterparts. Besides economic benefits, the data center project may lead to some economic issues. Notably, communities in both counties identify higher utility bills as their top economic worry. A finding unique to Loudoun County is a high level of concern about public service strain. This reflects the county's existing experience, as Loudoun already hosts numerous data centers that have strained local infrastructure and public services. This finding is consistent with real-world scenarios in the region. According to a report of Loudoun County Supervisor\cite{loudoun2025datacenter}, the rapid proliferation of data centers has reportedly placed essential public services like the local power grid under severe stress and generated public resistance to new development. 

\begin{figure}[htbp!]
    \centering
    \subfloat[Overall attitudes]{\includegraphics[height=0.15\textwidth, valign=b]{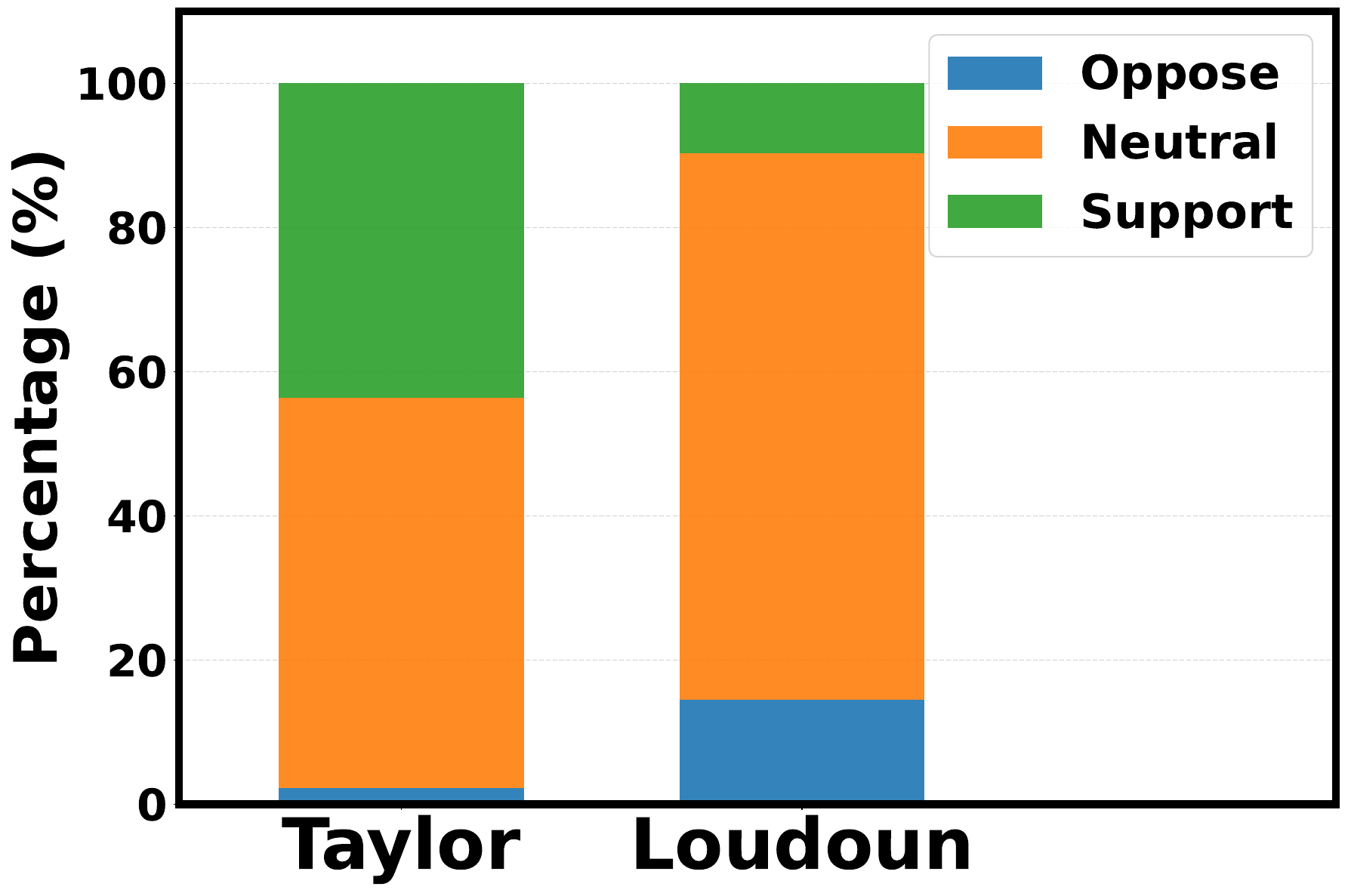}
    }
    \subfloat[Top-3 support condition]
    {\includegraphics[height=0.15\textwidth, valign=b]{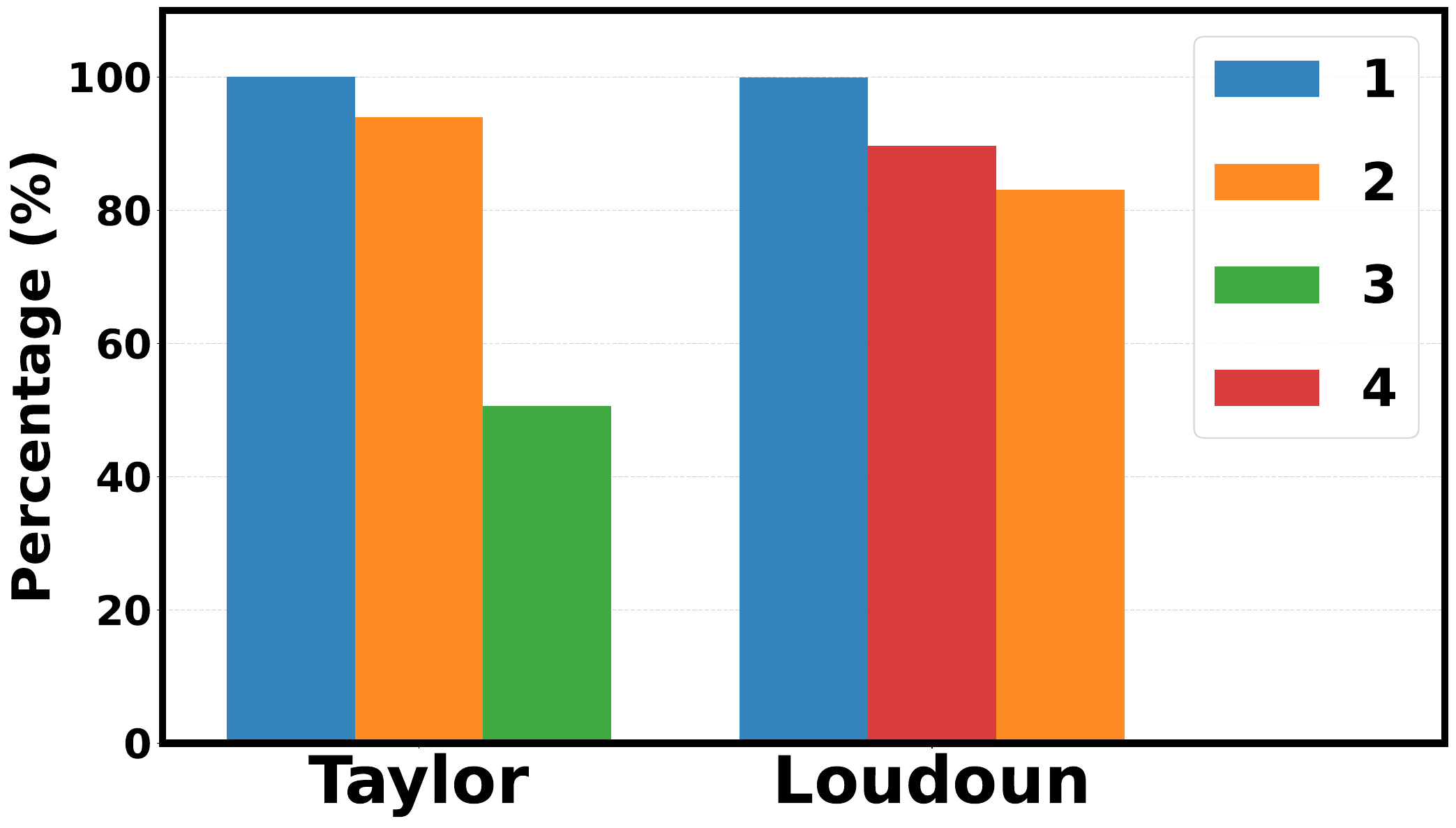}
    }
    \subfloat[Economic attitudes]{\includegraphics[height=0.15\textwidth, valign=b]{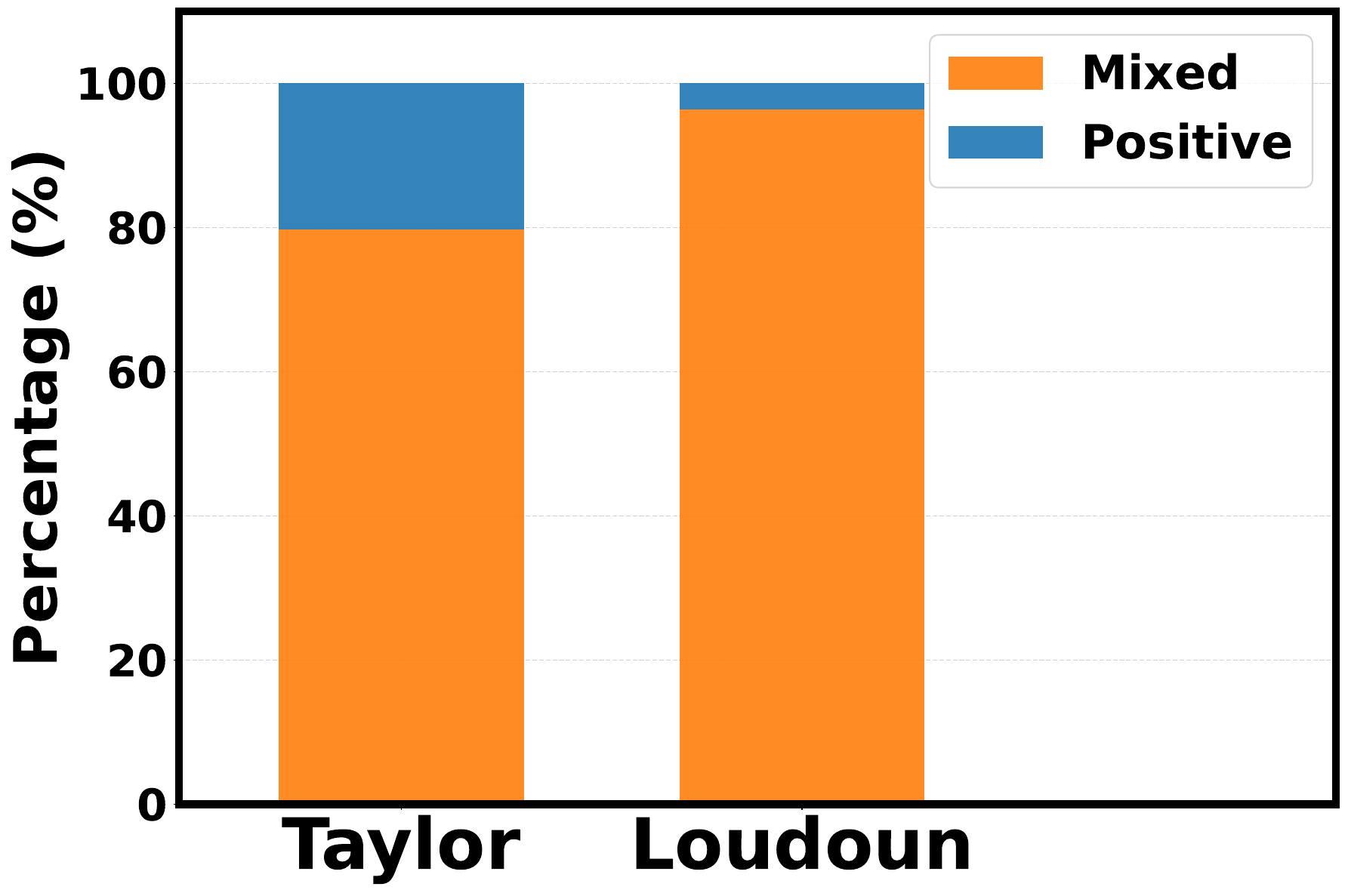}
    }
    \subfloat[Top-3 economic concerns]
    {\includegraphics[height=0.15\textwidth, valign=b]{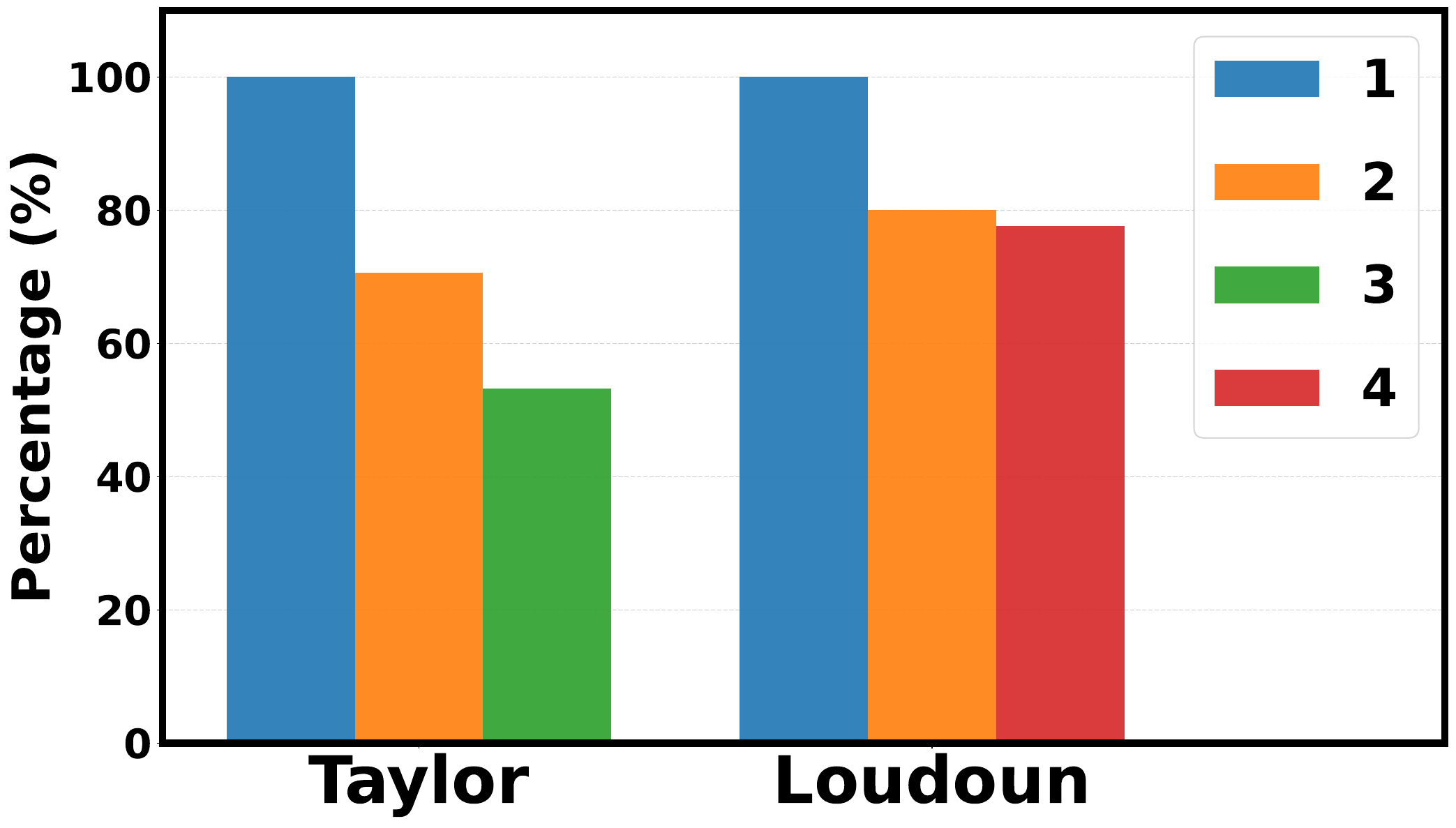}
    }
    \caption{\textbf{Cross-regional polling results ($n$=1000)}. (a) Overall attitudes for proposal data center. (b) The top conditions that would increase AI agents' support for the project (1: Environmental Protections, 2: Lower Utility Bills, 3: Local Job Guarantees, 4: Stricter Oversight). (c) Overall economic attitudes towards the project. (d) The community's top economic concerns regarding the proposed data center (1: Higher Utility Bills, 2: Benefits to Outsiders, 3: Housing Cost Inflation, 4: Public Service Strain). Note: Selected categories; see Appendix ~\ref{appendix} for complete data.}
    \label{fig:cross_region_overall_attitude}
\end{figure}

\subsection{Comparison with Real-World Human Polls} \label{sec:comparison_human_polls}
To contextualize our agent polling results, we compare them with a recent human poll by Heatmap News~\cite{heatmap_poll}. However, significant methodological differences (e.g., surveyed population and instrument design) render a direct quantitative comparison inappropriate. Accordingly, this section first details these key differences to %transparently 
acknowledge the inherent limitations of the comparison. Subsequently, we conduct a qualitative analysis, comparing the findings from our AI agent experiments with those of the human poll.

\subsubsection{Methodological Differences}\label{method_differences}
Firstly, our study focuses on localized public sentiment at the county level, where agent responses are grounded in specific demographic and economic contexts. In contrast, the Heatmap poll has a national scope, surveying voters across all 50 states and Washington, D.C.. Secondly, our research focuses exclusively on a specific data center proposal, whereas the Heatmap survey contextualizes data centers by comparing their public acceptance against other energy infrastructures and analyzes sentiments along political affiliations. A third distinction lies in the questionnaire design. Our survey instrument is more granular, structured around core domains including economic, environmental, and governance issues. The Heatmap survey is more concise, gauging overall support or opposition and evaluating common arguments for each stance. Crucially, the response options differ; for instance, our survey includes ``Neutral'' as a distinct middle option for overall attitude, an option absent from the Heatmap poll's.

Considering these differences, we conduct the following comparison. We examine common patterns identified in our cross-regional analysis against the national results of the Heatmap poll. Our comparison focuses on similar questions while excluding unrelated survey items. Furthermore, due to differences in response options, we conduct qualitative rather than quantitative analysis.

\subsubsection{Topical Comparison and Findings}
Table ~\ref{tab:human_ai_comparison} summarizes the key topical alignments and differences. It should be noted that the quantitative figures shown are for illustrative purposes within their respective contexts and are not directly comparable, given the significant methodological distinctions previously discussed.

Regarding overall attitudes toward data centers, our county-level experiments demonstrate substantial geographic heterogeneity, with net support of approximately 41\% in Taylor County, Texas, versus -5\% in Loudoun County, Virginia. The national Heatmap poll, by comparison, shows net support of approximately 2\%. Although the two approaches are not directly comparable due to differences in geographic scope, they may nonetheless reveal a consistent underlying pattern: while national-level polling reflects broadly neutral attitudes toward data centers, actual support appears highly location-dependent, with significant local variation.

Regarding expected benefits brought by data centers, the Heatmap poll identifies tax revenue and job creation as the most popular reasons. Our experimental results rank tax revenue and infrastructure upgrades as the most important economic benefits among AI agents. A similar pattern can be seen: tax revenue is highly selected by both human and AI agents. The divergence lies in the other priority: the human poll indicates a preference for job creation, while the AI agents prioritize infrastructure upgrades. This difference is likely attributable to the survey design. Our study provides a more granular set of choices, which allows agents to prioritize 'infrastructure upgrades' as a distinct benefit. This specific option is not available in the Heatmap poll, whose list of potential benefits focuses on broader arguments such as 'create high-paying jobs' and 'power digital economy'. It is plausible that the human poll's results would have shown greater similarity to our findings had these more detailed options been available.

A strong alignment is also evident regarding the arguments for opposing data centers. The Heatmap poll finds that ``water usage'' and ``electricity usage'', which can lead to higher utility costs, are the most persuasive to the public. Our experiments reveal similar patterns, identifying potential water consumption as a primary environmental worry and higher utility bills as the top economic concern among AI agents. As mentioned by Heatmap, these results are consistent with real-world cases. For example, residents in Tucson, Arizona, have expressed opposition to a proposed data center project due to concerns over water consumption~\cite{tucson_datacenter_water}.

\begin{table}[!htbp]
%\tiny
\scriptsize
    \captionsetup{width=\linewidth}
    \caption{Comparison between human poll and AI agent polling}
    \renewcommand{\arraystretch}{1.5}
    \centering
    \begin{tabular}{l|l|l|l}
    \toprule
    \textbf{Metric} & \textbf{Heatmap Poll} & \multicolumn{2}{c}{\textbf{AI Agent Polling (County-Level)}} \\
     & \textbf{(National Voters)} & \textbf{Taylor County, TX} & \textbf{Loudoun County, VA} \\
    \hline
    \begin{tabular}[c]{@{}l@{}}Overall\\Attitudes\end{tabular} 
    & \begin{tabular}[c]{@{}l@{}}Net support +2\%\\(44\% support vs. 42\% oppose)\end{tabular} 
    & \begin{tabular}[c]{@{}l@{}}Net support +41\%\\(Support 44\%,\\Oppose 2\%)\end{tabular}
    & \begin{tabular}[c]{@{}l@{}}Net support -5\%\\(Support 10\%,\\Oppose 15\%)\end{tabular} \\
    \hline
    \begin{tabular}[c]{@{}l@{}}Benefits\end{tabular} 
    & \begin{tabular}[c]{@{}l@{}}Tax revenue \&\\Create high-paying jobs\end{tabular} 
    & \multicolumn{2}{c}{\begin{tabular}[c]{@{}c@{}}Tax revenue \&\\Infrastructure upgrades\end{tabular}} \\
    \hline
    \begin{tabular}[c]{@{}l@{}}Concerns\end{tabular} 
    & \begin{tabular}[c]{@{}l@{}}Water usage \&\\Electricity usage\end{tabular} 
    & \multicolumn{2}{c}{\begin{tabular}[c]{@{}c@{}}Water consumption \&\\Higher utility bills\end{tabular}} \\
    \bottomrule
    \end{tabular}
    
    \vspace{0.3cm}
    \begin{flushleft}
    \footnotesize
    \textit{Note:} Net support is calculated as the sum of all support levels minus the sum of all opposition levels. Methodological differences (Section ~\ref{method_differences}) mean percentage values reflect different contexts and should be interpreted thematically rather than as direct equivalents. Concerns of AI agent polling represent top environmental (water) and economic (utility bills) issues.
    \end{flushleft}
    \label{tab:human_ai_comparison}
\end{table}

\section{Conclusion and Discussion}\label{main：conclusion}
We propose a novel and scalable method to assess community feedback on data center projects, guiding the siting and design of responsible AI data centers. We introduce an AI agent polling framework that leverages foundation models (i.e., LLM) to gauge nuanced and mixed public attitudes toward these facilities. This approach is beneficial for responsible AI infrastructure deployment because it can allow the local community voice to be taken into consideration in a scalable way. Our results show key concerns and priorities within the targeted community. Cross-regional analysis demonstrates that the framework's results reflect distinct local contexts, revealing its potential to inform data center site selection. Furthermore, comparison with human polls confirms a clear topical alignment between our experimental results and real-world survey data. 

\textbf{Limitations.} While our methodology includes conformal prediction calibration, its implementation requires (small) real-world survey data and financial resources beyond the scope of this study and our budget. Our current validation relies on comparison with existing national polls (Section~\ref{sec:comparison_human_polls}). Our analysis focuses on two counties with existing data center infrastructure; applicability to other community contexts warrants further investigation. The framework also has known boundaries. The AI agents model a generalized cognitive response and may not fully capture specific human nuances, such as neurodiverse perspectives or the phenomenon of ``engagement silence.'' Research also shows that LLMs exhibit inherent bias, such as racial biases~\cite{llmbi, ling2025bias, dialectpred, breakingbias, inherentbiasllm, llmbias}. 

\textbf{Future work.} Our research opens up multiple  directions for future work, including implementing the conformal prediction calibration with real-world surveys, extending this framework to other infrastructure domains, developing site selection algorithms that incorporate community sentiment, simulating human deliberative processes and social interactions using multi-agent systems, and establishing hybrid approaches that combine AI polling with traditional engagement methods.

\section*{Acknowledgement}
This work was supported in part by the U.S. NSF under
the grants
 CCF-2324941, CNS-2326598, and CCF-2324916.

\newpage
\bibliographystyle{plain}

\appendix 
\section*{Appendix} \label{appendix}
\section{Detailed Experimental Settings}
\subsection{Regional Context Modeling}
To construct the region context required for simulation, we integrate data from multiple sources. All demographic and statistical data used are based on 2023 estimates.

\subsubsection{State-level Context}
The electricity consumption figures, are obtained from publicly available reports by the Electric Power Research Institute (EPRI) and lawrence berkeley national laboratory (LBNL)~\cite{EPRIreport, DOEreport}. 

\subsubsection{County-level Profiling}
Detailed profiles of county-level communities are generated programmatically through the American Community Survey (ACS) 5-Year Data Profile API provided by the U.S. Census Bureau~\cite{census_acs_5year}. The variable codes we use for the API calls are listed below. Note that the demographic variables here are high-level summaries for a concise community profile; a more granular set of variables is used for the agent construction.

\begin{itemize}
    \item \textbf{Demographic Variables} \quad DP05\_0001E (Total population), DP05\_0002E (Male), DP05\_0003E (Female), DP05\_0018E (Median age), DP05\_0069E (White), DP05\_0070E (Black or African American), DP05\_0071E (American Indian and Alaska Native), DP05\_0072E (Asian), DP05\_0073E (Native Hawaiian and Other Pacific Islander), DP05\_0074E (Other race), DP05\_0076E (Hispanic or Latino), DP05\_0081E (Not Hispanic or Latino).
    \item \textbf{Social Variables} \quad DP02\_0001E (Total households), DP02\_0016E (Average household size), DP02\_0060E to DP02\_0066E (Educational attainment), DP02\_0153E (Households with a computer).
    \item \textbf{Economic Variables} \quad DP03\_0008E (Civilian labor force), DP03\_0033E to DP03\_0045E (Employment by industry), DP03\_0006E (Armed forces), DP03\_0062E (Median household income), DP03\_0088E (Per capita income).
    \item \textbf{Housing Variables} \quad DP04\_0045E (Occupied housing units), DP04\_0046E (Owner-occupied units), DP04\_0047E (Renter-occupied units), DP04\_0089E (Median value of owner-occupied housing units), DP04\_0134E (Median gross rent).
\end{itemize}

\subsubsection{Proposed Data Center Project Description}
The proposed data center project is designed as a specific and standardized case study, with the following core technical specifications:
\begin{itemize}
    \item \textbf{Power Specifications} \quad Rated capacity of 100 MW, with a Capacity Factor of 70\%. The power usage effectiveness (PUE) is set to 1.1, based on values from Google's data center operations~\cite{google_datacenter_efficiency}. This represents a conservative estimate given the absence of state-specific PUE data. The annual data center energy consumption is calculated as: rated capacity × capacity factor × annual hours × PUE.
    
    \item \textbf{Environmental Impacts} 
    \begin{itemize}
        \item Carbon emissions are calculated by multiplying the facility's annual energy consumption by the state-specific emission factor from the U.S. Energy Information Administration~\cite{eia_state_emissions_2025}.

        \item Air pollution estimates for nitrogen oxides (NO$_x$), volatile organic compounds (VOCs), particulate matter (PM$_{2.5}$), and sulfur dioxide (SO$_2$) are calculated by multiplying data center energy consumption by emission intensities, which are derived from data of diesel backup generators at Virginia data centers~\cite{virginia_deq_data_centers_2024, piedmont_environmental_council_2024, han2024unpaid}. Based on permitted annual emission limits for Northern Virginia data centers, the total allowable emissions are 13,000 tons of NO$_x$, 1,400 tons of VOCs, 50 tons of SO$_2$, and 600 tons of PM$_{2.5}$. We assume actual emissions represent 10\% of the permitted limits. The resulting emission quantities are then divided by the total annual energy consumption of data centers in the region to derive pollutant-specific emission intensity factors.
        
        \item Water consumption encompasses both on-site and off-site components. For on-site part, the water usage effectiveness (WUE) of 0.36 L/kWh, representing the national average from Lawrence Berkeley National Laboratory~\cite{DOEreport}, is multiplied by the annual IT energy consumption (calculated as total facility energy consumption divided by PUE). For off-site consumption, the electricity water intensity factor (EWIF) of 3.14 L/kWh from the World Resources Institute~\cite{reig2022guidance} is multiplied by the total facility energy consumption.
    \end{itemize}
    \item \textbf{Economic Impacts} To estimate the economic impacts of the data center, we reference a case study by the U.S. Chamber of Commerce~\cite{pham2017datacenters}. During the construction phase (18-24 months), the project supports approximately 1,700 temporary local jobs, and generates around \$240 million in local economic activity and \$10 million in taxes. Once operational, it supports nearly 160 permanent local jobs annually, with an average salary of about \$50k, and contributes over \$32 million in local economic activity and \$1.1 million in taxes each year. Additionally, we draw upon ~\cite{nguyen2025datacenterstown} to ensure objectivity of our prompt design.
\end{itemize}

\subsection{Virtual Agent Construction}
To ensure consistency and control across experiments, generated agents are saved and reused for all simulations pertaining to that specific community.
\subsubsection{Demographic Data Acquisition}
To construct AI agents that statistically mirror real residents, we acquire detailed demographic distributions from the U.S. Census Bureau's American Community Survey (ACS) 5-Year API~\cite{census_acs_5year}.

The variable codes used for the API calls are listed below.
\begin{itemize}
    \item \textbf{DP05: Demographic Variables} \quad
    Age distribution (DP05\_0005E--DP05\_0017E covering 13 age groups from "Under 5 years" to "85 years and over"), sex distribution (DP05\_0002E--DP05\_0003E for male and female), detailed race categories (DP05\_0037E--DP05\_0067E including single race and multiracial combinations), and ethnicity classification (DP05\_0076E, DP05\_0081E for Hispanic/Latino status).
    \item \textbf{DP02: Social Variables} \quad
    Citizenship status (DP02\_0091E--DP02\_0097E covering native-born and foreign-born categories), language spoken at home (DP02\_0113E--DP02\_0122E for English, Spanish, and other language groups), educational attainment (DP02\_0060E--DP02\_0066E from "Less than 9th grade" to "Graduate or professional degree"), and marital status by gender (DP02\_0026E--DP02\_0030E for males, DP02\_0032E--DP02\_0036E for females).
    
    \item \textbf{DP03: Economic Variables} \quad
    Employment by industry sectors (DP03\_0033E--DP03\_0045E covering 13 major industries from agriculture to public administration), unemployment and labor force status (DP03\_0005E--DP03\_0007E), and household income distribution (DP03\_0052E--DP03\_0061E spanning 10 income brackets from "Less than \$10,000" to "\$200,000 or more").
    
    \item \textbf{DP04: Housing Variables} \quad
    Housing tenure and value for owner-occupied units (DP04\_0081E--DP04\_0088E across 8 value ranges), rental costs for renter-occupied units (DP04\_0127E--DP04\_0135E covering 8 rent brackets), and household vehicle availability (DP04\_0058E--DP04\_0061E from "No vehicles" to "3 or more vehicles").
\end{itemize}

A specific preprocessing step is necessary for age data. As our simulated community survey targets only adults, agents must be 18 years or older. Therefore, the ACS age bracket ``15-19 years" is partitioned. Assuming a uniform distribution within this bracket, we proportionally allocate the population to isolate an ``18-19 years'' subgroup, allowing for the accurate sampling of the adult population.

\subsubsection{IPF implementation}
Key parameters for the IPF implementation are as follows:
\begin{itemize}
    \item Maximum iterations: 10
    \item Convergence threshold ($\epsilon$): $10^{-9}$
\end{itemize}

\textbf{IPF implementation} \quad IPF is a numerical method that constructs multidimensional arrays matching specified marginal constraints~\cite{estimate_matrices_from_marginal}. The algorithm iteratively adjusts a multidimensional array $\mathbf{M}^{(k)}$ by scaling each dimension to match target marginal distributions. At iteration $k$, for dimension $d$, the scaling factor is $\frac{\mathbf{t}_d}{\mathbf{m}_d^{(k)}}$ where $\mathbf{t}_d$ is the target marginal and $\mathbf{m}_d^{(k)}$ is the current marginal. The process continues until convergence: $\|\mathbf{m}_d^{(k)} - \mathbf{t}_d\| < \epsilon$ for all dimensions $d$. We initialize the process with a uniform distribution and use the ACS marginals as constraints. Then we apply the IPF algorithm to iteratively construct joint distributions. The algorithm iterates until convergence or reaches the maximum iteration limit (set at 10). In all cases, convergence is achieved before reaching the maximum iteration threshold. The resulting joint distribution enables sampling of the desired number of agents with realistic demographic correlations. 

\textbf{Post-sampling adjustments} \quad Following the sampling, two post-processing steps are performed. First, any sampled agents under 18 years of age are discarded, ensuring the final population consists solely of adults. Second, since marital status distributions are gender-dependent, we assign marital status after initial agent construction based on each agent's gender characteristics rather than incorporating it directly into the IPF algorithm.

\textbf{Verification} \quad To verify that generated agent demographics conform to Census population distributions, We employ chi-square goodness-of-fit tests. The test compares observed frequencies (generated agents) against expected frequencies (Census data). P-values above 0.05 indicate successful preservation of demographic structure; otherwise, significant deviations require regeneration.

\subsubsection{Post-sampling adjustments} 
Besides excluding agents aged below 18 years, we make two adjustments.

First, we store sex-dependent marital status distributions in advance and assign marital status after IPF implementation, since agents' sex is already allocated. Because marital status is not incorporated in IPF implementation, we may fail to capture correlations between marital status and other variables like age. Therefore, we adjust marital distributions based on age, increasing unmarried probability for young people and widowed probability for elderly.

Second, we assign education levels after agent generation to avoid unrealistic educational assignments. The U.S. Census Bureau provides two types of educational data: general enrollment status for ages 3 and above and detailed educational attainment for ages 25 and above. Using only the general enrollment data could result in unrealistic scenarios (e.g., 30-year-olds in elementary school or 19-year-olds with doctoral degrees). Therefore, we apply age-appropriate educational distributions: for agents aged 19 to 24 years, we use college enrollment data and classify them as "Attending some college or graduate school" or "Not attending any college"; for agents aged 25 and above, we use detailed educational attainment categories from census data.

These adjustments are not perfect but provide reasonable estimates in our cases.

\subsubsection{Verification results}\label{appendix:chi_square}
To validate that the demographic distribution of the sampled agents is statistically consistent with the U.S. Census data, we perform a Chi-square goodness-of-fit test for each of the 10 demographic dimensions for two regions. Note that marital status and education level are assigned probabilistically after IPF implementation based on Census distributions, ensuring automatic conformity with Census data. The null hypothesis ($H_0$) for each test is that the observed frequency distribution of the sampled agents is not significantly different from the expected frequency distribution derived from the Census data. We use a significance level of $\alpha$=0.05.

\begin{table}[!htbp]
\small
\captionsetup{width=\linewidth}
    \caption{Chi-square Goodness-of-Fit Test Results for Taylor County, Texas}
    \label{tab:taylor_chi_square}
    \renewcommand{\arraystretch}{1.5}
    \centering
    \begin{tabular}{lcccc}
    \toprule
    \textbf{Demographic Attribute} & \textbf{Chi-square ($\chi^2$)} & \textbf{Degree of Freedom} & \textbf{P-value} & \textbf{Result} (at $\alpha$=0.05)\\
    \hline
    Age Group & 2.8606 & 8 & 0.9428 & Fail to reject $H_0$\\
    Sex & 0.3500 & 1 & 0.5541 & Fail to reject $H_0$ \\
    Race & 4.1485 & 7 & 0.7625 & Fail to reject $H_0$ \\
    Ethnicity & 0.3047 & 1 & 0.5809 & Fail to reject $H_0$ \\
    Citizenship & 1.1907 & 4 & 0.8796 & Fail to reject $H_0$ \\
    Language at Home & 7.2699 & 4 & 0.1223 & Fail to reject $H_0$ \\
    Employment Status & 8.0744 & 15 & 0.9208 & Fail to reject $H_0$ \\
    Household Income & 1.6218 & 9 & 0.9961 & Fail to reject $H_0$ \\
    Housing & 7.2419 & 15 & 0.9506 & Fail to reject $H_0$ \\
    Vehicles & 1.3800 & 3 & 0.7102 & Fail to reject $H_0$ \\
    \bottomrule
    \end{tabular}
\end{table}

\begin{table}[!htbp]
\small
\captionsetup{width=\linewidth}
    \caption{Chi-square Goodness-of-Fit Test Results for Loudoun County, Virginia}
    \label{tab:loudoun_chi_square}
    \renewcommand{\arraystretch}{1.5}
    \centering
    \begin{tabular}{lcccc}
    \toprule
    \textbf{Demographic Attribute} & \textbf{Chi-square ($\chi^2$)} & \textbf{Degree of Freedom} & \textbf{P-value} & \textbf{Result} (at $\alpha$=0.05)\\
    \hline
    Age Group & 8.3377 & 8 & 0.4012 & Fail to reject $H_0$\\
    Sex & 0.0081 & 1 & 0.9281 & Fail to reject $H_0$ \\
    Race & 2.3274 & 7 & 0.9395 & Fail to reject $H_0$ \\
    Ethnicity & 0.3392 & 1 & 0.5603 & Fail to reject $H_0$ \\
    Citizenship & 4.6569 & 4 & 0.3243 & Fail to reject $H_0$ \\
    Language at Home & 2.9878 & 4 & 0.5599 & Fail to reject $H_0$ \\
    Employment Status & 16.8381 & 15 & 0.3286 & Fail to reject $H_0$ \\
    Household Income & 7.0488 & 9 & 0.6320 & Fail to reject $H_0$ \\
    Housing & 8.1773 & 15 & 0.9165 & Fail to reject $H_0$ \\
    Vehicles & 0.5687 & 3 & 0.9035 & Fail to reject $H_0$ \\
    \bottomrule
    \end{tabular}
\end{table}

\subsection{AI Agent Polling}
\subsubsection{LLM-Driven Topic Analysis}
Our LLM-driven topic analysis modifies traditional BERTopic~\cite{bertopic} by replacing vector embedding and clustering steps with LLM operations, leveraging its superior semantic understanding. The LDTA pipeline consists of three stages: (1) an LLM extracts at most three key phrases from each individual response via batch API; (2) these phrases are aggregated by an LLM to generate overarching themes; and (3) the frequency of each theme is quantified across the entire set of responses. This approach enables capture of nuanced perspectives that may not emerge through structured multiple-choice questions alone. The LLM we use here is Gemini-2.5-Pro with default configuration such as temperature and token limit. Since a single response can address multiple themes, the cumulative frequency of all identify themes can exceed 100\%.

\subsection{Conformal Prediction Calibration}
\textbf{Conformal Prediction} \quad Conformal prediction is a practical method that produces statistically valid uncertainty intervals/sets for any model's predictions, regardless of whether the model is interpretable or a black-box~\cite{conformal_prediction_intro}. The workflow proceeds in two phases: (1) Calibration: Using held-out data $\{(X_i, Y_i)\}_{i=1}^n$, one would compute nonconformity scores $s_i = s(X_i, Y_i)$ for each example. The larger the nonconformity scores are, the greater the deviation between the model's prediction and the ground truth value, and the greater the prediction uncertainty. The score threshold $\hat{q}$ would then be calculated as the $\lceil(n+1)(1-\alpha)\rceil/n$ empirical quantile of $\{s_i\}_{i=1}^n$, where $\alpha$ is the target error rate (or equivalently, $1-\alpha$ is the desired confidence level). (2) Deployment: For a new input $X_{\text{test}}$, the prediction set would be constructed as $C(X_{\text{test}}) = \{y : s(X_{\text{test}}, y) \leq \hat{q}\}$, including all candidate outputs whose scores fall below the threshold. Under the exchangeability assumption, which is weaker than the independent and identically distributed (i.i.d.) assumption, this provides theoretical guarantees that $\mathbb{P}(Y_{\text{test}} \in C(X_{\text{test}})) \geq 1-\alpha$ ~\cite{conformal_prediction_intro}.

In this setting, each community poll would serve as a single sample. In addition to AI agent polling results, one would collect real-world survey results at a relatively small scale. For each question option, the AI agent polling would produce a selection probability $\hat{y}$ based on the community's context (agent profiles, demographics, socioeconomic features, etc.). The true selection probability $y$ for that option would then be obtained from the collected real-world survey, which serves as the ground truth for calibration. The nonconformity score would be computed as $s = |y - \hat{y}|$, measuring the absolute deviation between the simulated and true probabilities. By repeating this process across $n$ communities, one would obtain a series of calibration scores $\{s_i\}_{i=1}^n$. Given a target confidence level (e.g., 95\%, corresponding to $\alpha=0.05$), the threshold $\hat{q}$ would be computed as the $\lceil(n+1)(1-\alpha)\rceil/n$ quantile of these scores.

During the deployment phase, for each survey question-option pair, let $\hat{y}_{\text{new}}$ denote the agent-predicted probability. The conformal prediction interval $[\hat{y}_{\text{new}} - \hat{q}, \hat{y}_{\text{new}} + \hat{q}]$ would then be guaranteed to contain the true selection probability with at least $(1-\alpha)$ probability under the exchangeability assumption.

Conformal prediction offers three key advantages. First, it establishes theoretically grounded confidence intervals for AI agent polling results, converting black-box outputs into statistically principled uncertainty quantification. Second, by calibrating against real-world survey data, the method aligns synthetic simulations to empirical observations, enhancing prediction reliability. Third, in comparison to conventional large-scale surveys, this approach offers significant economic efficiency, as it requires only a modest number of samples for calibration.

\clearpage

\section{Supplementary Experimental Analysis and Results}

\textbf{Basic settings} \quad We select Taylor County, Texas (Abilene) as our baseline case, which houses the first ``Stargate" AI data center ~\cite{openai_stargate_oracle_2025}. For cross-regional comparison, we contrast this with Loudoun County, Virginia, known as ``Data Center Alley,'' which hosts many data centers including major technology companies ~\cite{microsoft_northern_virginia_2025, datacentermap_aws_loudoun_2025}. Each experiment polls 1000 virtual agents responding to a hypothetical 100 MW data center proposal through a 13-question survey covering economic, environmental, and community engagement attitudes. 

\textbf{Data collection} \quad Demographic data is collected via the U.S. Census Bureau's American Community Survey (ACS) 2023 5-year estimates API ~\cite{census_acs_5year} based on location of the targeted community, retrieving individual-level characteristics across 12 features and county-level statistics. The data generates representative populations of \textbf{1,000 virtual residents} using IPF algorithm, maintaining statistical consistency with regional demographics. Survey responses are collected through batch LLM API calls using standardized prompts incorporating regional context and individual agent profiles. 

\subsection{Baseline Case Analysis}
\subsubsection{Baseline Results}
\begin{figure}[htbp!]
    \centering
    \subfloat[Top-3 economic benefits]{
    \includegraphics[width=0.4\linewidth]{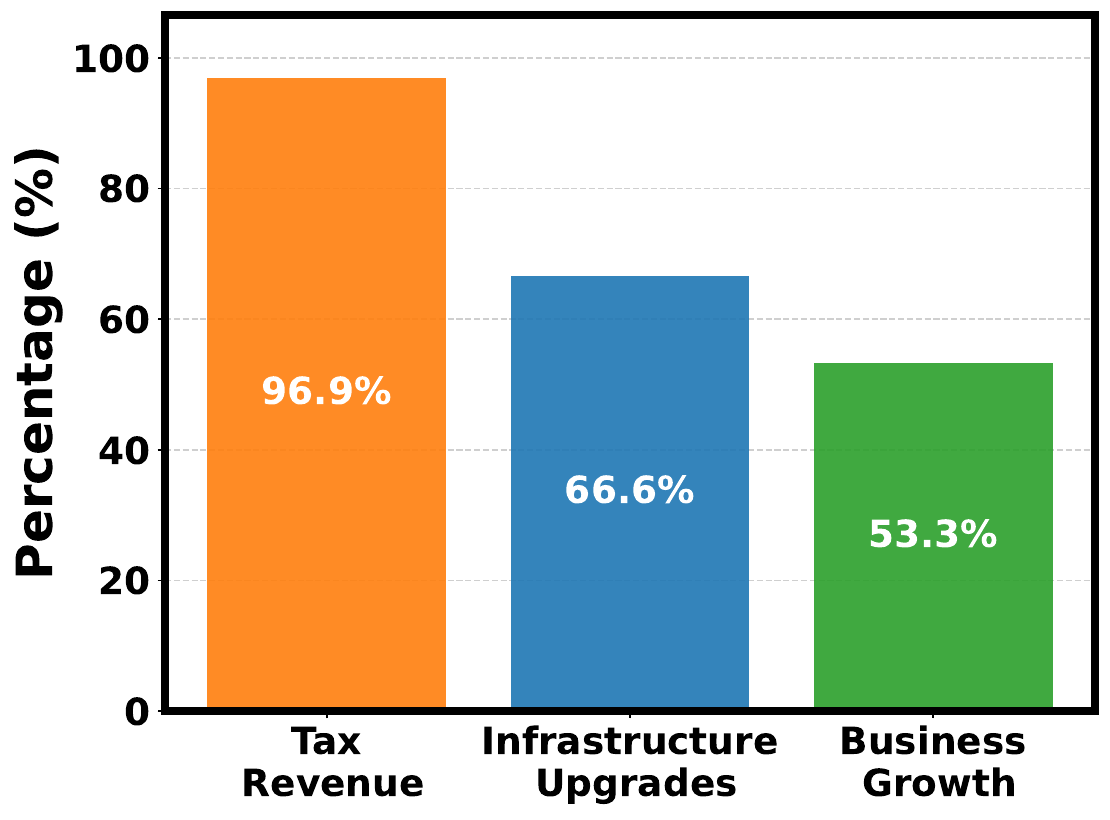}
    \label{fig:economic_benefits_base}
    }
    \subfloat[Top-3 economic concerns]{
    \includegraphics[width=0.4\linewidth]{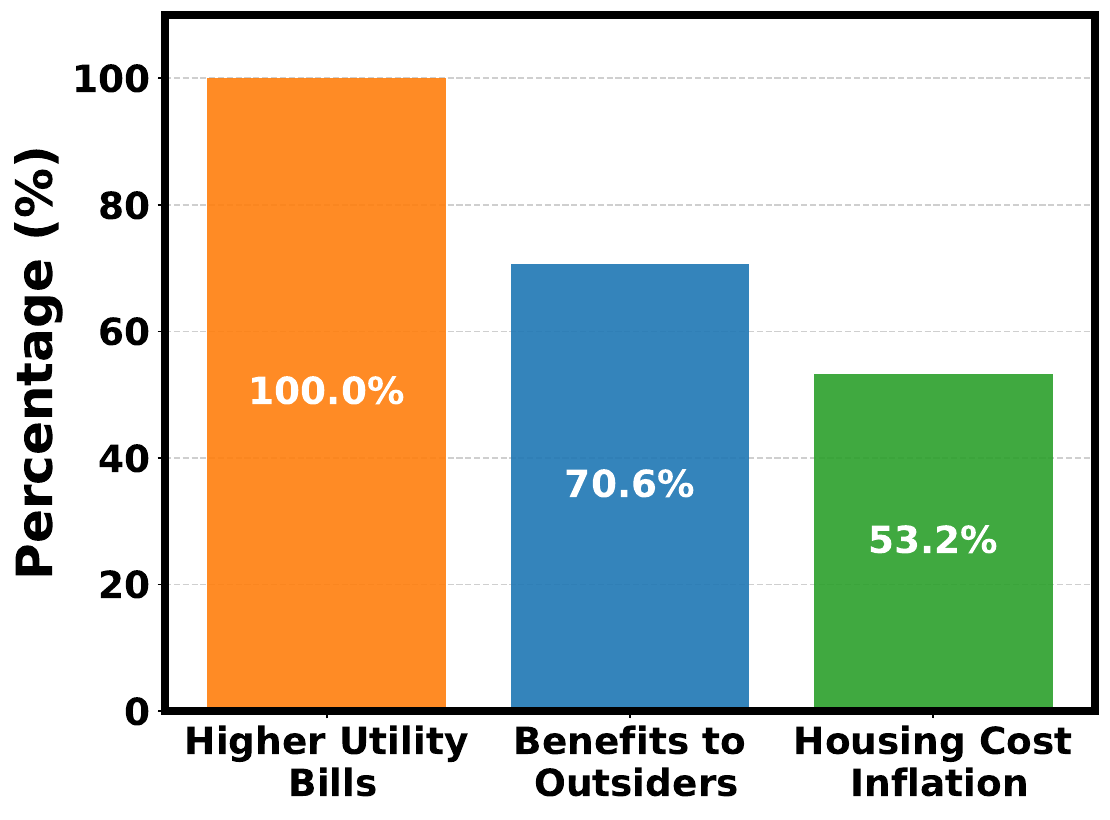}
    \label{fig:economic_concerns_base}
    }

    \caption{Taylor County results using GPT-5 on economic issues. (a) Community opinions about the most important economic benefits brought by the data center project. (b) Community economic concerns. Note: figures showing distribution for all response options in Appendix ~\ref{appendix}.}
    \label{fig:taylor_county_economic}
\end{figure}

\begin{figure}[htbp!]
    \centering
    \subfloat[Top-3 environmental concern]{
    \includegraphics[width=0.4\linewidth]{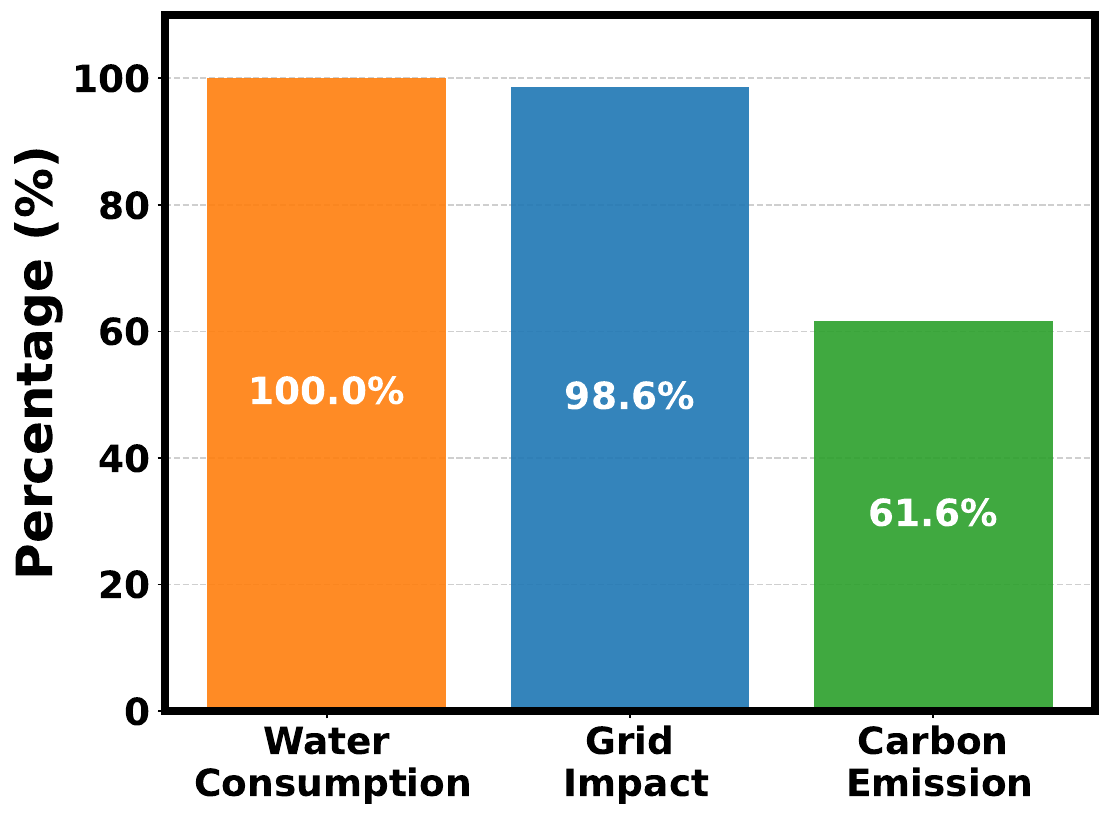}
    \label{fig:environmental_concern_base}
    }
    \subfloat[Top-3 environmental protection]{
    \includegraphics[width=0.4\linewidth]{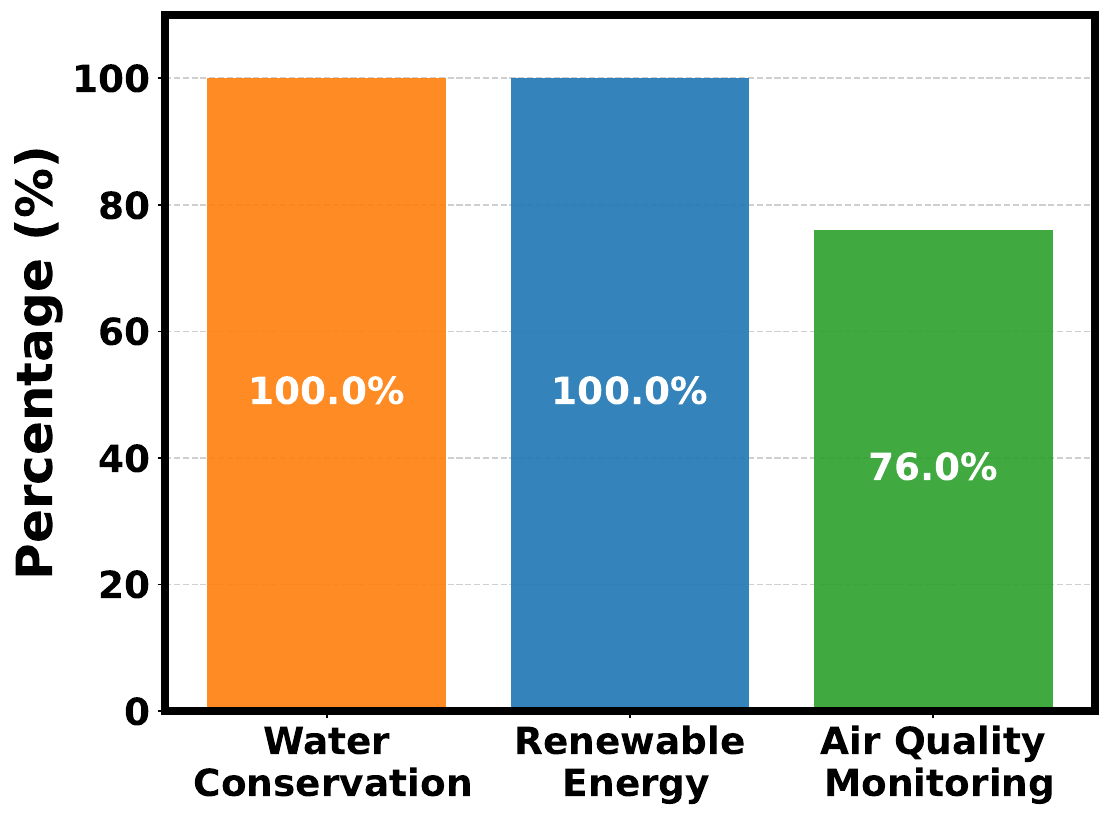}
    \label{fig:environment_protection_base}
    }
    \caption{Taylor County results using GPT-5 on environmental issues. (a) The community's top environmental concerns regarding the data center project. (b) The most frequently requested environmental protections for the project. Note: figures showing distribution for all response options in Appendix ~\ref{appendix}.}
    \label{fig:taylor_county_environmental}
\end{figure}

\begin{figure}[htbp!]
    \centering
    \subfloat[Top-3 information sources]{
    \includegraphics[width=0.4\linewidth]{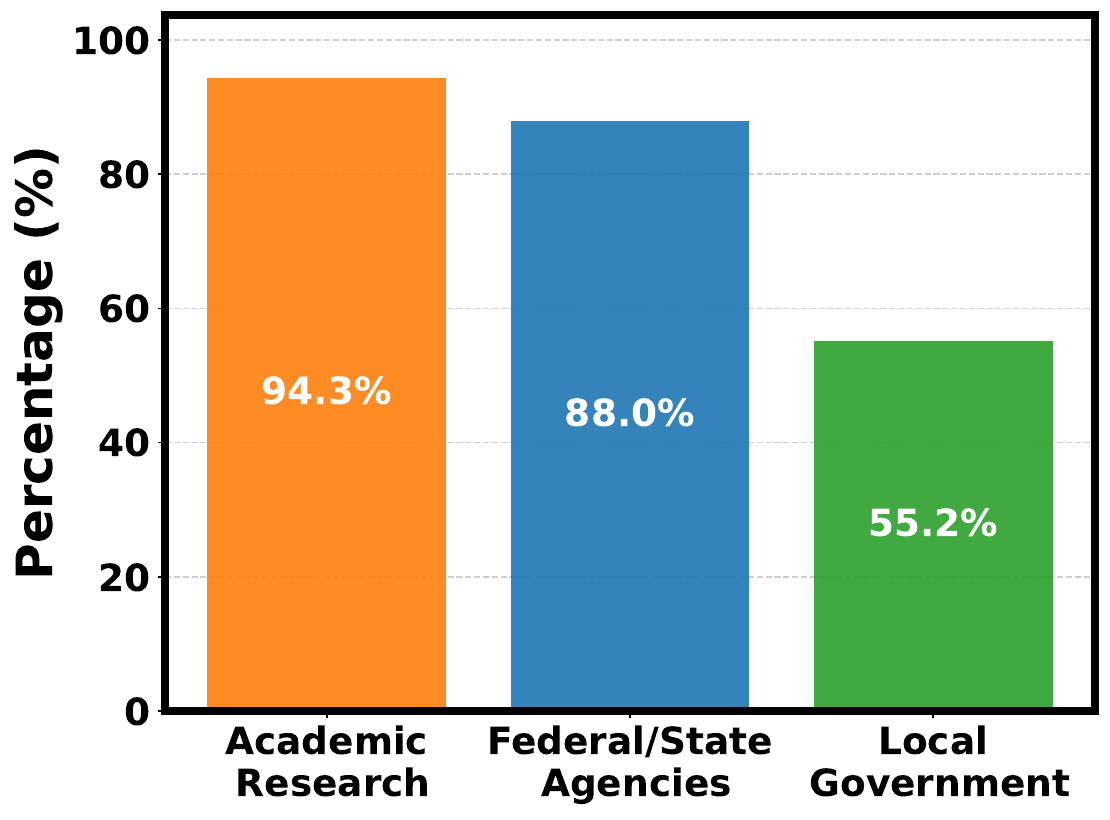}
    \label{fig:information_source_base}
    }
    \subfloat[Top-3 support condition]{
    \includegraphics[width=0.4\linewidth]{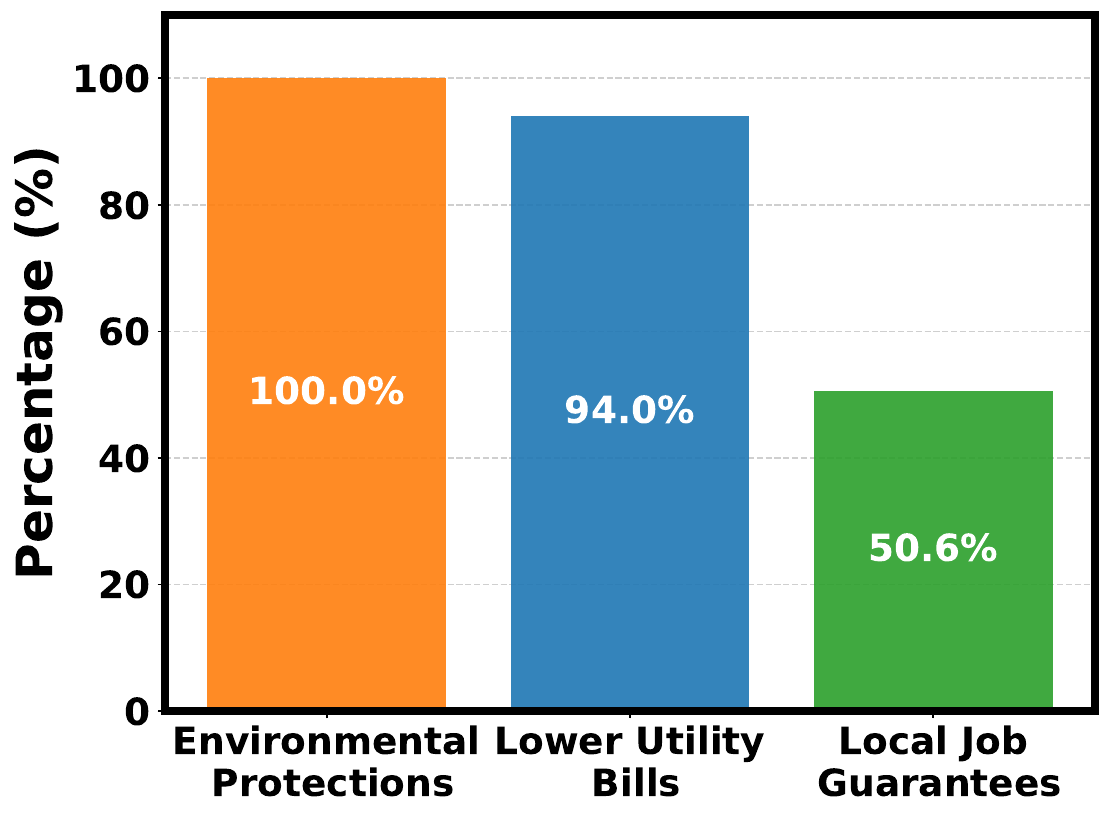}
    \label{fig:support_condition_base}
    }
    \caption{Taylor County results using GPT-5 on preferred information sources and support conditions. (a) The community's most trusted sources of information regarding the data center project. (b) The top conditions that would increase community support for the project. Note: figures showing distribution for all response options in Appendix ~\ref{appendix}.}
    \label{fig:taylor_county_information_protection}
\end{figure}

A more detailed examination of the survey results is shown in Figures \ref{fig:taylor_county_economic}, \ref{fig:taylor_county_environmental}, and \ref{fig:taylor_county_information_protection}. In the economic domain, agents identify tax revenue (96.9\%) as the project's most important benefit, followed by infrastructure upgrades (66.6\%) and business growth (53.3\%). Conversely, the predominant economic concern shared by all agents is the prospect of higher utility bills, likely due to the immediate and tangible relevance of such costs to residents' household budgets. A strong correspondence exists between the agents' environmental concerns and their requested protections. Water is the paramount issue, with water consumption ranking as the top concern and water conservation as the most demanded protection. Similarly, energy-related impacts are a significant priority, which is understandable given the project's scale. Regarding trusted information sources, AI agents consider academic research the most trustworthy source. Consistent with their previously expressed concerns, the agents' conditions for supporting the project are clear: the environmental protections is the most critical condition (100\%), followed by lower utility bills (94.0\%), directly mirroring their primary environmental and economic worries.

\begin{figure}[htbp!]
    \centering
    \subfloat[Economic benefits]{
    \includegraphics[width=0.42\linewidth]{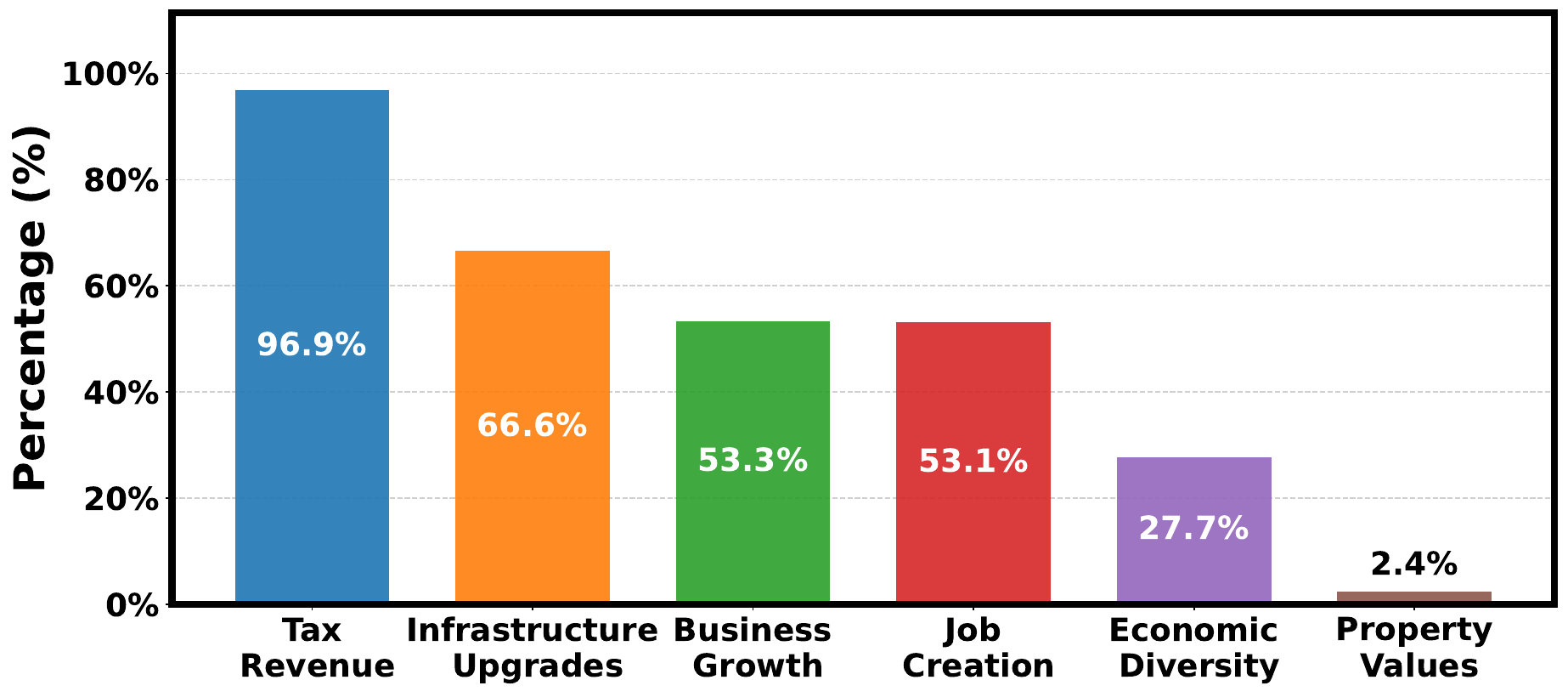}
    \label{fig:economic_benefits_base_full}
    }
    \subfloat[Economic concerns]{
    \includegraphics[width=0.42\linewidth]{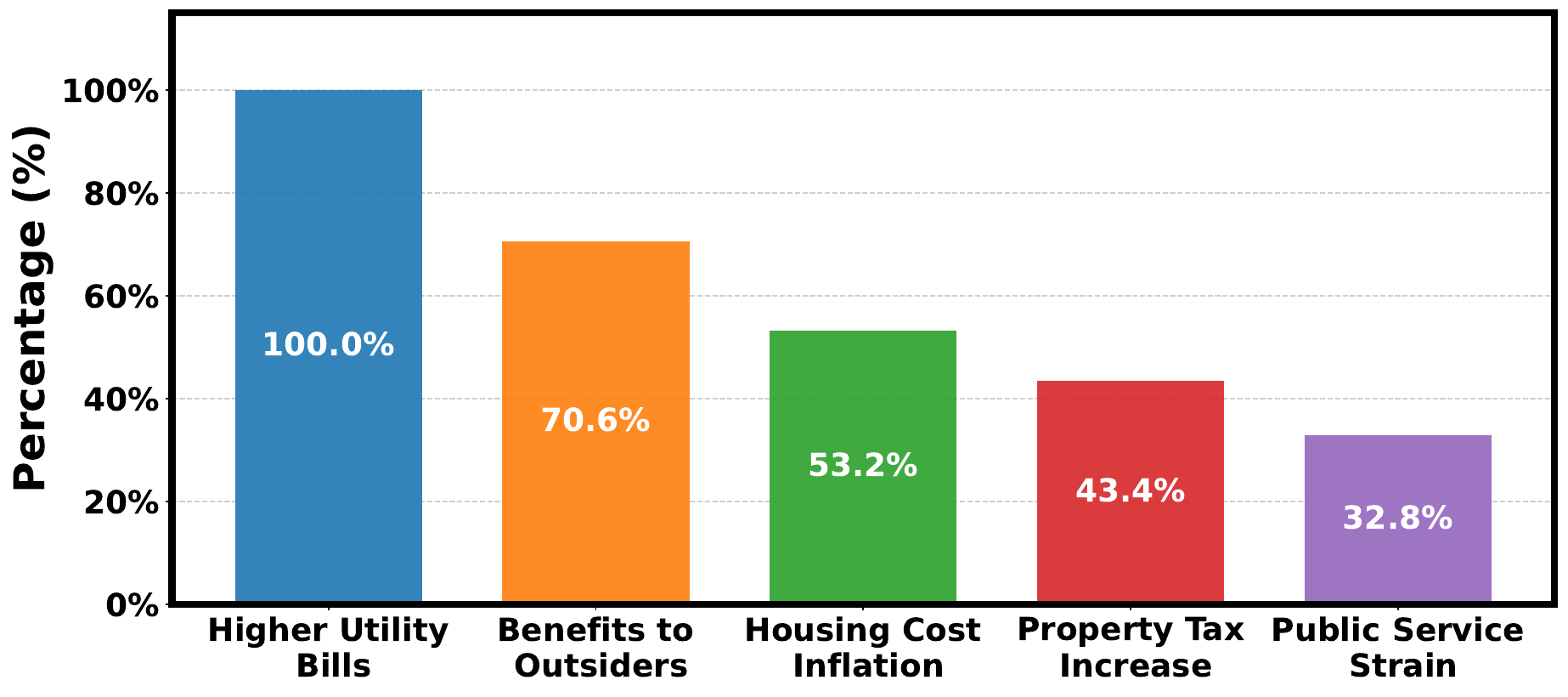}
    \label{fig:economic_concerns_base_full}
    }

    \caption{Taylor County results using GPT-5. (a) Community opinions about the most important economic benefits brought by the data center project. (b) Community economic concerns.}
    \label{fig:taylor_county_economic_full}
\end{figure}

\begin{figure}[htbp!]
    \centering
    \subfloat[Environmental concern]{
    \includegraphics[width=0.42\linewidth]{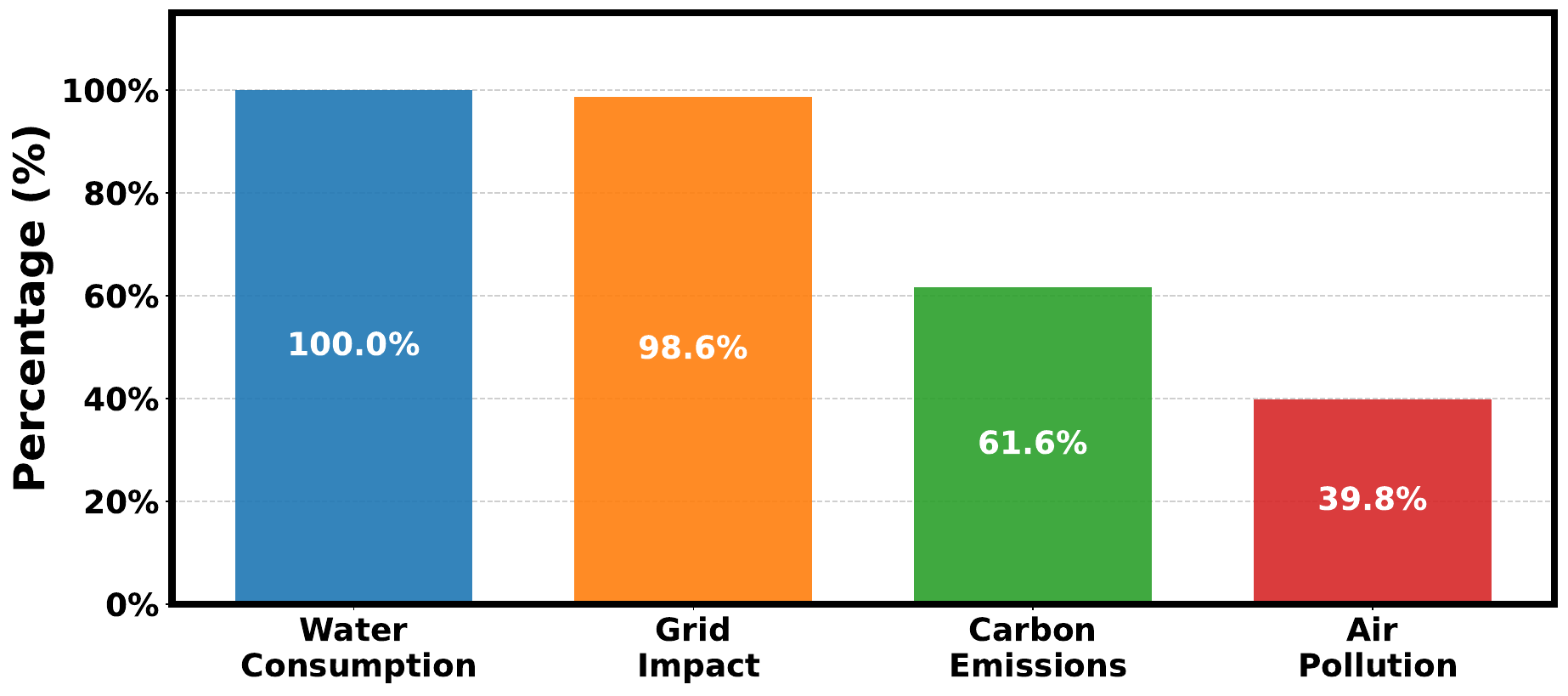}
    \label{fig:environmental_concern_base_full}
    }
    \subfloat[Environmental protection]{
    \includegraphics[width=0.42\linewidth]{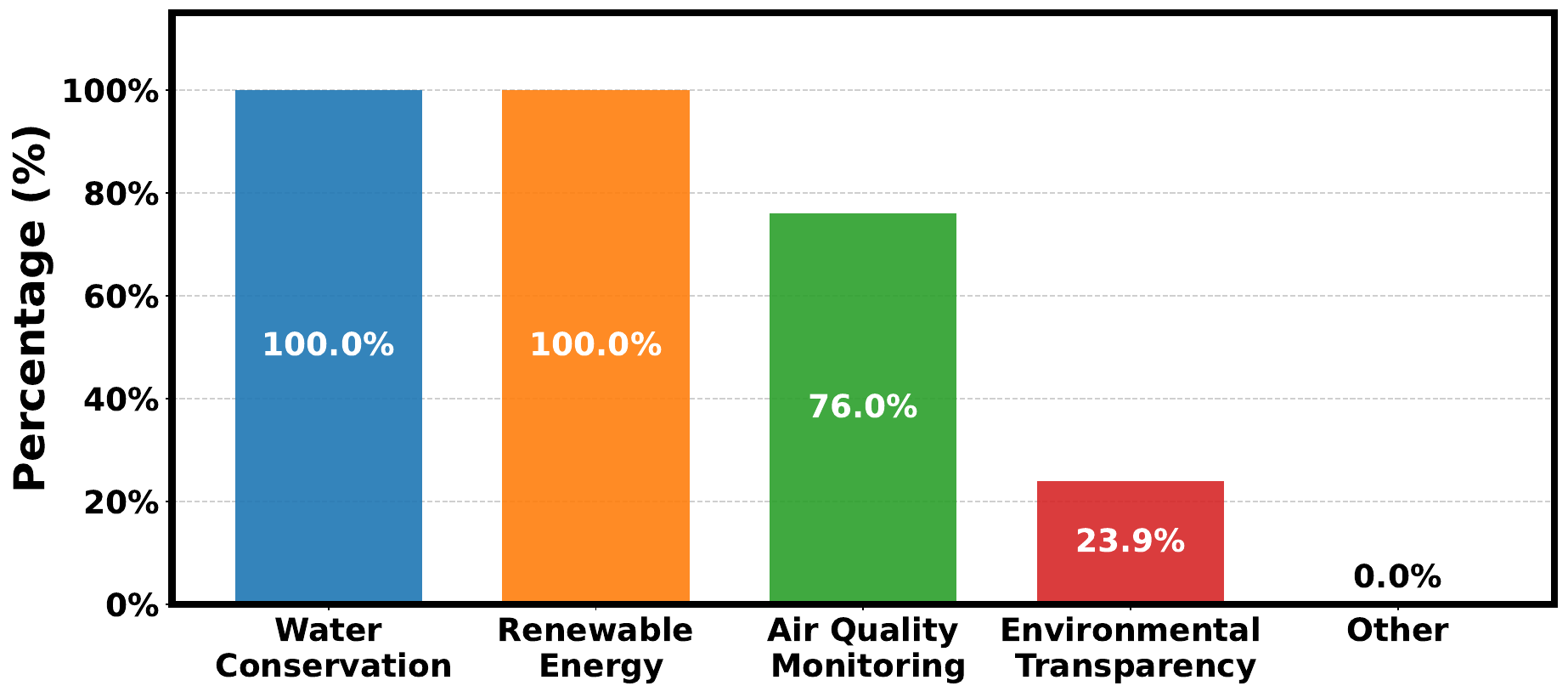}
    \label{fig:environment_protection_base_full}
    }
    \caption{Taylor County results using GPT-5. (a) The community's top environmental concerns regarding the data center project. (b) The most frequently requested environmental protections for data center project. }
    \label{fig:taylor_county_environmental_full}
\end{figure}

\begin{figure}[htbp!]
    \centering
    \subfloat[Information sources]{
    \includegraphics[width=0.4\linewidth]{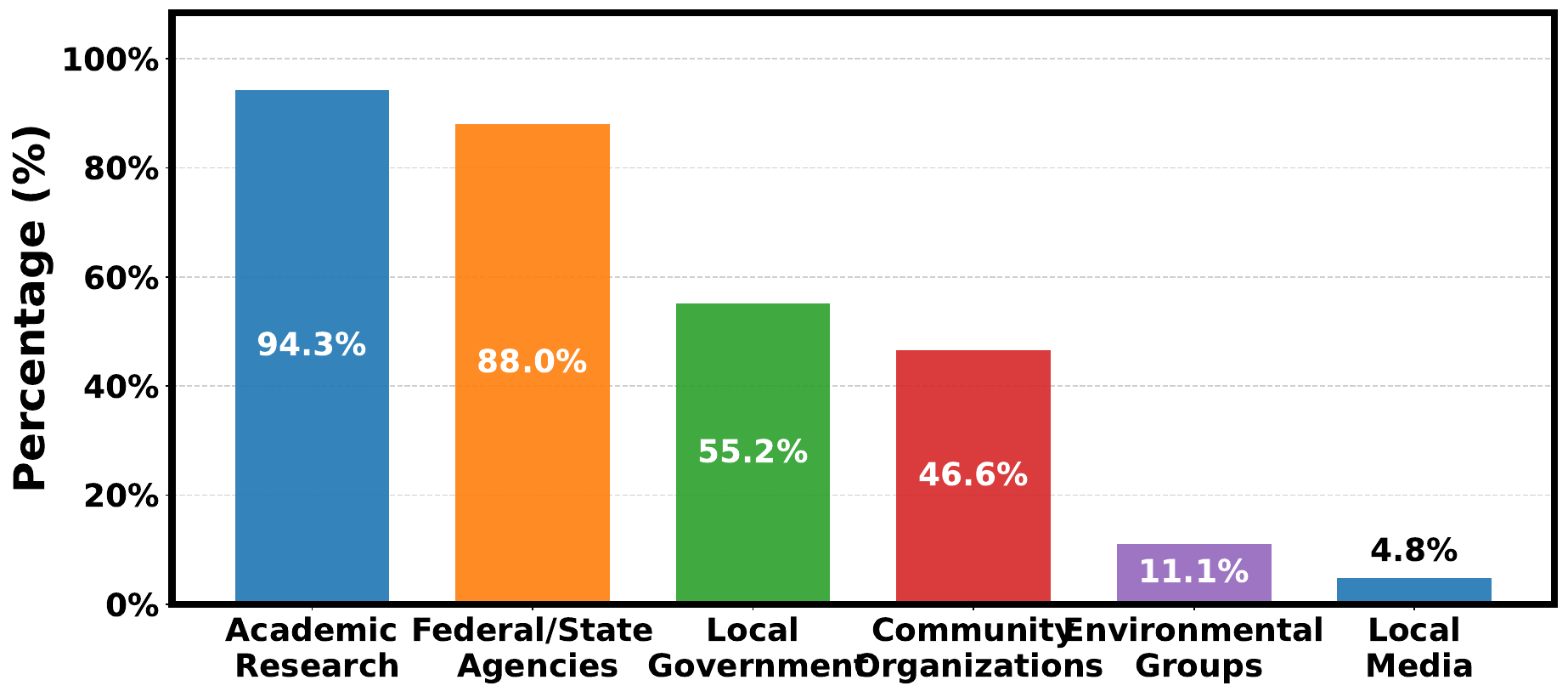}
    \label{fig:information_source_base_full}
    }
    \subfloat[Support condition]{
    \includegraphics[width=0.4\linewidth]{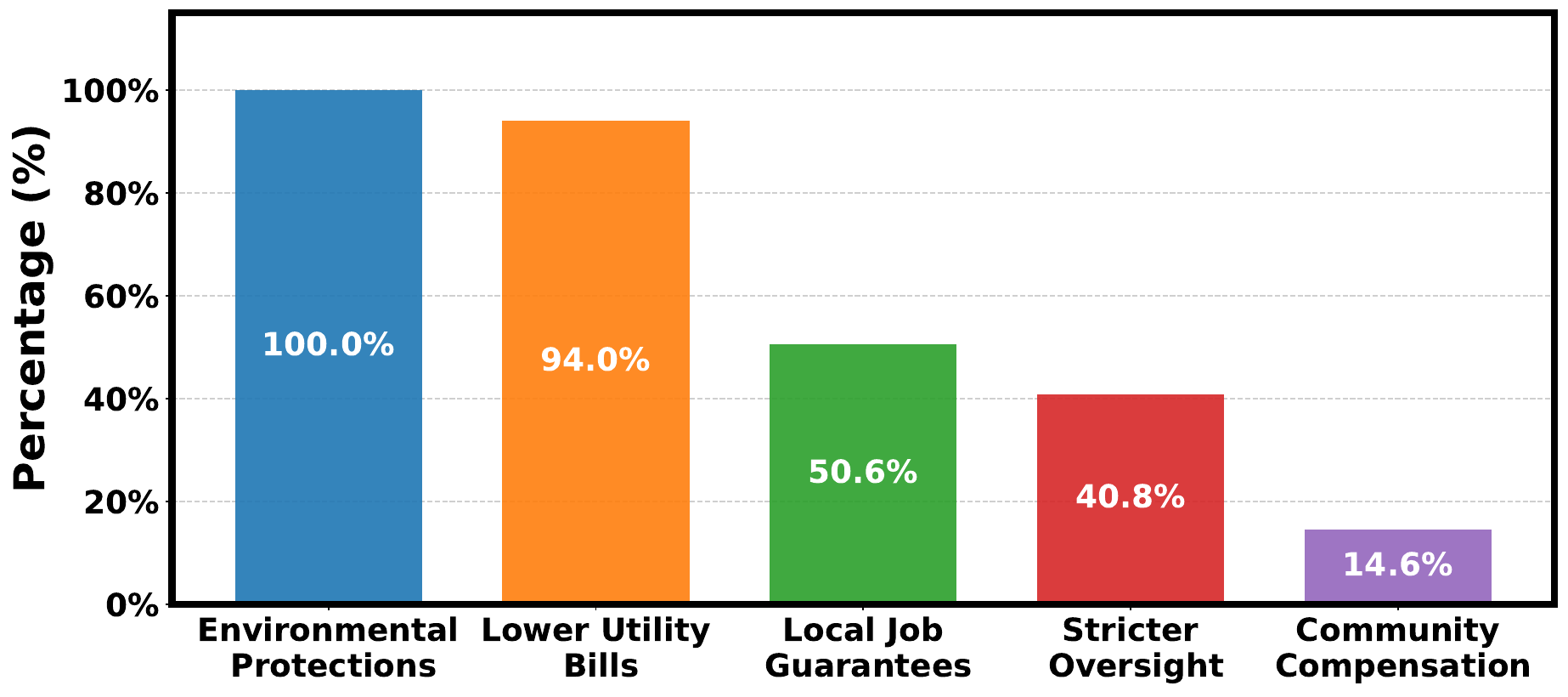}
    \label{fig:support_condition_base_full}
    }
    \caption{Taylor County results using GPT-5. (a) The community's most trusted sources of information regarding the data center project. (b) The top conditions that would increase community support for the project. }
    \label{fig:taylor_county_information_protection_full}
\end{figure}

\subsubsection{Cross-Model Comparison}
\begin{figure}[htbp!]
    \centering
    \subfloat[Environmental concern levels]{\includegraphics[width=0.33\textwidth, valign=b]{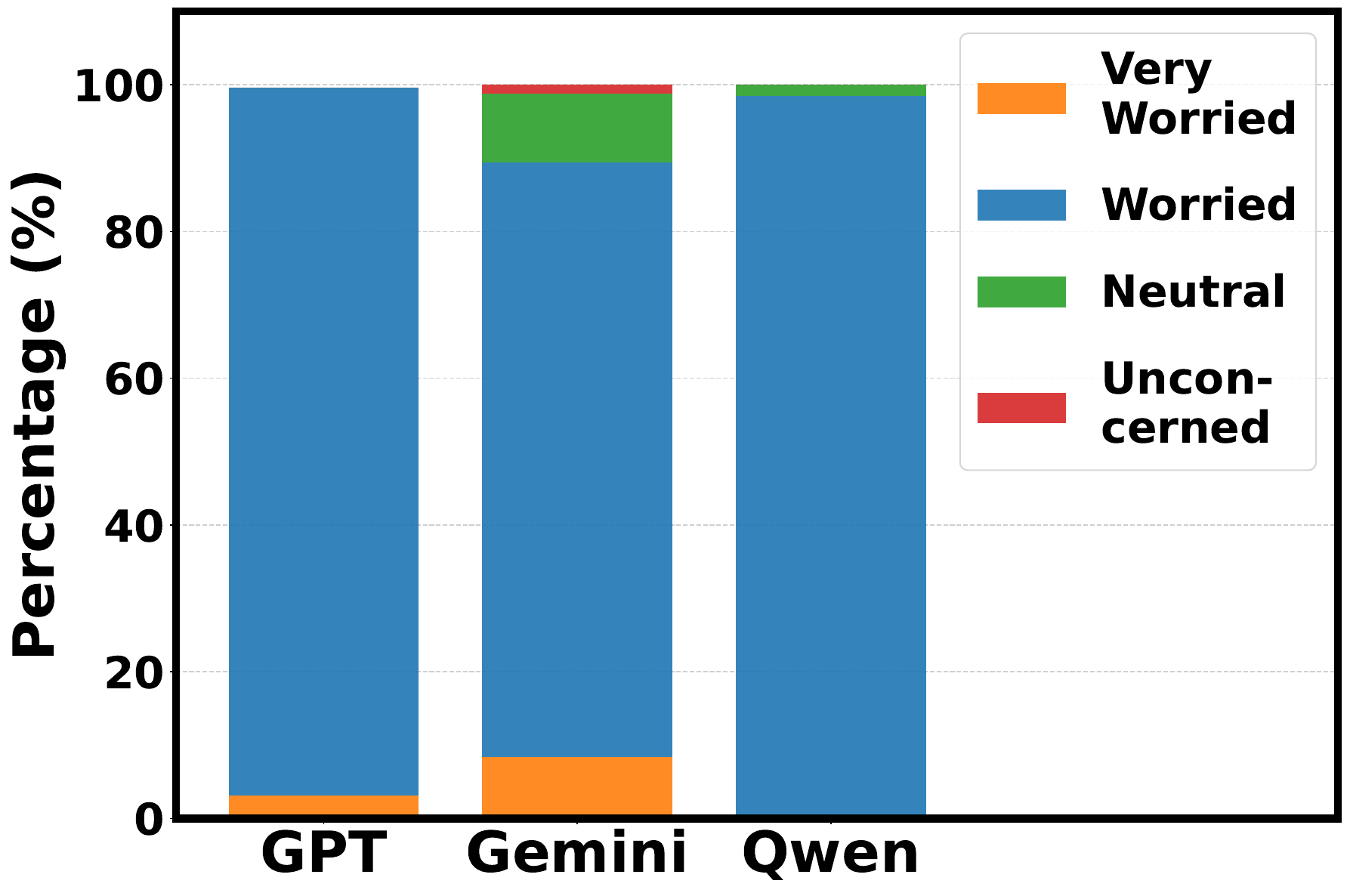}\label{fig:environmental_worry_models}
    }
    \subfloat[Engagement willingness]
    {\includegraphics[width=0.33\textwidth, valign=b]{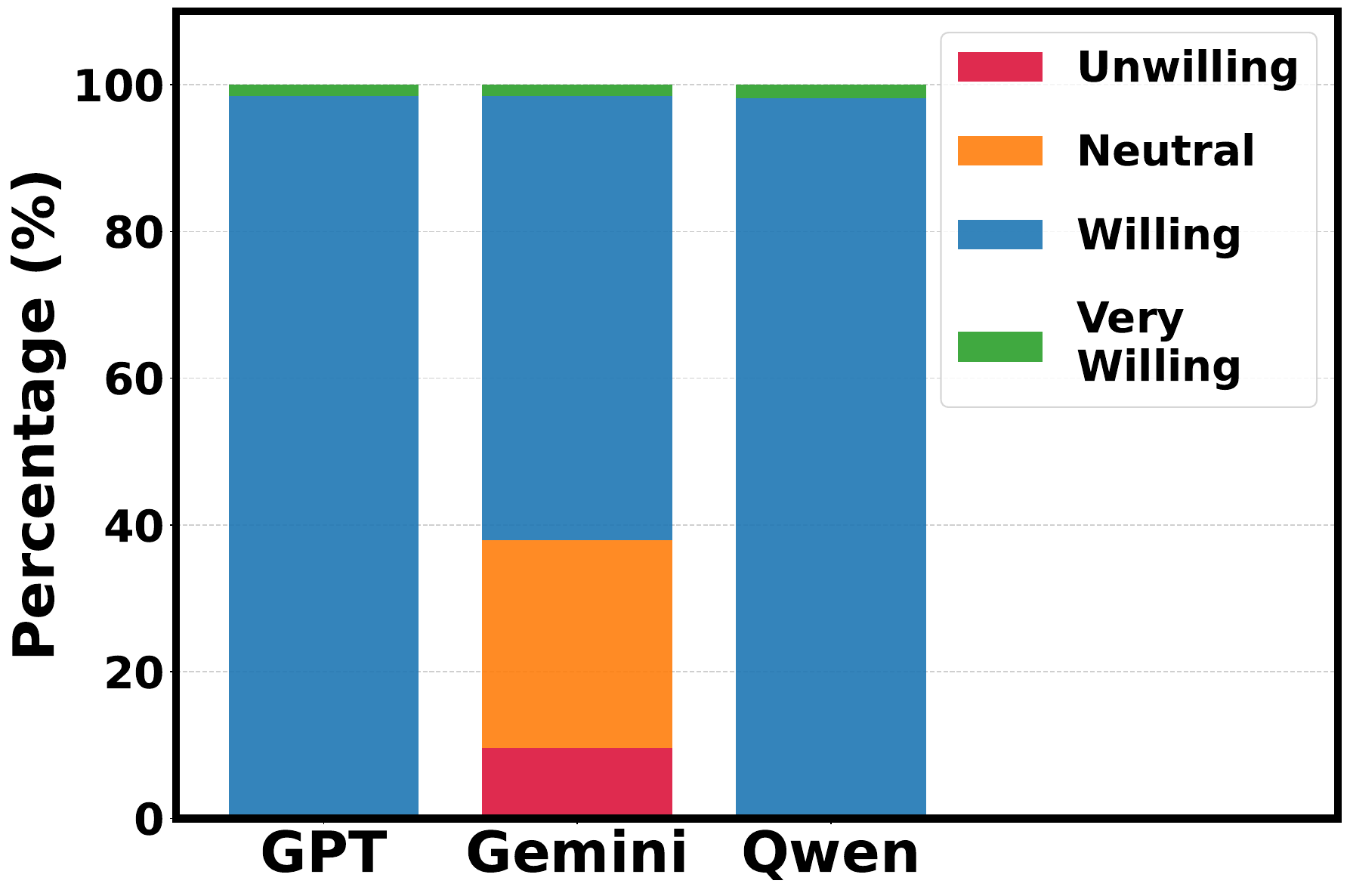}\label{fig:engagement_willingness_models}
    }
    \subfloat[Personal impacts]
    {\includegraphics[width=0.33\textwidth, valign=b]{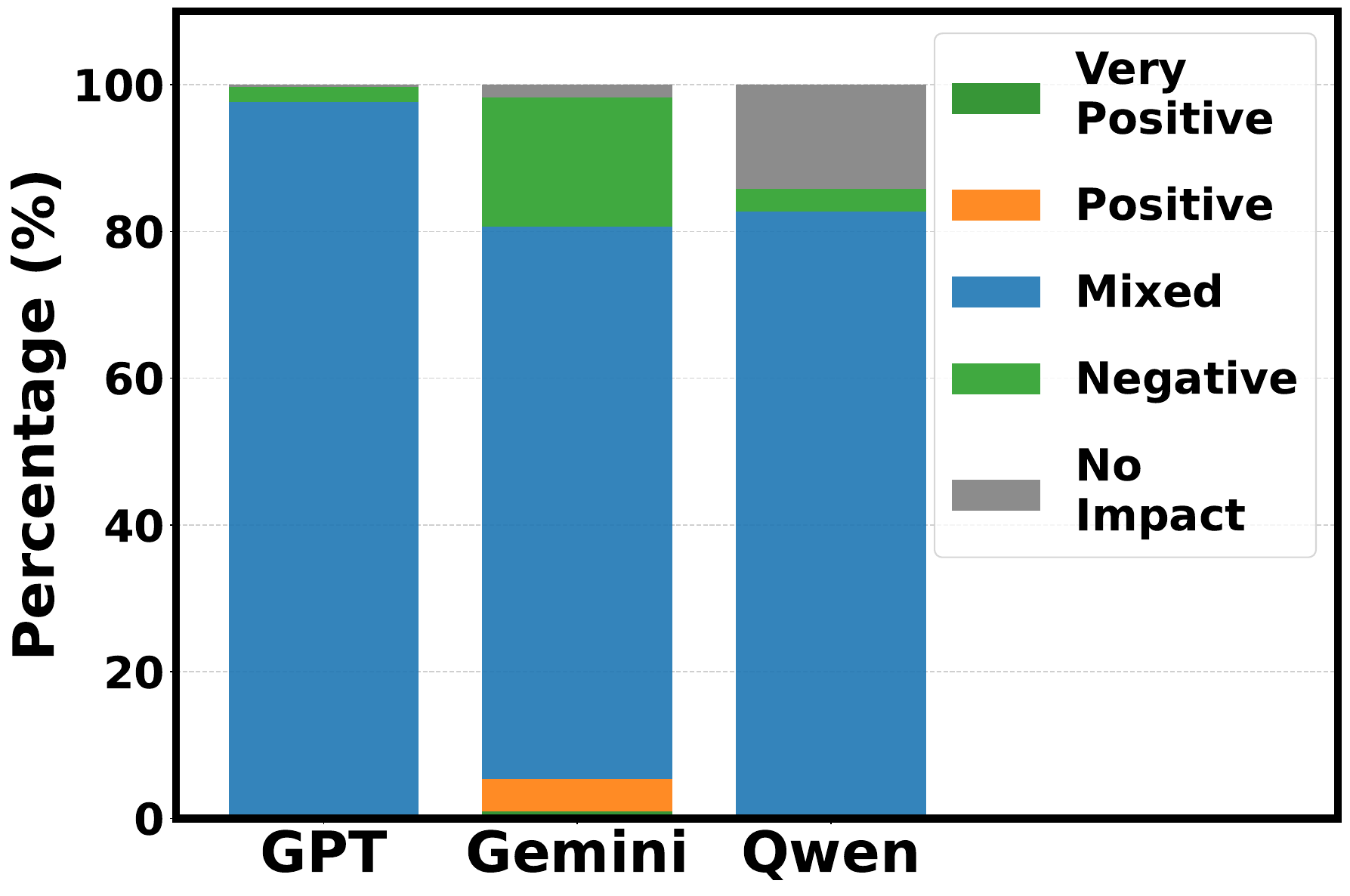}\label{fig:personal_impacts_models}
    }
    \caption{Common patterns in cross-model polling results. (a) Level of concern expressed by agents regarding the project's potential environmental impacts. (b) Agents' willingness to participate in community planning discussions about the project. (c) Agents' expectations of how the project will personally affect their households. Note: Charts display only selected response categories. Complete survey options are in Appendix ~\ref{appendix}.}
    \label{fig:cross_model_commonalities}
\end{figure}

\paragraph{Common Results}
Cross-model analysis reveals several consistent patterns in polling results (see Figure~\ref{fig:cross_model_commonalities}). Specifically, a high level of environmental concern is also a consistent finding, with the vast majority of agents in each model simulation reporting being worried about the project's impacts. Furthermore, a strong willingness to participate in community planning discussions is observed, although agents simulated by Gemini-2.5 exhibit a greater degree of neutrality and unwillingness compared to the others. Finally, agents across all models primarily anticipate that the project will have a mixed personal impact on their households.

\begin{figure}[htbp!]
    \centering
    \subfloat[Economic benefits (Gemini)]{
    \includegraphics[width=0.42\linewidth]{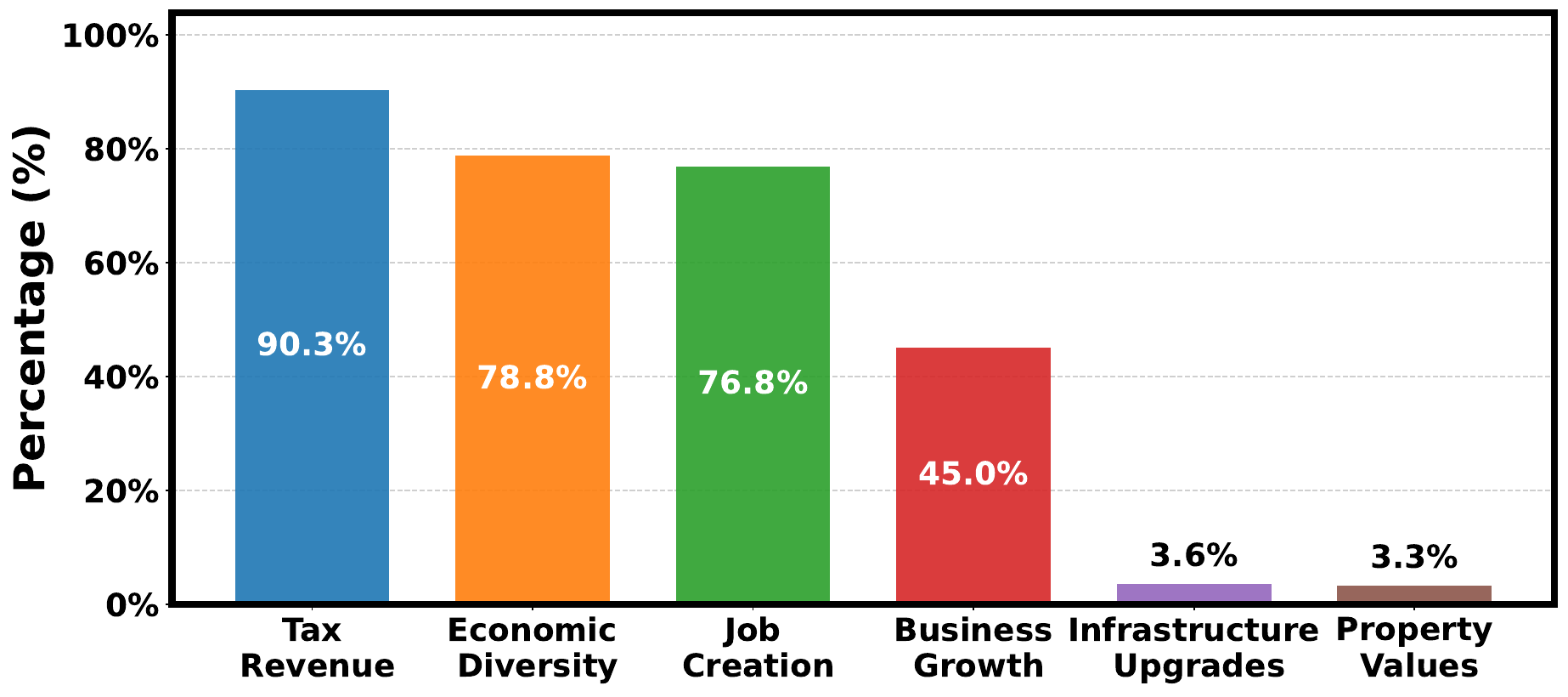}
    }
    \subfloat[Economic benefits (Qwen)]{
    \includegraphics[width=0.42\linewidth]{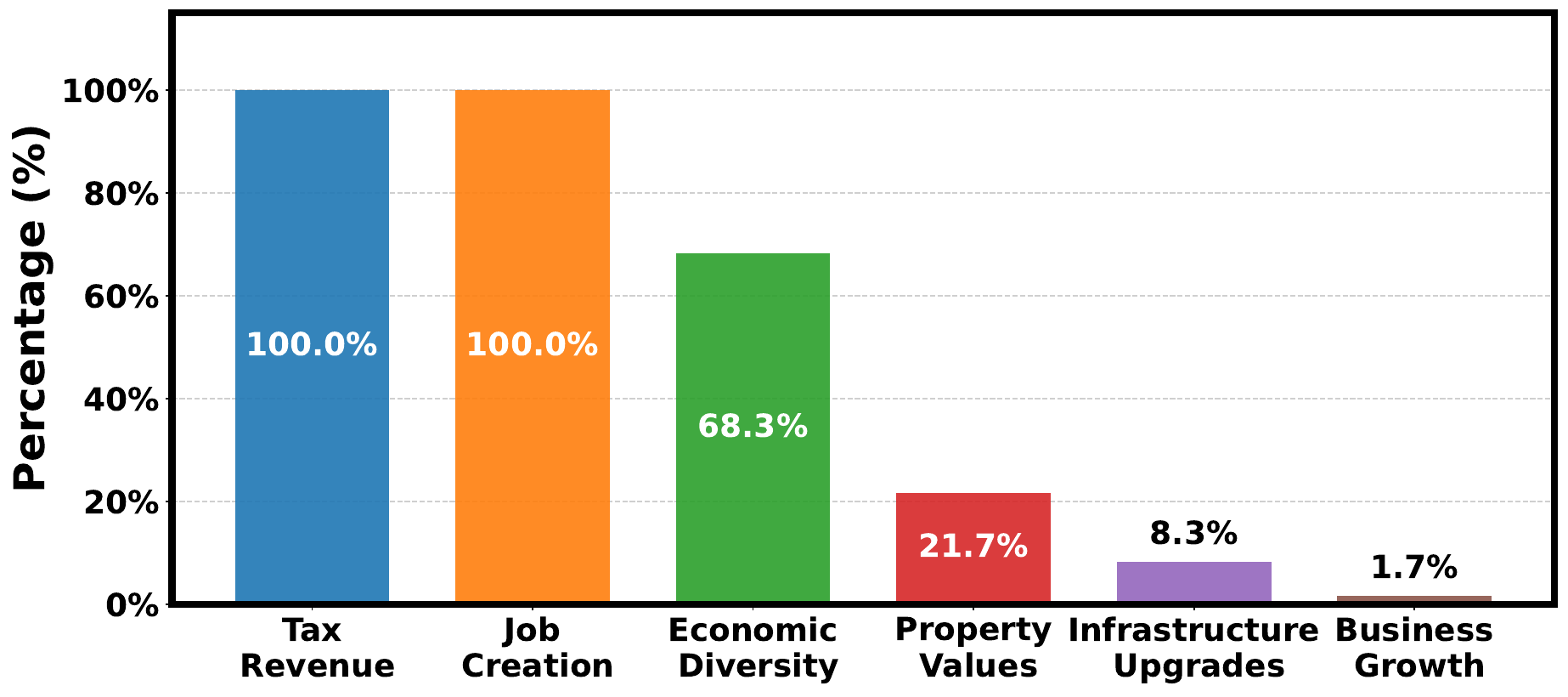}
    }

    \caption{Taylor County results using Gemini and Qwen. (a) Community opinions about the most important economic benefits brought by the data center project using Gemini. (b) The most important economic benefits using Qwen.}
    \label{fig:taylor_county_economic_full_gemini_qwen}
\end{figure}

\begin{figure}[htbp!]
    \centering
    \subfloat[Information sources (Gemini)]{
    \includegraphics[width=0.42\linewidth]{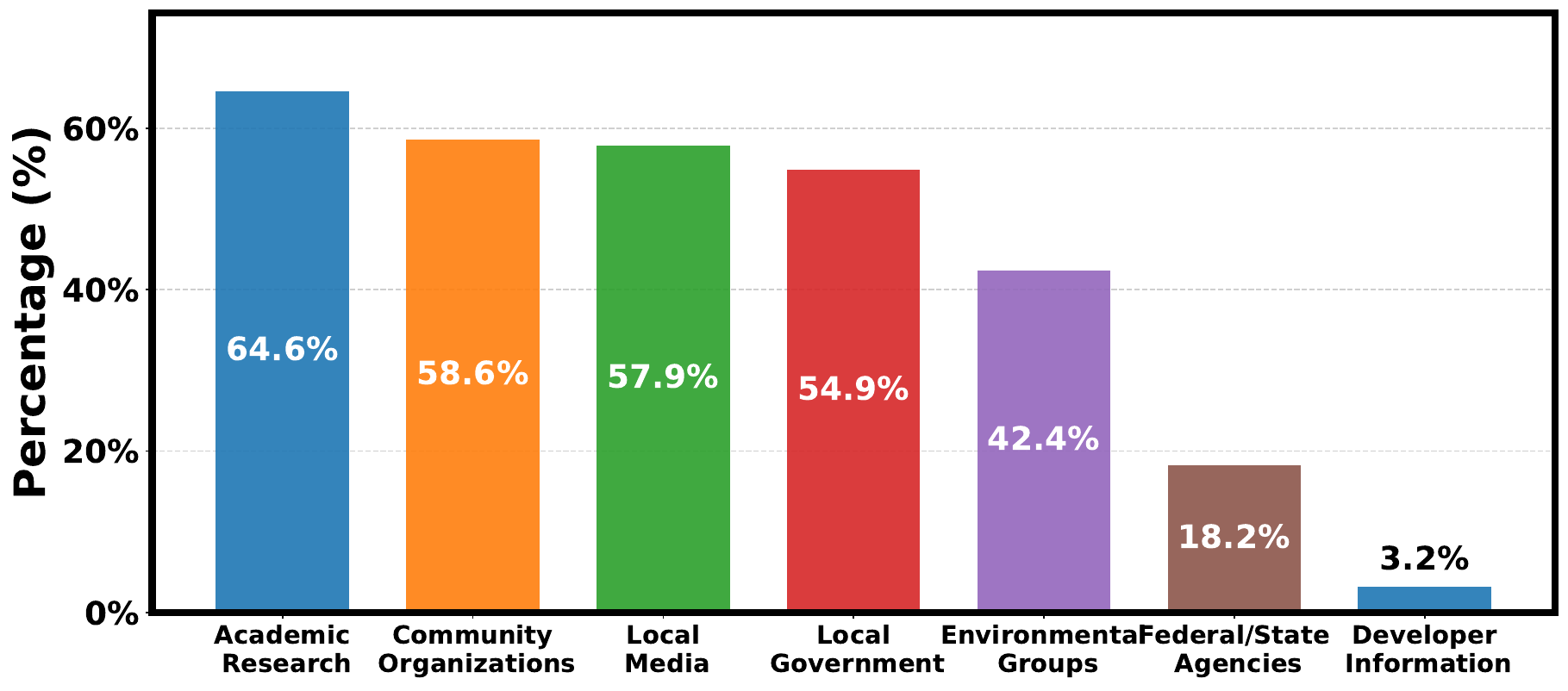}
    \label{fig:information_source_base_full_gemini}
    }
    \subfloat[Information sources (Qwen)]{
    \includegraphics[width=0.42\linewidth]{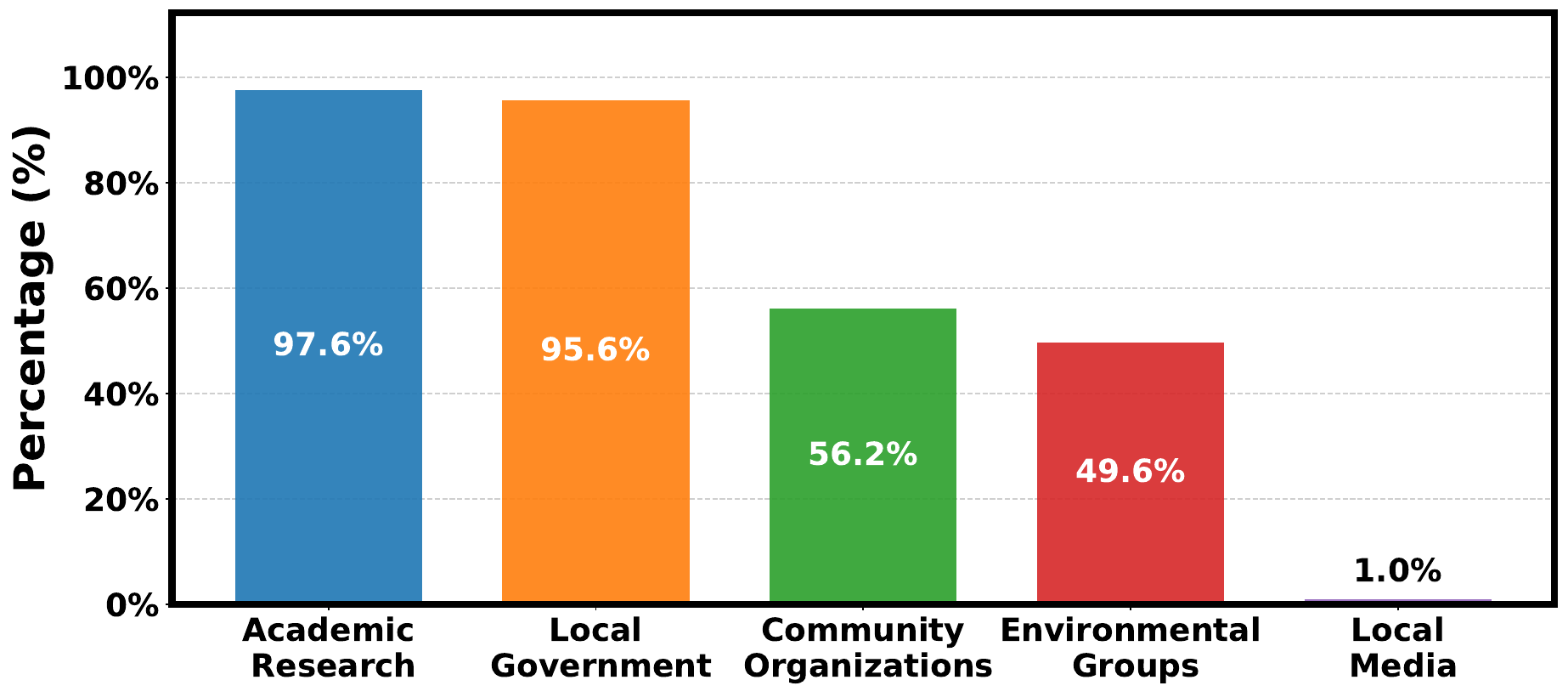}
    \label{fig:information_sources_base_full_qwen}
    }

    \caption{Taylor County results using Gemini and Qwen. (a) The community’s most trusted sources of information regarding the data center project using Gemini. (b) The trusted information source using Qwen.}
    \label{fig:taylor_county_info_full_gemini_qwen}
\end{figure}

\begin{figure}[htbp!]
    \centering
    \subfloat[Support conditions (Gemini)]{
    \includegraphics[width=0.42\linewidth]{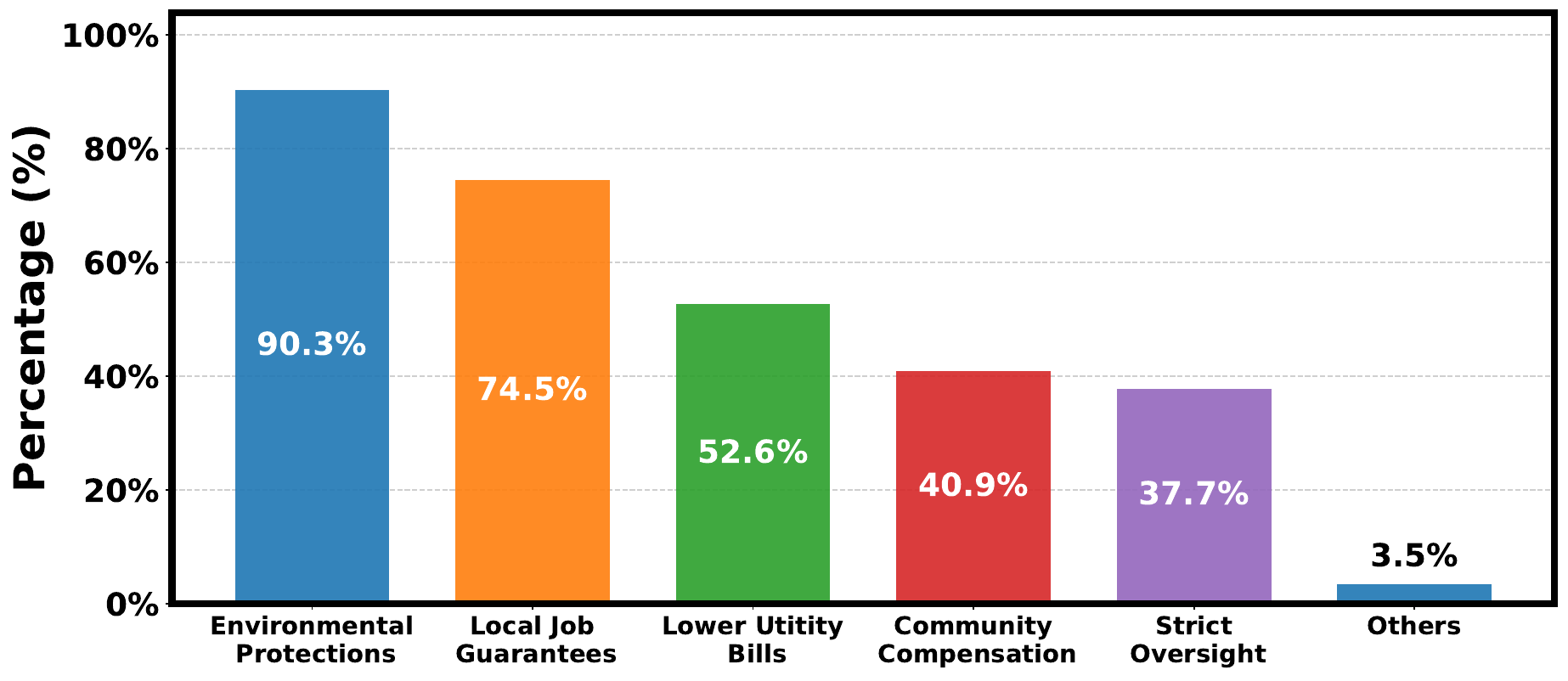}
    \label{fig:support_condition_base_full_gemini}
    }
    \subfloat[Support conditions (Qwen)]{
    \includegraphics[width=0.42\linewidth]{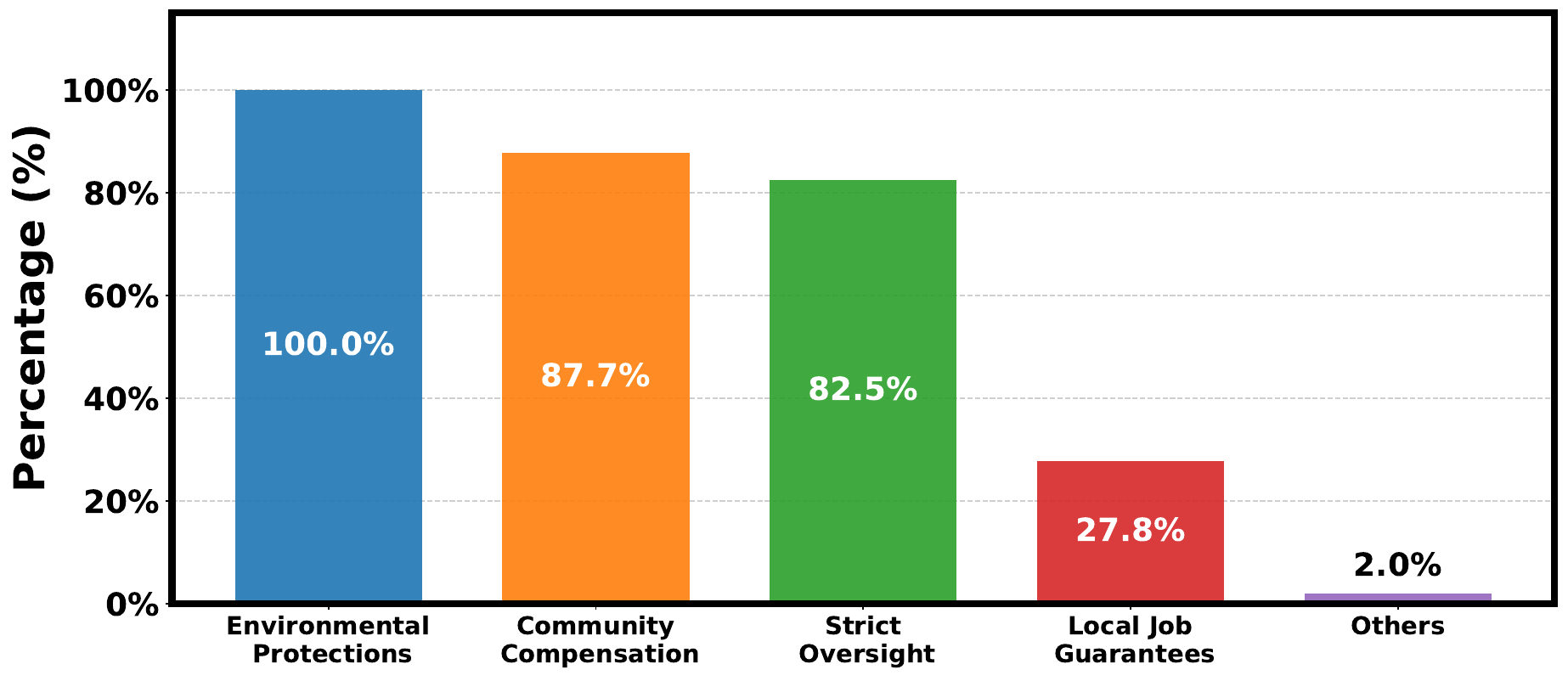}
    \label{fig:support_condition_base_full_qwen}
    }

    \caption{Taylor County results using Gemini and Qwen. (a) The conditions that would increase AI agents’ support for the project using Gemini. (b) The conditions using Qwen.}
    \label{fig:taylor_county_support_full_gemini_qwen}
\end{figure}

\paragraph{Different Results}
Despite common patterns, significant divergences emerge in key areas:
\begin{itemize}
    \item \textbf{Overall attitudes}: Qwen exhibits notably stronger opposition compared to GPT-5 and Gemini-2.5, despite viewing economic impacts more positively. This apparent contradiction can be explained by examining the conditions under which agents would support the project. Qwen agents strongly demand community compensation and stricter oversight, which are far less frequently mentioned by the other models. Since these conditions are not explicitly guaranteed in the current proposal, it is reasonable that Qwen agents oppose the project despite recognizing its economic benefits. The oversight requirement aligns with Qwen's high government trust, reflecting an expectation that trusted authorities should exercise rigorous regulatory control. This pattern likely stems from training data emphasizing government-led infrastructure development with explicit compensation mechanisms.
\end{itemize}

\begin{figure}[htbp!]
    \centering
    \subfloat[Overall attitudes]{\includegraphics[width=0.36\textwidth, valign=b]{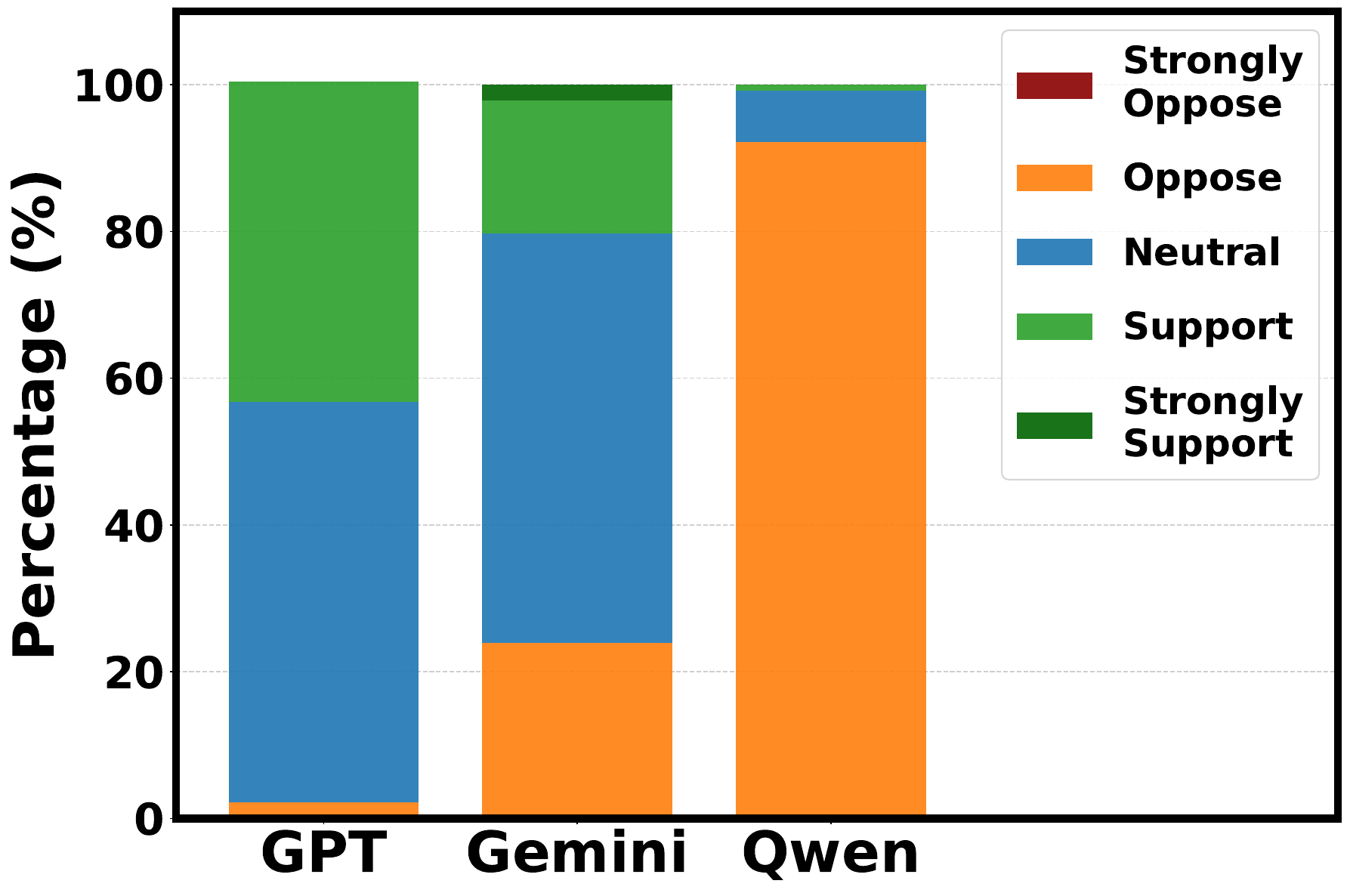}
    }
    \subfloat[Top-3 support condition]
    {\includegraphics[width=0.54\textwidth, valign=b]{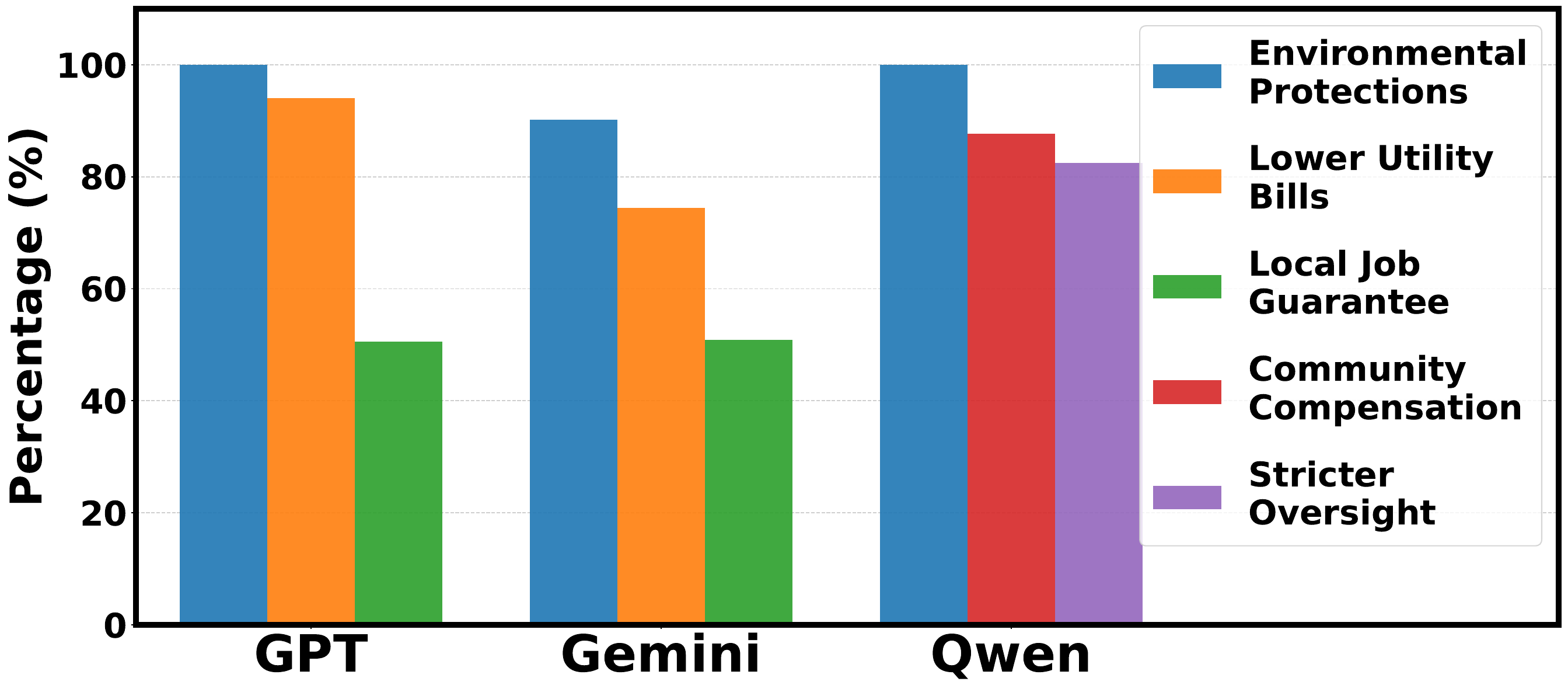}
    }
    \caption{Key differences in cross-model polling results on overall attitudes and support conditions. (a) Overall attitudes for proposed data center. (b) The top conditions that would increase AI agents’ support for the project. Note: See Figure~\ref{fig:cross_model_commonalities} regarding response category display conventions.}    \label{fig:cross_model_attitudes_condition}
\end{figure}

\paragraph{Key Insights}
\begin{itemize}
    \item \textbf{Multi-model analysis reveals complementary viewpoints}: Cross-model comparison reveals that data center project evaluation benefits from incorporating different AI viewpoints. For example, Qwen's emphasis on employment creation and institutional trust offers valuable counterpoints to GPT-5 and Gemini's perspectives.
    
    \item \textbf{Model-specific biases in agent polling}: The observed variations suggest that different LLMs may produce different responses for certain questions, reflecting distinct cultural, economic, and institutional perspectives embedded in their architectures. These differences demonstrate the sensitivity of polling results to model selection in AI-based social research.
\end{itemize}

\subsection{Cross-Regional Analysis}

Regarding environmental aspects, agents in the two regions show highly similar trends. When asked about their concern levels toward potential environmental impacts, over 90\% of agents in both regions express that they are worried, with a minority being "Very Worried". In terms of specific concerns, water consumption and grid impact are the predominant factors for both communities. The next most common concern is carbon emissions, which is chosen by around half of the agents in both Taylor and Loudoun counties.
\begin{figure}[htbp!]
    \centering
    \subfloat[Environmental concern levels]{\includegraphics[width=0.36\textwidth, valign=b]{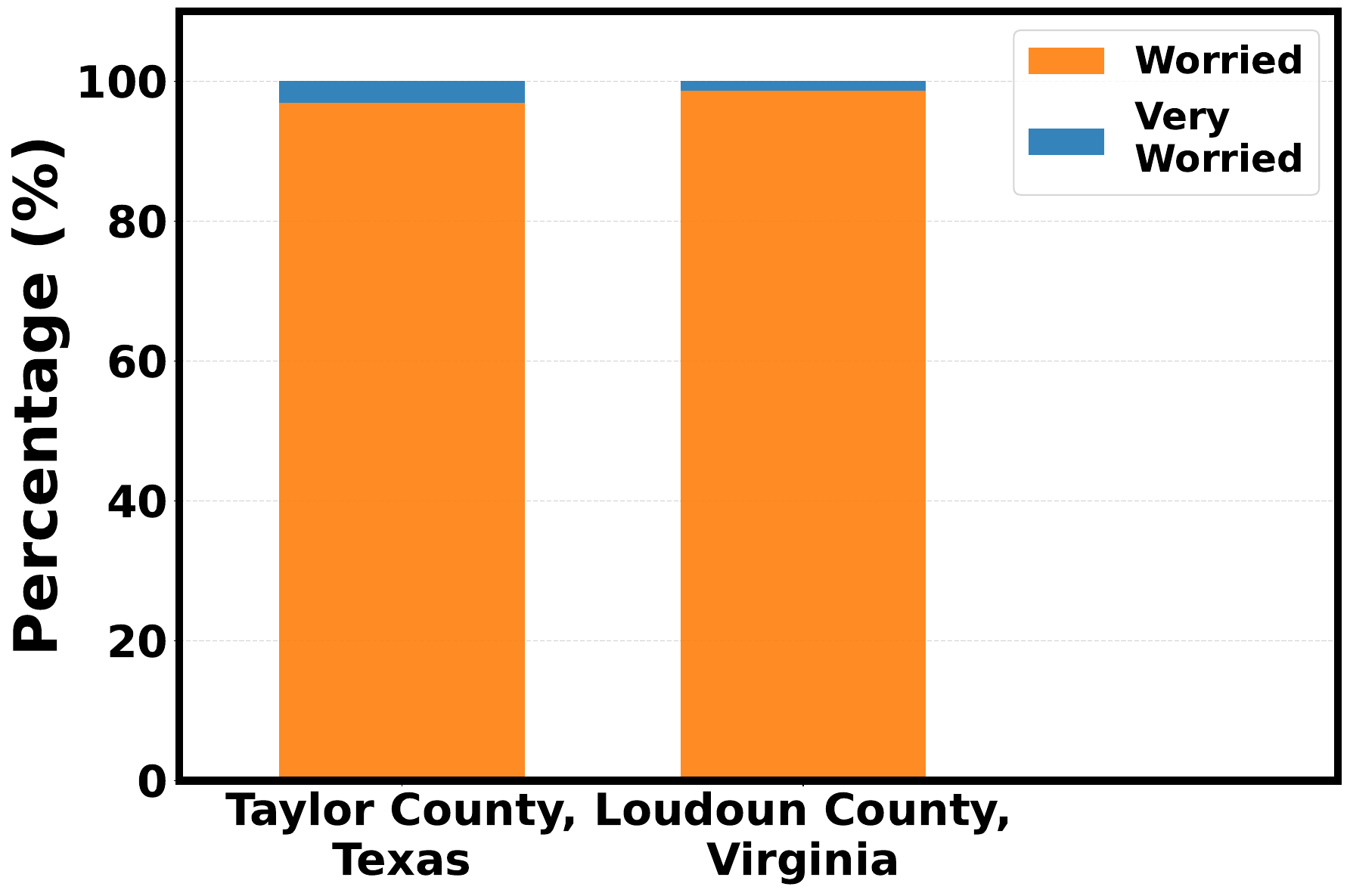}
    }
    \subfloat[Top-3 environmental concerns]
    {\includegraphics[width=0.42\textwidth, valign=b]{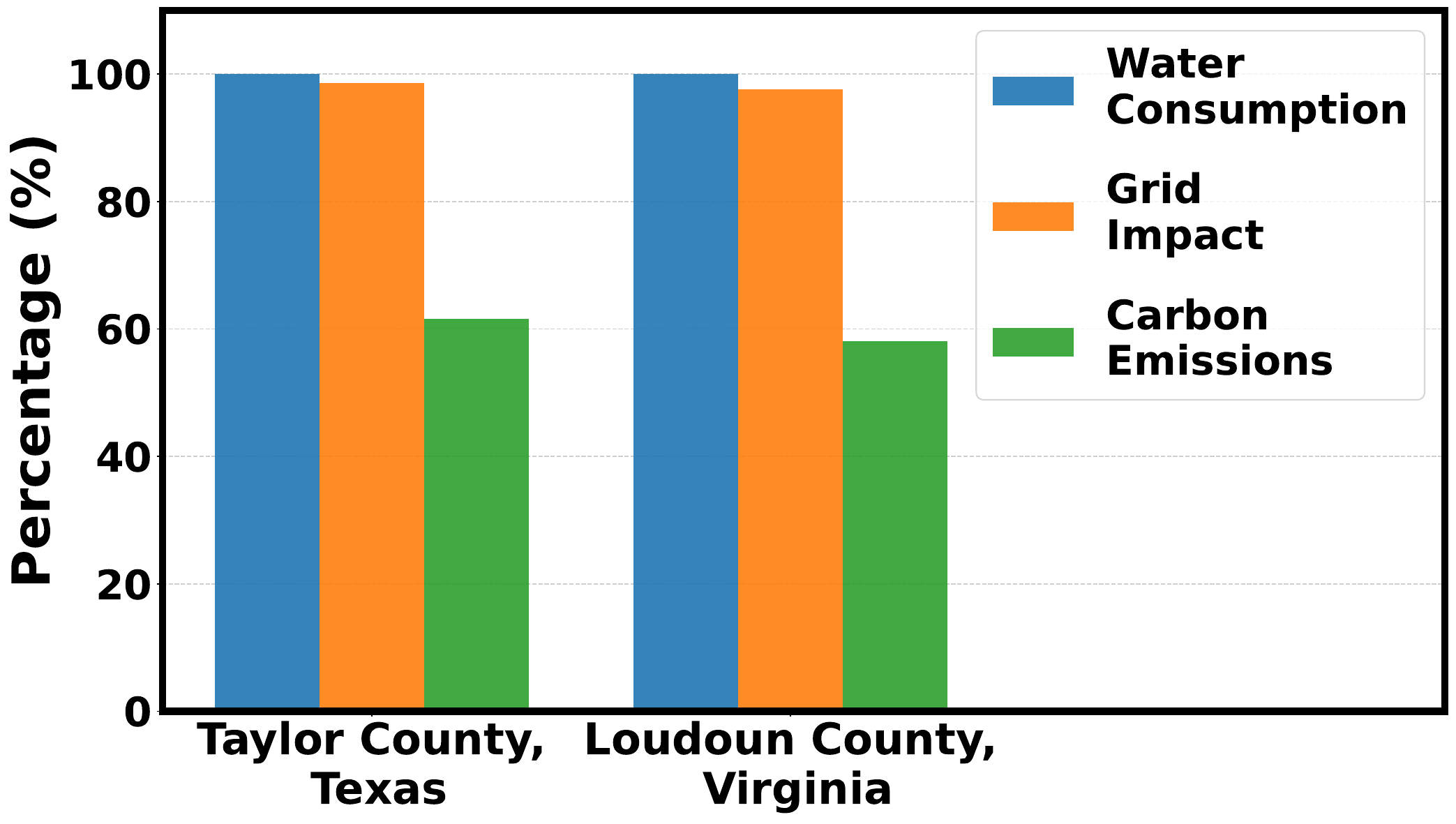}
    }
    \caption{Cross-regional polling results on environmental aspects. (a) Level of concern expressed by agents regarding the project’s potential environmental impacts. (b) The community’s top environmental concerns regarding the data center project.  Note: See Figure~\ref{fig:cross_model_commonalities} regarding response category display conventions.}
    \label{fig:cross_region_environmental}
\end{figure}

Finally, in the domain of governance, we analyze agents' trust in government and their preferred information sources. Regarding trust in government regulation, agents in both counties are predominantly neutral or distrusted. However, a notable difference is that agents in Loudoun County exhibit lower levels of distrust (29\%) compared to those in Taylor County (40\%). This pattern corresponds with their preferred information sources. While agents in both regions identify academic research as the most trustworthy source, a slightly greater proportion of agents in Loudoun County also express trust in federal and local government. This higher trust in official institutions may be explained by two factors. First, the region's extensive experience with data center regulation has established relatively mature oversight frameworks. Second, Loudoun County's historically high government satisfaction ratings~\cite{loudoun2025residentsSurvey} suggest that, on average, residents rate local institutional performance more favorably than in many other jurisdictions.

\begin{figure}[htbp!]
    \centering
    \subfloat[Government trust]{\includegraphics[width=0.36\textwidth, valign=b]{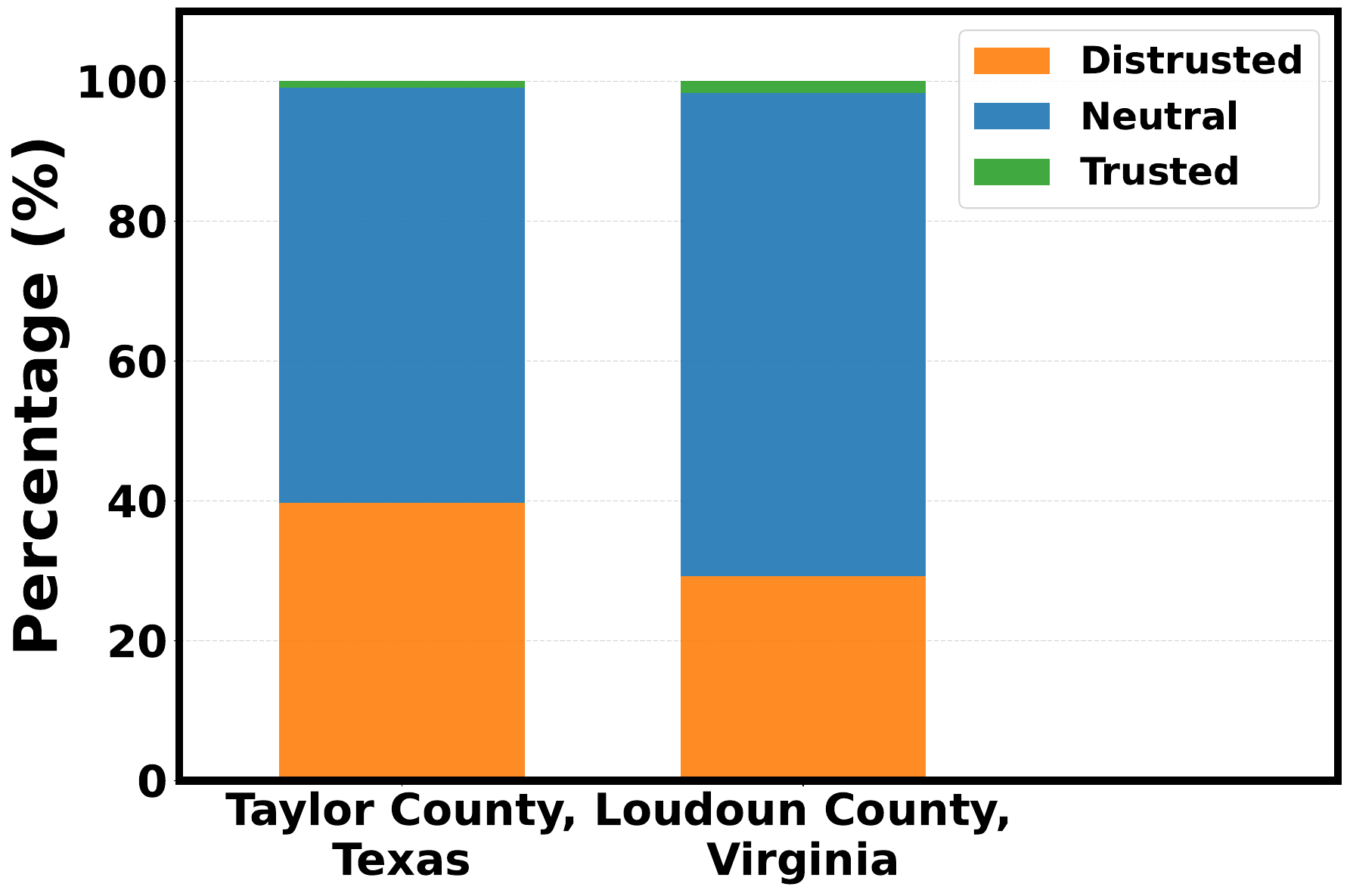}
    }
    \subfloat[Top-3 information source]
    {\includegraphics[width=0.42\textwidth, valign=b]{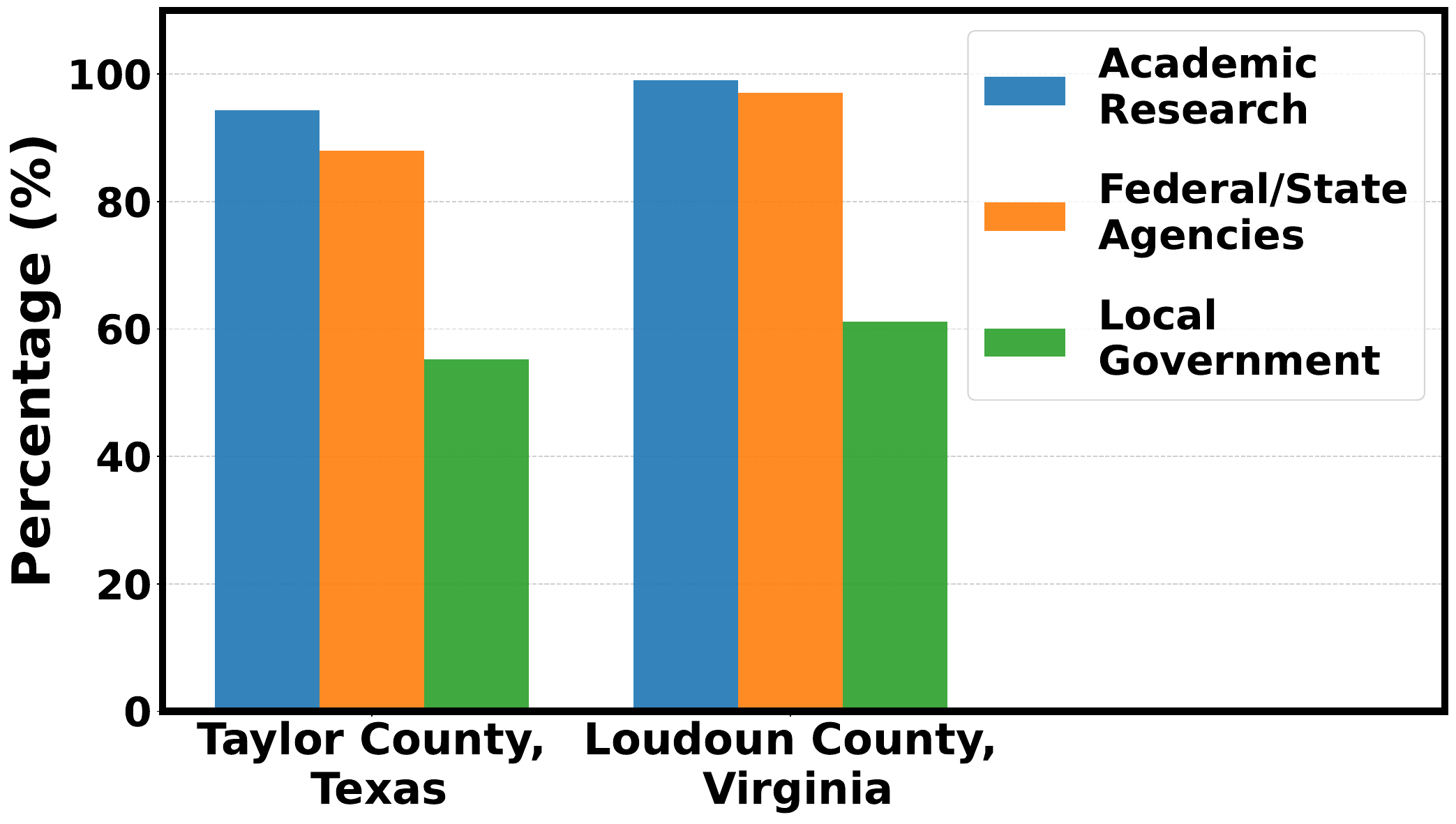}
    }
    \caption{Cross-regional polling results on governance and preferred information sources. (a) Distribution of agents' trust in the government's ability to regulate the data center project. (b) The most trustworthy information sources about the data center project. Note: See Figure~\ref{fig:cross_model_commonalities} regarding response category display conventions.}
    \label{fig:cross_region_governance_info_sources}
\end{figure}

\textbf{Implications} \quad These differences highlight the need for location-specific engagement strategies. For example, communities prioritizing job creation may require stronger employment guarantees and economic development assurances. Varying government trust levels suggest tailored communication approaches are essential, with higher-distrust communities potentially requiring more independent oversight and transparency mechanisms.

\begin{figure}[htbp!]
    \centering
    \subfloat[Support condition (Taylor, Texas)]{
    \includegraphics[width=0.42\linewidth]{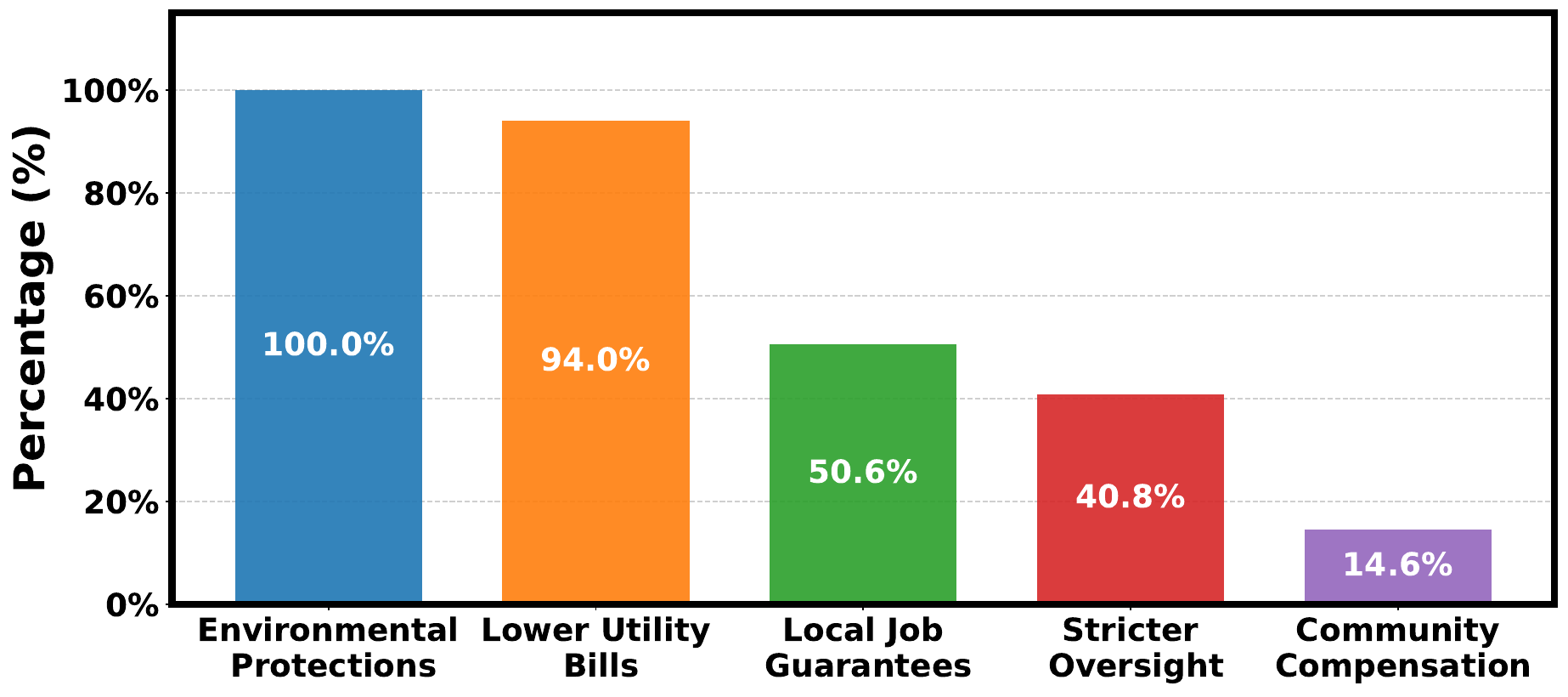}
    \label{fig:support_condition_base_full_taylo}
    }
    \subfloat[Support condition (Loudoun, Virginia)]{
    \includegraphics[width=0.42\linewidth]{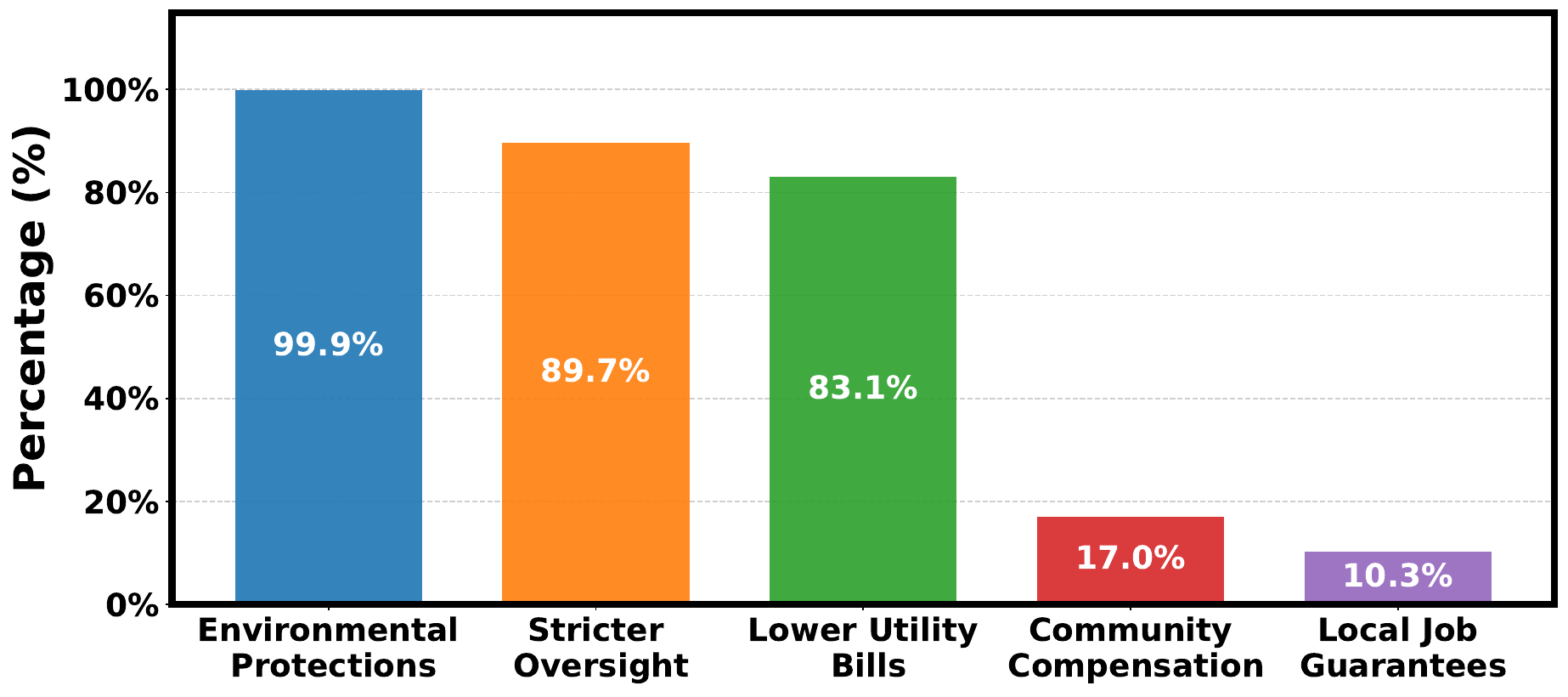}
    \label{fig:support_condition_base_full_loudoun}
    }

    \caption{The top conditions that would increase community support for the project using GPT-5. (a) Taylor County, Texas. (b) Loudoun County, Virginia.}
    \label{fig:support_condition_taylor_loudoun_full}
\end{figure}

\begin{figure}[htbp!]
    \centering
    \subfloat[Economic concern (Taylor, Texas)]{
    \includegraphics[width=0.42\linewidth]{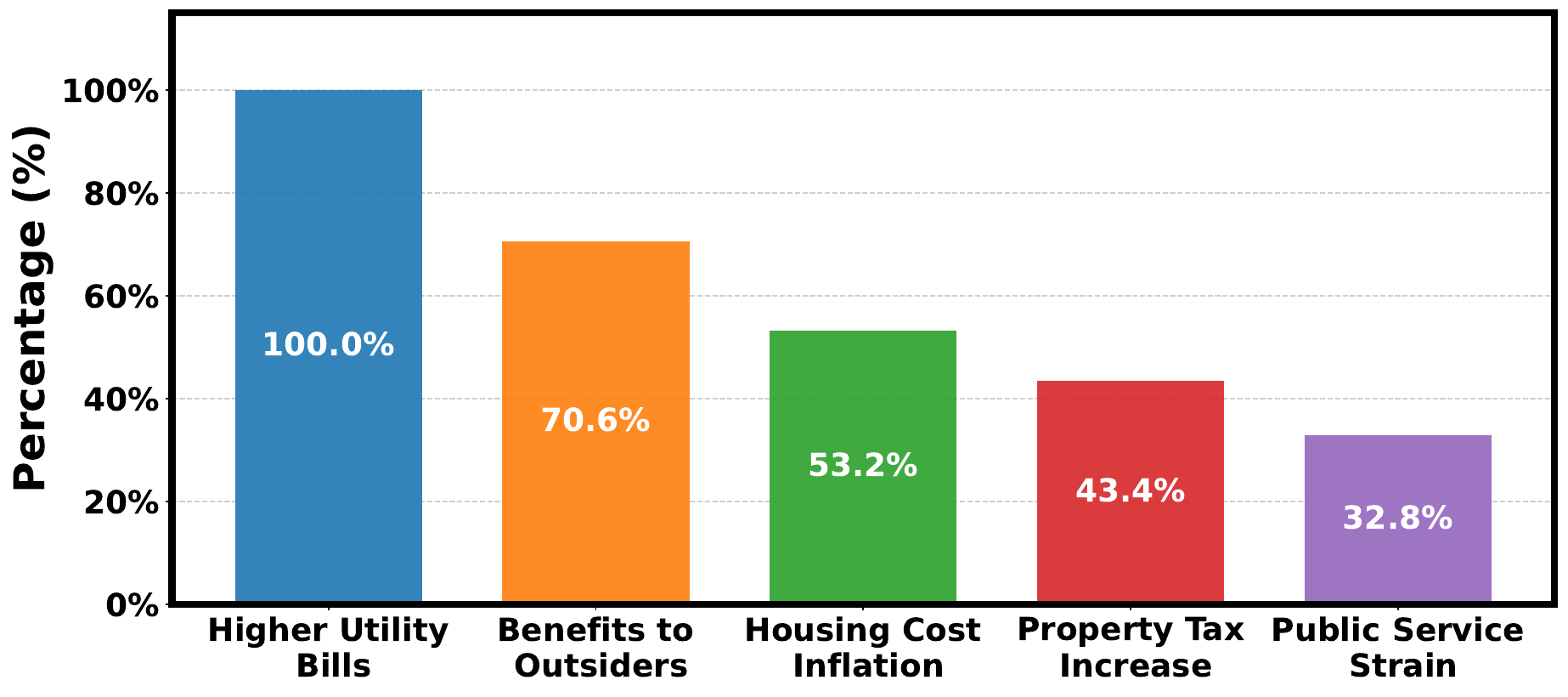}
    \label{fig:economic_concern_base_full_taylor}
    }
    \subfloat[Economic concern (Loudoun, Virginia)]{
    \includegraphics[width=0.42\linewidth]{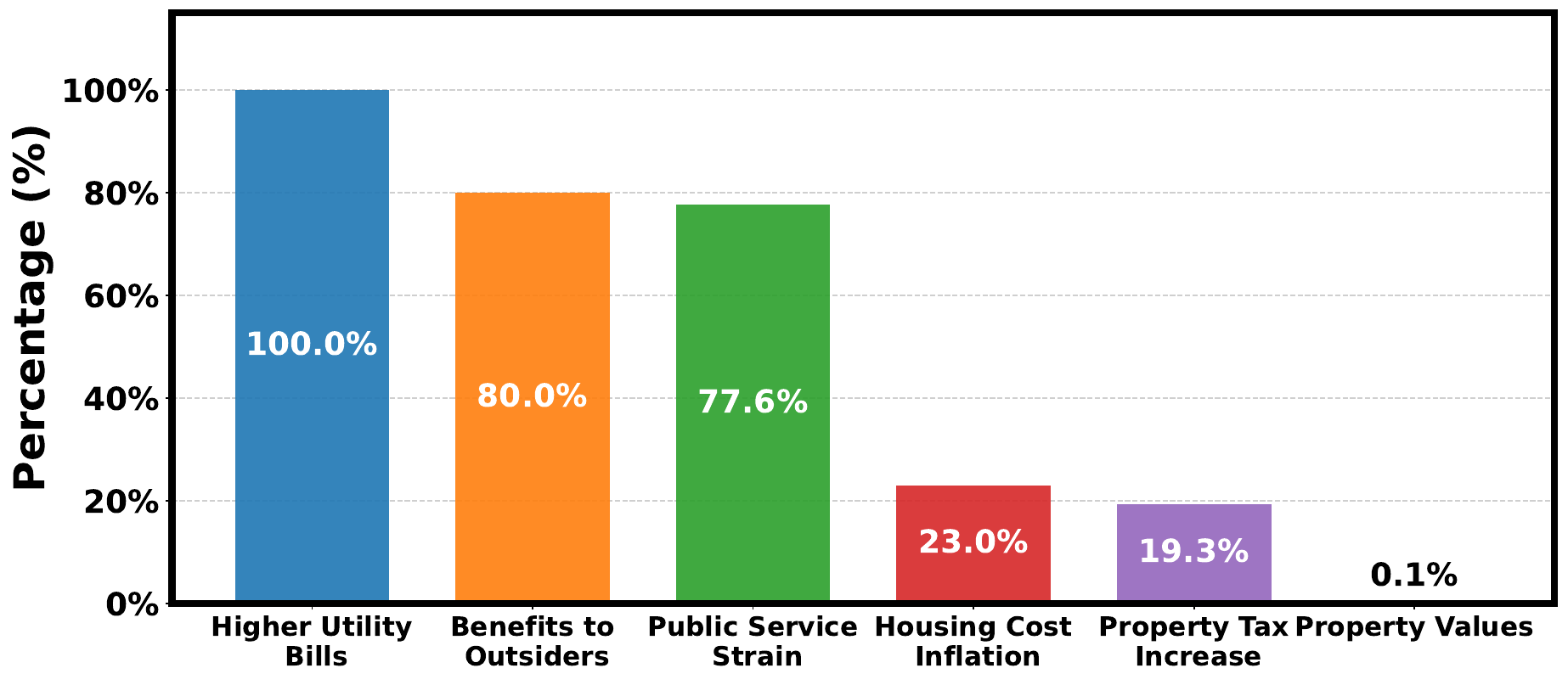}
    \label{fig:economic_concern_base_full_loudoun}
    }

    \caption{Community economic concerns for the project using GPT-5. (a) Taylor County, Texas. (b) Loudoun County, Virginia.}
    \label{fig:economic_concern_taylor_loudoun_full}
\end{figure}

\begin{figure}[htbp!]
    \centering
    \subfloat[Environmental concern (Taylor, Texas)]{
    \includegraphics[width=0.42\linewidth]{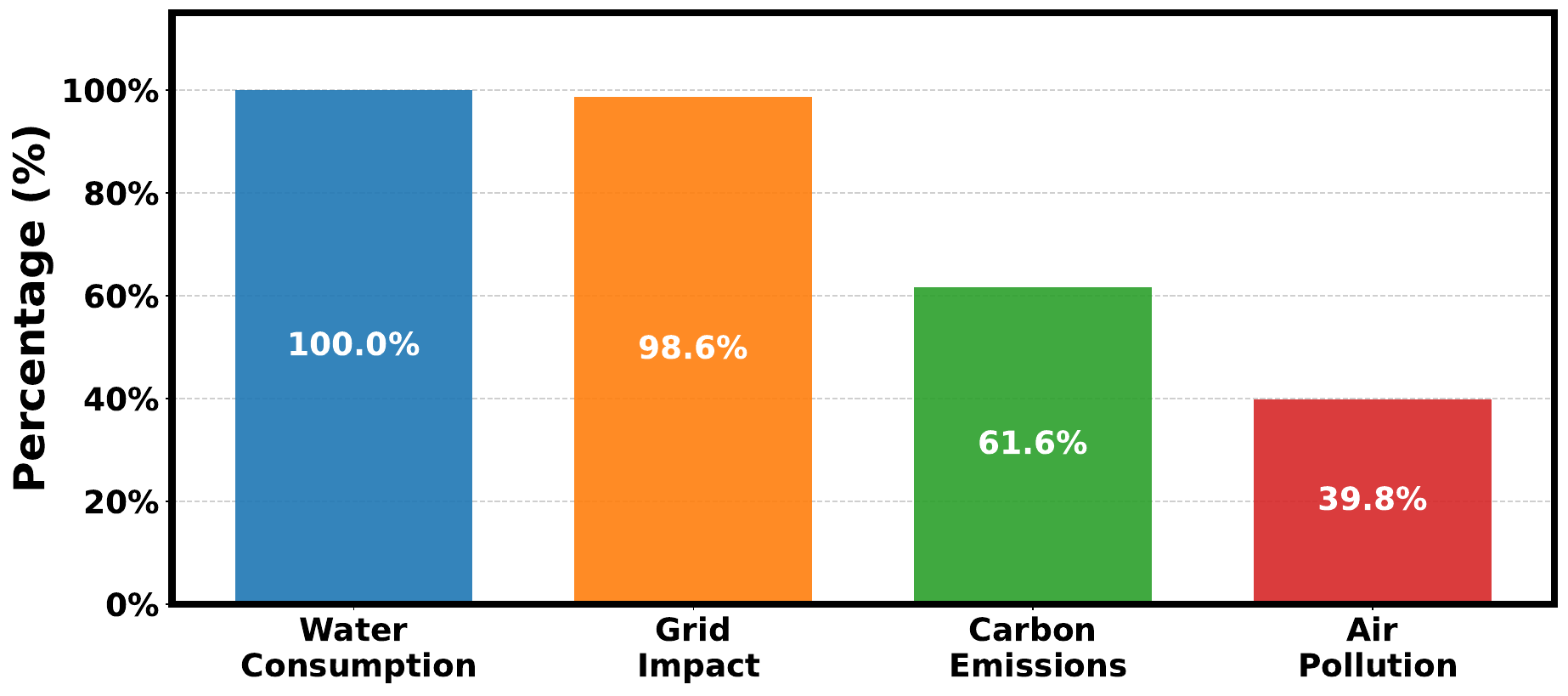}
    \label{fig:environmental_concern_base_full_taylor}
    }
    \subfloat[Environmental concern (Loudoun, Virginia)]{
    \includegraphics[width=0.42\linewidth]{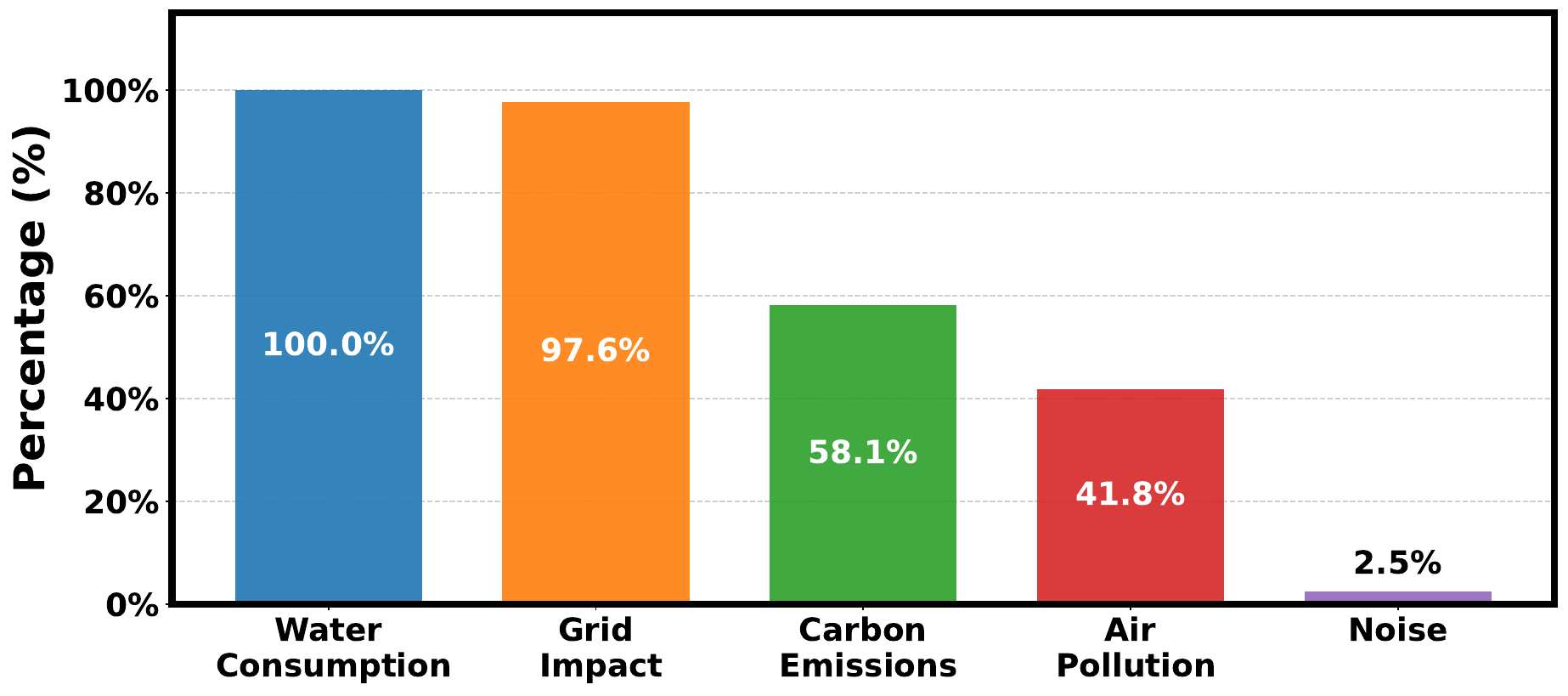}
    \label{fig:environmental_concern_base_full_loudou}
    }

    \caption{The community’s top environmental concerns regarding the data center project using GPT-5. (a) Taylor County, Texas. (b) Loudoun County, Virginia.}
    \label{fig:environment_taylor_loudoun_full}
\end{figure}

\begin{figure}[htbp!]
    \centering
    \subfloat[Information sources (Taylor, Texas)]{
    \includegraphics[width=0.42\linewidth]{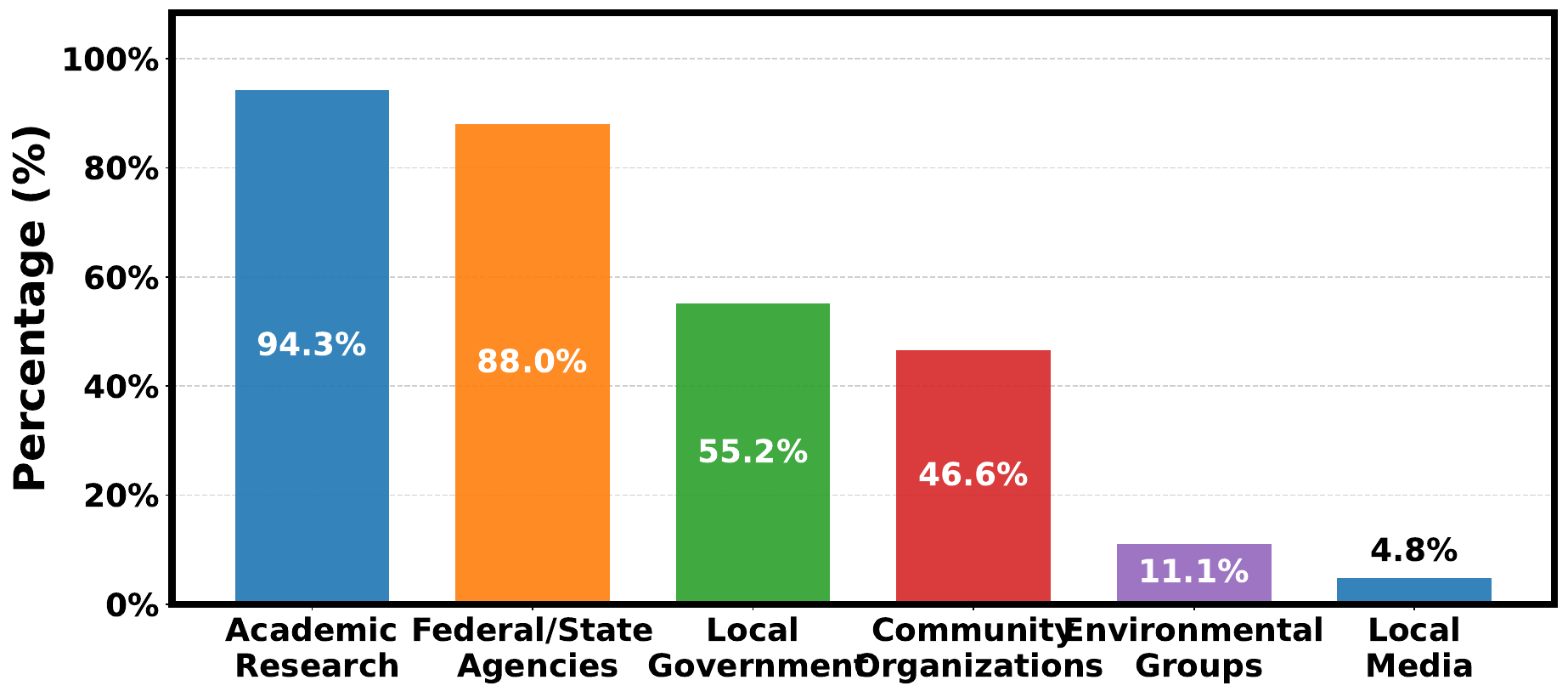}
    \label{fig:information_sources_base_full_taylor}
    }
    \subfloat[Information sources (Loudoun, Virginia)]{
    \includegraphics[width=0.42\linewidth]{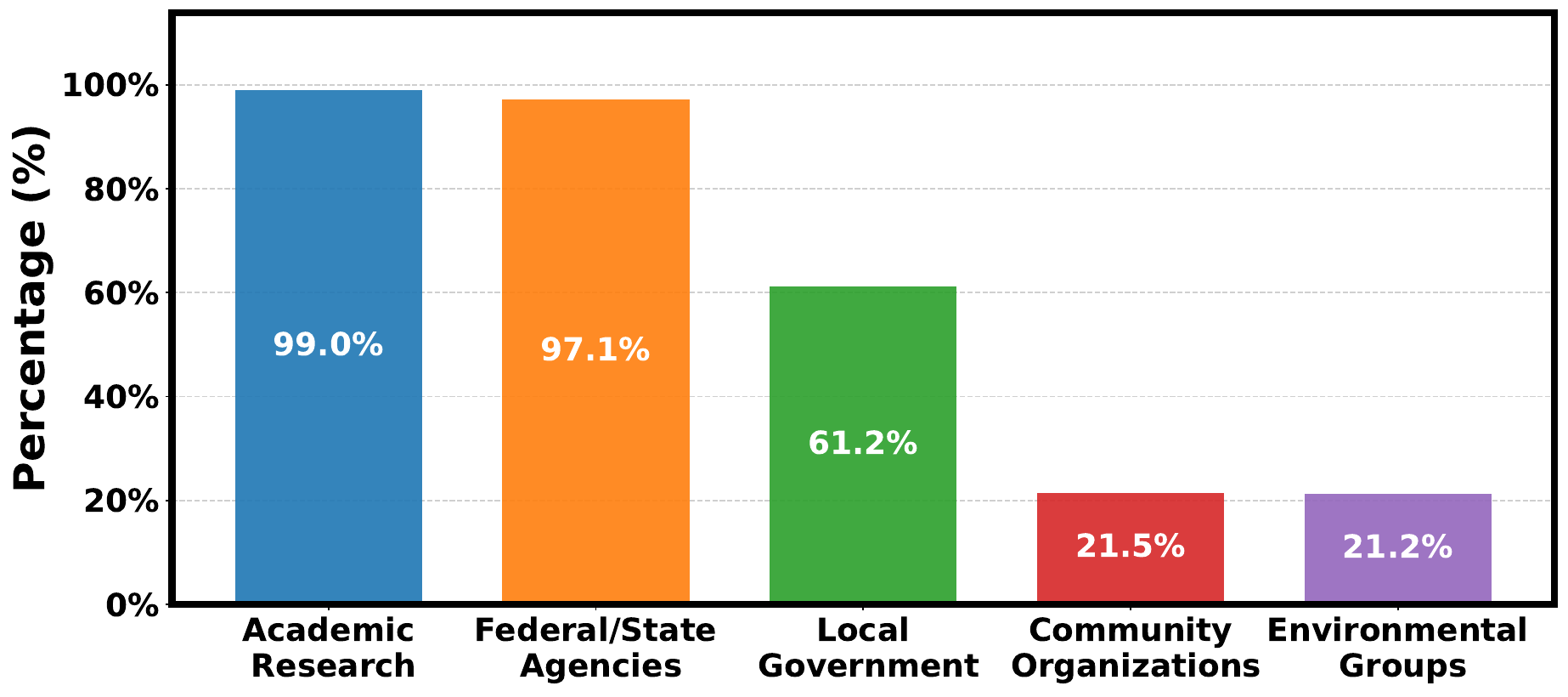}
    \label{fig:information_sources_base_full_loudoun.pdf}
    }

    \caption{The community’s most trusted sources of information regarding the data center project using GPT-5. (a) Taylor County, Texas. (b) Loudoun County, Virginia.}
    \label{fig:info_sources_taylor_loudoun_full}
\end{figure}

\subsubsection{Counterfactual Analysis}
To disentangle the relative influence of the system prompt versus the agents' demographic features on polling results, we conduct a counterfactual analysis. In this experiment, we utilize the AI agent profiles from Taylor County, Texas (identical to the baseline experiment), but provide them with the regional context (system prompt) of Loudoun County, Virginia. The results exhibit a hybrid pattern: while most of attitudes align with the agents' original demographic profiles, specific responses shift to reflect the characteristics of the substituted regional context.

A detailed comparison between this counterfactual experiment (Taylor agents with Loudoun context) and the baseline (Taylor agents with Taylor context) reveals several key divergences. Regarding economic benefits, while tax revenue remains the primary factor in both scenarios, agents in the counterfactual setting demonstrate a significantly higher focus on infrastructure upgrades (97\% vs. 68\%). Similarly, regarding economic concerns, although higher utility bills remain the top consideration, agents attach considerably more importance to public service strain (45\% vs. 29\%). These shifts are directly attributable to the Loudoun County system prompt. Because Loudoun County hosts a high density of data centers, the local community experiences strain on public services and infrastructure~\cite{loudoun2025datacenter}, a factor that directly influences the agents' responses.

In terms of the environmental impacts, agents continue to view water consumption and grid impact as the top two considerations, a pattern consistent across both the Taylor and Loudoun baselines. However, significant differences emerge regarding the conditions required to increase support for the project. In the counterfactual setting, agents demand stricter oversight more frequently (60\% vs. 40\%) and place less emphasis on local job guarantees (29\% vs. 51\%). The increased demand for stricter oversight likely stems from the Loudoun context; given the region's extensive experience with data centers and higher government satisfaction metrics, agents may perceive oversight as both a necessary and effective mechanism. Conversely, the reduced demand for local job guarantees is explained by the economic context of Loudoun County, which is characterized by higher income levels and a more robust economy. This enhanced economic backdrop increases the agents' confidence in general economic prospects, thereby reducing the perceived urgency for specific employment guarantees.

\subsection{Comparison with Real-World Human Polls}

\begin{table}[!htbp]
    \captionsetup{width=\linewidth}
    \caption{Methodological Differences Between Human Poll and AI Agent Polling}
    \renewcommand{\arraystretch}{1.5}
    \centering
    \begin{tabular}{lll}
    \toprule
    \textbf{Aspect} & \textbf{Heatmap Poll} & \textbf{AI Agent Polling} \\
    \hline
    \begin{tabular}[c]{@{}l@{}}Scope\end{tabular} 
    & \begin{tabular}[c]{@{}l@{}}National-level\end{tabular} & \begin{tabular}[c]{@{}l@{}}County-level\end{tabular} \\
    \hline
    \begin{tabular}[c]{@{}l@{}}Context\end{tabular} 
    & \begin{tabular}[c]{@{}l@{}}Data centers vs. other\\energy infrastructures\end{tabular} & \begin{tabular}[c]{@{}l@{}}Data center projects\\only\end{tabular} \\
    \hline
    \begin{tabular}[c]{@{}l@{}}Survey\\Design\end{tabular} 
    & \begin{tabular}[c]{@{}l@{}}Overall assessment\end{tabular} & \begin{tabular}[c]{@{}l@{}}Granular multi-domain\\(13 questions, 5 domains)\end{tabular} \\
    \hline
    \begin{tabular}[c]{@{}l@{}}Population\end{tabular} 
    & \begin{tabular}[c]{@{}l@{}}National voters\end{tabular} & \begin{tabular}[c]{@{}l@{}}Demographically-grounded\\AI agents\end{tabular} \\
    \hline
    \begin{tabular}[c]{@{}l@{}}Others\end{tabular} 
    & \begin{tabular}[c]{@{}l@{}}Political affiliation\\analysis included\end{tabular} & \begin{tabular}[c]{@{}l@{}}No political affiliation\\analysis\end{tabular} \\
    \bottomrule
    \end{tabular}
    \label{tab:human_ai_methodological_differences}
\end{table}

\subsubsection{Key Insights}

The topical comparison with real-world human polls yields several insights into our AI agent polling framework's performance and characteristics.

From the comparison, we find that agent-based polls exhibit subject-based trends highly consistent with real-world human surveys, suggesting the framework can capture meaningful patterns in public opinion. However, the comparison also reveals limitations: AI agents show less diversity in responses than human participants.

Our framework can serve as a valuable supplement to traditional polling. The topical alignment with human surveys, combined with its scalable and customizable nature, suggests it could help decision-makers predict core community concerns regarding data center development.

\clearpage

\section{Prompting Framework}
\subsection{Regional Context Prompt}
\begin{tcolorbox}[breakable, colback=white, colframe=black, boxrule=1pt]
\textbf{State Data Center Context}

[STATE\_NAME] state data center status ([YEAR]): total annual electricity consumption is around [ENERGY\_CONSUMPTION] MWh.

\bigskip

\textbf{Community Profile}

[COUNTY\_NAME] County, [STATE\_NAME] is a community with a total population of [POPULATION]. The population is [FEMALE\_PCT]\% female and [MALE\_PCT]\% male, with a median age of [MEDIAN\_AGE] years.
The racial composition includes [WHITE\_PCT]\% White residents, [ASIAN\_PCT]\% Asian residents, [BLACK\_PCT]\% African American residents, and [HISPANIC\_PCT]\% Hispanic or Latino residents of any race.
The community consists of [TOTAL\_HOUSEHOLDS] households with an average household size of [AVG\_HOUSEHOLD\_SIZE] people per household.
In terms of educational attainment, [BACHELOR\_OR\_HIGHER\_PCT]\% of adults hold a bachelor's degree or higher, including [GRADUATE\_PCT]\% with graduate or professional degrees. Additionally, [COMPUTER\_PCT]\% of households have access to a computer.
The economic profile shows a median household income of \$[MEDIAN\_HOUSEHOLD\_INCOME] and a per capita income of \$[PER\_CAPITA\_INCOME].
The major employment sectors include [TOP\_INDUSTRIES].
For housing, [HOMEOWNERSHIP\_RATE]\% of households own their homes, with a median home value of \$[MEDIAN\_HOME\_VALUE]. Rental households pay a median rent of \$[MEDIAN\_RENT].

\bigskip

\textbf{Proposed Data Center Project and Its Estimated Impact}

The annual electricity consumption of data center project is around [YEARLY\_ENERGY\_CONSUMPTION] MWh.

During the construction phase ([CONSTRUCTION\_DURATION] months), the project is estimated to support approximately [CONSTRUCTION\_JOBS] temporary local jobs, and generates around \$[CONSTRUCTION\_ECONOMIC\_ACTIVITY] million in local economic activity and \$[CONSTRUCTION\_TAX] million in taxes.

Once operational, it is estimated to support nearly [OPERATIONAL\_JOBS] permanent local jobs annually, with an average salary of about \$[SALARY]k, and generates over \$[OPERATIONAL\_ECONOMIC\_ACTIVITY] million in local economic activity and \$[OPERATIONAL\_TAX] million in taxes each year.

The annual water consumption includes: [ONSITE\_WATER] million liters for on-site water consumption for data center cooling, and [OFFSITE\_WATER] million liters for off-site electricity generation.

The annual carbon emissions is [CARBON\_EMISSIONS] million short tons.

The annual air pollutants generated by on-site backup generators includes: NO$_x$ [NOX], VOCs [VOCS], PM$_{2.5}$ [PM25], SO$_2$ [SO2] short tons.

\bigskip

\textbf{Survey Instructions}

You will be asked for your opinions about this proposed data center project. Consider the various impacts of the project on you and your community,including economic factors (such as economic growth, jobs, and tax revenue) and environmental factors (such as energy usage, carbon emissions, water consumption, and air pollution). 
\end{tcolorbox}

\subsection{Demographic Agent Prompt}  
The following is the verbatim prompt used to instruct the LLM agents in our study. The instructional language is designed to be direct and explicit to elicit strong persona adoption from the model.

\begin{tcolorbox}[breakable, colback=white, colframe=black, boxrule=1pt]
ASSUME THE ROLE of this resident: [AGE\_GROUP], [SEX], [RACE], [ETHNICITY], [EDUCATION\_LEVEL] education, [MARITAL\_STATUS], [LANGUAGE\_SPOKEN\_AT\_HOME], [CITIZENSHIP], [EMPLOYMENT\_STATUS], [HOUSEHOLD\_INCOME] household yearly income, [HOUSING], household has [VEHICLES]

Put yourself completely in this person's position. Answer ALL questions from your perspective:

[SURVEY\_QUESTIONS]

Give short, clear answers. Be honest and share your real thoughts even if they're critical.

Please respond in JSON format with the following structure:
\{
    "question\_1\_id": "your\_answer",
    "question\_2\_id": "your\_answer",
    ...
\}
\end{tcolorbox}

\subsection{Community Survey Questionnaire}\label{appendix:questionnaire}
\begin{tcolorbox}[breakable, colback=white, colframe=black, boxrule=1pt]
% \textbf{Section 1: Economic Impacts}

\textbf{1.} What do you think will be the overall economic impact of this data center on the local community? Select the one option that best represents your view.
\begin{itemize}
\item Very Positive
\item Positive  
\item Mixed
\item Negative
\item Very Negative
\item Unsure
\end{itemize}

\textbf{2.} Which economic benefits are most important for your community? Select up to three that you consider most important. Separate your answers with a comma only. 
\begin{itemize}
\item Job Creation
\item Tax Revenue
\item Infrastructure Upgrades
\item Business Growth
\item Property Values
\item Economic Diversity
\item Other (please specify)
\end{itemize}

\textbf{3.} What economic costs or burdens concern you the most about this data center? Select up to three that you consider most important. Separate your answers with a comma only.
\begin{itemize}
\item Higher Utility Bills
\item Property Tax Increases
\item Job Competition
\item Housing Cost Inflation
\item Public Service Strain
\item Benefits to Outsiders
\item No Major Concerns
\item Other (please specify)
\end{itemize}

% \textbf{Section 2: Environmental Concerns}

\textbf{4.} How worried are you about the potential environmental impacts of the data center? Select the one option that best represents your view.
\begin{itemize}
\item Very Worried
\item Worried
\item Neutral
\item Unconcerned
\item Very Unconcerned
\end{itemize}

\textbf{5.} Which potential environmental impacts of this data center concern you the most? Select up to three that you consider most important. Separate your answers with a comma only.
\begin{itemize}
\item Water Consumption
\item Carbon Emissions
\item Air Pollution
\item Grid Impact
\item Heat Generation
\item Noise
\item No Major Concerns
\item Other (please specify)
\end{itemize}

\textbf{6.} What environmental protections should be required for this data center? Select up to three that you consider most important. Separate your answers with a comma only.
\begin{itemize}
\item Water Conservation
\item Renewable Energy
\item Air Quality Monitoring
\item Noise Limits
\item Green Building
\item Environmental Transparency
\item No Special Requirements
\item Other (please specify)
\end{itemize}

% \textbf{Section 3: Community Engagement \& Decision Making}

\textbf{7.} If given the opportunity to participate in planning discussions for the data center project, would you be willing to participate? Select the one option that best represents your view.
\begin{itemize}
\item Very Willing
\item Willing
\item Neutral
\item Unwilling
\item Very Unwilling
\end{itemize}

\textbf{8.} How much do you trust the government and relevant departments' ability to regulate data center operations? Select the one option that best represents your view.
\begin{itemize}
\item Very Trusted
\item Trusted
\item Neutral
\item Distrusted
\item Very Distrusted
\end{itemize}

\textbf{9.} Which sources of information would you trust most for this project? Select up to three that you consider most important. Separate your answers with a comma only.
\begin{itemize}
\item Environmental Groups
\item Local Government
\item Community Organizations
\item Academic Research
\item Developer Information
\item Federal/State Agencies
\item Local Media
\item Other (please specify)
\end{itemize}

% \textbf{Section 4: Personal \& Community Impact}

\textbf{10.} How do you expect this data center to personally affect you and your household? Select the one option that best represents your view.
\begin{itemize}
\item Very Positive
\item Positive
\item Mixed
\item No Impact
\item Negative
\item Very Negative
\item Other
\end{itemize}

% \textbf{Section 5: Conditions \& Alternatives}

\textbf{11.} What would make you more supportive of this data center project? Select up to three that you consider most important. Separate your answers with a comma only.
\begin{itemize}
\item Already Support
\item Lower Utility Bills
\item Environmental Protections
\item Local Job Guarantees
\item Community Compensation
\item Stricter Oversight
\item Smaller Scale
\item Nothing Would Help
\item Other (please specify)
\end{itemize}

% \textbf{Section 6: Overall Assessment}

\textbf{12.} Would you support or oppose a data center built near your community? Select the one option that best represents your view.
\begin{itemize}
\item Strongly Support
\item Support
\item Neutral
\item Oppose
\item Strongly Oppose
\end{itemize}

\textbf{13.} What is the most important thing decision-makers should know about your views on this data center project? Share your key message or main concern.
\begin{itemize}
\item No additional thoughts
\item Other (please specify)
\end{itemize}
\end{tcolorbox}

\newpage
\section*{NeurIPS Paper Checklist}

\begin{enumerate}

\item {\bf Claims}
    \item[] Question: Do the main claims made in the abstract and introduction accurately reflect the paper's contributions and scope?
    \item[] Answer: \answerYes{} % Replace by \answerYes{}, \answerNo{}, or \answerNA{}.
    \item[] Justification: The main claim is to provide a scalable way to gauge public engagement on data centers. We achieve that through Section~\ref{main:exp}.

\item {\bf Limitations}
    \item[] Question: Does the paper discuss the limitations of the work performed by the authors?
    \item[] Answer: \answerYes{} % Replace by \answerYes{}, \answerNo{}, or \answerNA{}.
    \item[] Justification: We have discussion about limitation, see Section ~\ref{main：conclusion}.

\item {\bf Theory assumptions and proofs}
    \item[] Question: For each theoretical result, does the paper provide the full set of assumptions and a complete (and correct) proof?
    \item[] Answer: \answerNA{} % Replace by \answerYes{}, \answerNo{}, or \answerNA{}.
    \item[] Justification: The paper does not include theoretical results. 

    \item {\bf Experimental result reproducibility}
    \item[] Question: Does the paper fully disclose all the information needed to reproduce the main experimental results of the paper to the extent that it affects the main claims and/or conclusions of the paper (regardless of whether the code and data are provided or not)?
    \item[] Answer: \answerYes{} % Replace by \answerYes{}, \answerNo{}, or \answerNA{}.
    \item[] Justification: We provide details in the main body for understanding results (Section ~\ref{main:exp}), while giving full details in Appendix.

\item {\bf Open access to data and code}
    \item[] Question: Does the paper provide open access to the data and code, with sufficient instructions to faithfully reproduce the main experimental results, as described in supplemental material?
    \item[] Answer: \answerYes{} % Replace by \answerYes{}, \answerNo{}, or \answerNA{}.
    \item[] Justification: Data and code will be released upon publication of the paper.

\item {\bf Experimental setting/details}
    \item[] Question: Does the paper specify all the training and test details (e.g., data splits, hyperparameters, how they were chosen, type of optimizer, etc.) necessary to understand the results?
    \item[] Answer: \answerYes{} % Replace by \answerYes{}, \answerNo{}, or \answerNA{}.
    \item[] Justification: API costs of foundation models are listed in Section ~\ref{main:exp}, and the research does not involve traditional training and test processes.

\item {\bf Experiment statistical significance}
    \item[] Question: Does the paper report error bars suitably and correctly defined or other appropriate information about the statistical significance of the experiments?
    \item[] Answer: \answerNo{} % Replace by \answerYes{}, \answerNo{}, or \answerNA{}.
    \item[] Justification: Due to time and resource constraints (Section~\ref{main:exp}), we do not conduct repeated experiments. We instead establish robustness via multi-model (GPT-5, Gemini-2.5-Pro, Qwen-Max) and cross-regional comparison (2 counties), and alignment with human polls. All sampled agents will be released upon publication to enable reproducibility.

\item {\bf Experiments compute resources}
    \item[] Question: For each experiment, does the paper provide sufficient information on the computer resources (type of compute workers, memory, time of execution) needed to reproduce the experiments?
    \item[] Answer: \answerYes{} % Replace by \answerYes{}, \answerNo{}, or \answerNA{}.
    \item[] Justification: The resources used are mainly the API calls to LLMs, and we provide detailed costs in Section~\ref{main:exp}. 
    
\item {\bf Code of ethics}
    \item[] Question: Does the research conducted in the paper conform, in every respect, with the NeurIPS Code of Ethics \url{https://neurips.cc/public/EthicsGuidelines}?
    \item[] Answer: \answerYes{}{} % Replace by \answerYes{}, \answerNo{}, or \answerNA{}.
    \item[] Justification: The research conform with the NeurIPS Code of Ethics.

\item {\bf Broader impacts}
    \item[] Question: Does the paper discuss both potential positive societal impacts and negative societal impacts of the work performed?
    \item[] Answer: \answerYes{} % Replace by \answerYes{}, \answerNo{}, or \answerNA{}.
    \item[] Justification: We discuss potential positive impacts in the Section~\ref{main:intro} and Section~\ref{main：conclusion}, and possible negative impacts in disclosure part of Section~\ref{main:intro}.
    
\item {\bf Safeguards}
    \item[] Question: Does the paper describe safeguards that have been put in place for responsible release of data or models that have a high risk for misuse (e.g., pretrained language models, image generators, or scraped datasets)?
    \item[] Answer: \answerNA{} % Replace by \answerYes{}, \answerNo{}, or \answerNA{}.
    \item[] Justification: The paper poses no such risks.

\item {\bf Licenses for existing assets}
    \item[] Question: Are the creators or original owners of assets (e.g., code, data, models), used in the paper, properly credited and are the license and terms of use explicitly mentioned and properly respected?
    \item[] Answer: \answerYes{} % Replace by \answerYes{}, \answerNo{}, or \answerNA{}.
    \item[] Justification: All foundation models and data sources used are publicly available and credited throughout the paper.

\item {\bf New assets}
    \item[] Question: Are new assets introduced in the paper well documented and is the documentation provided alongside the assets?
    \item[] Answer: \answerNA{} % Replace by \answerYes{}, \answerNo{}, or \answerNA{}.
    \item[] Justification: The paper does not release new assets.

\item {\bf Crowdsourcing and research with human subjects}
    \item[] Question: For crowdsourcing experiments and research with human subjects, does the paper include the full text of instructions given to participants and screenshots, if applicable, as well as details about compensation (if any)? 
    \item[] Answer: \answerNA{} % Replace by \answerYes{}, \answerNo{}, or \answerNA{}.
    \item[] Justification: The paper does not involve crowdsourcing nor research with human subjects.

\item {\bf Institutional review board (IRB) approvals or equivalent for research with human subjects}
    \item[] Question: Does the paper describe potential risks incurred by study participants, whether such risks were disclosed to the subjects, and whether Institutional Review Board (IRB) approvals (or an equivalent approval/review based on the requirements of your country or institution) were obtained?
    \item[] Answer: \answerNA{} % Replace by \answerYes{}, \answerNo{}, or \answerNA{}.
    \item[] Justification: The paper does not involve crowdsourcing nor research with human subjects.

\item {\bf Declaration of LLM usage}
    \item[] Question: Does the paper describe the usage of LLMs if it is an important, original, or non-standard component of the core methods in this research? Note that if the LLM is used only for writing, editing, or formatting purposes and does not impact the core methodology, scientific rigorousness, or originality of the research, declaration is not required.
    %this research? 
    \item[] Answer: \answerYes{} % Replace by \answerYes{}, \answerNo{}, or \answerNA{}.
    \item[] Justification: We use LLMs as the backbone of our AI agent system, and we list details in Section~\ref{main:exp} and the Appendix.

\end{enumerate}


\begin{thebibliography}{10}

\bibitem{conformal_prediction_intro}
Anastasios~N. Angelopoulos and Stephen Bates.
\newblock A gentle introduction to conformal prediction and distribution-free uncertainty quantification, 2022.

\bibitem{inherentbiasllm}
Noel~F. Ayoub, Karthik Balakrishnan, Marc~S. Ayoub, Thomas~F. Barrett, Abel~P. David, and Stacey~T. Gray.
\newblock Inherent bias in large language models: A random sampling analysis.
\newblock {\em Mayo Clinic Proceedings: Digital Health}, 2(2):186--191, 2024.

\bibitem{estimate_matrices_from_marginal}
Michael Bacharach.
\newblock Estimating nonnegative matrices from marginal data.
\newblock {\em International Economic Review}, 6(3):294--310, 1965.

\bibitem{measuring_public_opinion}
Adam~J. Berinsky.
\newblock Measuring public opinion with surveys.
\newblock {\em Annual Review of Political Science}, 20:309--329, 2017.

\bibitem{foundation_models}
Rishi Bommasani, Drew~A. Hudson, Ehsan Adeli, Russ Altman, Simran Arora, Sydney von Arx, Michael~S. Bernstein, Jeannette Bohg, Antoine Bosselut, Emma Brunskill, Erik Brynjolfsson, Shyamal Buch, Dallas Card, Rodrigo Castellon, Niladri Chatterji, Annie Chen, Kathleen Creel, Jared~Quincy Davis, Dora Demszky, Chris Donahue, Moussa Doumbouya, Esin Durmus, Stefano Ermon, John Etchemendy, Kawin Ethayarajh, Li~Fei-Fei, Chelsea Finn, Trevor Gale, Lauren Gillespie, Karan Goel, Noah Goodman, Shelby Grossman, Neel Guha, Tatsunori Hashimoto, Peter Henderson, John Hewitt, Daniel~E. Ho, Jenny Hong, Kyle Hsu, Jing Huang, Thomas Icard, Saahil Jain, Dan Jurafsky, Pratyusha Kalluri, Siddharth Karamcheti, Geoff Keeling, Fereshte Khani, Omar Khattab, Pang~Wei Koh, Mark Krass, Ranjay Krishna, Rohith Kuditipudi, Ananya Kumar, Faisal Ladhak, Mina Lee, Tony Lee, Jure Leskovec, Isabelle Levent, Xiang~Lisa Li, Xuechen Li, Tengyu Ma, Ali Malik, Christopher~D. Manning, Suvir Mirchandani, Eric Mitchell, Zanele Munyikwa, Suraj Nair,
  Avanika Narayan, Deepak Narayanan, Ben Newman, Allen Nie, Juan~Carlos Niebles, Hamed Nilforoshan, Julian Nyarko, Giray Ogut, Laurel Orr, Isabel Papadimitriou, Joon~Sung Park, Chris Piech, Eva Portelance, Christopher Potts, Aditi Raghunathan, Rob Reich, Hongyu Ren, Frieda Rong, Yusuf Roohani, Camilo Ruiz, Jack Ryan, Christopher Ré, Dorsa Sadigh, Shiori Sagawa, Keshav Santhanam, Andy Shih, Krishnan Srinivasan, Alex Tamkin, Rohan Taori, Armin~W. Thomas, Florian Tramèr, Rose~E. Wang, William Wang, Bohan Wu, Jiajun Wu, Yuhuai Wu, Sang~Michael Xie, Michihiro Yasunaga, Jiaxuan You, Matei Zaharia, Michael Zhang, Tianyi Zhang, Xikun Zhang, Yuhui Zhang, Lucia Zheng, Kaitlyn Zhou, and Percy Liang.
\newblock On the opportunities and risks of foundation models, 2022.

\bibitem{survey_llm_agents}
Shuaihang Chen, Yuanxing Liu, Wei Han, Weinan Zhang, and Ting Liu.
\newblock A survey on llm-based multi-agent system: Recent advances and new frontiers in application, 2025.

\bibitem{surveyllmbasedsystem}
Shuaihang Chen, Yuanxing Liu, Wei Han, Weinan Zhang, and Ting Liu.
\newblock A survey on llm-based multi-agent system: Recent advances and new frontiers in application, 2025.

\bibitem{IPFreview}
Abdoul-Ahad Choupani and Amir~Reza Mamdoohi.
\newblock Population synthesis using iterative proportional fitting (ipf): A review and future research.
\newblock {\em Transportation Research Procedia}, 17:223--233, 2016.
\newblock International Conference on Transportation Planning and Implementation Methodologies for Developing Countries (12th TPMDC) Selected Proceedings, IIT Bombay, Mumbai, India, 10-12 December 2014.

\bibitem{datacentermap_aws_loudoun_2025}
{Data Center Map}.
\newblock Amazon aws iad - loudoun county campus, 2025.
\newblock Data center facility listing; Part of AWS us-east-1 region.

\bibitem{social_media_public_opinion_analyses}
Xuefan Dong and Ying Lian.
\newblock A review of social media-based public opinion analyses: Challenges and recommendations.
\newblock {\em Technology in Society}, 67:101724, 2021.

\bibitem{EPRIreport}
{Electric Power Research Institute}.
\newblock Powering intelligence: Analyzing artificial intelligence and data center energy consumption.
\newblock Technical Report 3002028905, Electric Power Research Institute, Palo Alto, CA, 2024.

\bibitem{google_datacenter_efficiency}
Google.
\newblock {Google's Data Center Efficiency}.

\bibitem{bertopic}
Maarten Grootendorst.
\newblock Bertopic: Neural topic modeling with a class-based tf-idf procedure, 2022.

\bibitem{lifecycle_datacenter}
Fehmi {Görkem Üçtuğ} and Tayyar {Can Ünver}.
\newblock Life cycle assessment-based environmental impact analysis of a tier 4 data center: A case study in turkey.
\newblock {\em Sustainable Energy Technologies and Assessments}, 56:103076, 2023.

\bibitem{han2024unpaid}
Yuelin Han, Zhifeng Wu, Pengfei Li, Adam Wierman, and Shaolei Ren.
\newblock The unpaid toll: Quantifying the public health impact of ai, 2024.

\bibitem{dialectpred}
Valentin Hofmann, Pratyusha~Ria Kalluri, Dan Jurafsky, and Sharese King.
\newblock Dialect prejudice predicts ai decisions about people's character, employability, and criminality, 2024.

\bibitem{survey_methods_traditional}
Gerald Kosicki.
\newblock Survey methods, traditional, and public opinion polling.
\newblock pages 1--5, 09 2020.

\bibitem{Shaolei_Water_AI_Thirsty_arXiv_2023_li2023making}
Pengfei Li, Jianyi Yang, Mohammad~A. Islam, and Shaolei Ren.
\newblock Making {AI} less ``thirsty'': Uncovering and addressing the secret water footprint of {AI} models.
\newblock {\em Communications of the ACM}, 2025.

\bibitem{ling2025bias}
Lin Ling, Fazle Rabbi, Song Wang, and Jinqiu Yang.
\newblock Bias unveiled: Investigating social bias in llm-generated code.
\newblock In {\em Proceedings of the AAAI Conference on Artificial Intelligence}, volume~39, pages 27491--27499, 2025.

\bibitem{loudoun2025residentsSurvey}
{Loudoun County Government}.
\newblock Loudoun residents survey shows satisfaction with county services, quality of life.
\newblock County News \& Announcements.

\bibitem{breakingbias}
Sibo Ma, Alejandro Salinas, Julian Nyarko, and Peter Henderson.
\newblock Breaking down bias: On the limits of generalizable pruning strategies.
\newblock In {\em Proceedings of the 2025 ACM Conference on Fairness, Accountability, and Transparency}, pages 2437--2450, 2025.

\bibitem{microsoft_northern_virginia_2025}
{Microsoft Corporation}.
\newblock Northern virginia, 2025.
\newblock Datacenter Community Pledge and local operations information.

\bibitem{Ngata2025cloud_next_door}
Wacuka~M Ngata, Noman Bashir, Michelle Westerlaken, Laurent Liote, Yasra Chandio, and Elsa Olivetti.
\newblock The cloud next door: Investigating the environmental and socioeconomic strain of datacenters on local communities.
\newblock In {\em Proceedings of the 2025 ACM SIGCAS/SIGCHI Conference on Computing and Sustainable Societies}, COMPASS '25. ACM, 2025.

\bibitem{nguyen2025datacenterstown}
Terry Nguyen and Ben Green.
\newblock What happens when data centers come to town.
\newblock Technical report, Science, Technology, and Public Policy Program, Gerald R. Ford School of Public Policy, University of Michigan, July 2025.

\bibitem{NVTC2024datacenter_impact}
{Northern Virginia Technology Council}.
\newblock The impact of data centers on virginia's state and local economies: 5th biennial report.
\newblock Technical Report, prepared by Mangum Economics, April 2024.

\bibitem{Noveck2025governing_ai}
Beth~Simone Noveck.
\newblock Governing with ai - learning the how-to's of ai-enhanced public engagement.
\newblock The GovLab Blog, September 2025.

\bibitem{llmbi}
Abiodun~Finbarrs Oketunji, Muhammad Anas, and Deepthi Saina.
\newblock Large language model (llm) bias index--llmbi.
\newblock {\em arXiv preprint arXiv:2312.14769}, 2023.

\bibitem{examine_survey_costs}
K.~Olson, J.~Stevenson, N.~Assad, L.~Witt-Swanson, C.~P.~E. Jones, A.~Ganshert, and J.~Dykema.
\newblock Examining variation in survey costs across surveys.
\newblock {\em Sociological Methods \& Research}, 0(0):00491241241298914, 2024.

\bibitem{openai2024promptcaching}
{OpenAI}.
\newblock Prompt caching in the api, 2024.
\newblock Accessed: September 10, 2025.

\bibitem{openai_stargate_oracle_2025}
{OpenAI}.
\newblock Stargate advances with 4.5 gw partnership with oracle, July 2025.
\newblock Company blog post.

\bibitem{ai_town}
Joon~Sung Park, Joseph~C. O'Brien, Carrie~J. Cai, Meredith~Ringel Morris, Percy Liang, and Michael~S. Bernstein.
\newblock Generative agents: Interactive simulacra of human behavior, 2023.

\bibitem{datacenters_job}
Nam Pham.
\newblock Data centers: Jobs and opportunities in communities nationwide.
\newblock Technical report, SSRN, May 2017.

\bibitem{pham2017datacenters}
Nam~D. Pham.
\newblock Data centers: Jobs and opportunities in communities nationwide.
\newblock Technical report, U.S. Chamber Technology Engagement Center, May 2017.

\bibitem{piedmont_environmental_council_2024}
{Piedmont Environmental Council}.
\newblock Data centers, diesel generators and air quality -- {PEC} web map.
\newblock \url{https://www.pecva.org/uncategorized/data-centers-diesel-generators-and-air-quality-pec-web-map/}, 2024.
\newblock Accessed: December 2024.

\bibitem{reig2022guidance}
Paul Reig, Tianyi Luo, Eric Christensen, and Julie Sinistore.
\newblock Guidance for calculating water use embedded in purchased electricity.
\newblock Working paper, World Resources Institute, August 2022.
\newblock Collaboration with WSP USA.

\bibitem{llmbias}
Philip Resnik.
\newblock Large language models are biased because they are large language models.
\newblock {\em Computational Linguistics}, pages 1--21, 2025.

\bibitem{tutorialconformalprediction}
Glenn Shafer and Vladimir Vovk.
\newblock A tutorial on conformal prediction, 2007.

\bibitem{Shaik_2023}
Thanveer Shaik, Xiaohui Tao, Christopher Dann, Haoran Xie, Yan Li, and Linda Galligan.
\newblock Sentiment analysis and opinion mining on educational data: A survey.
\newblock {\em Natural Language Processing Journal}, 2:100003, March 2023.

\bibitem{DOEreport}
Arman Shehabi, Alex Newkirk, Sarah~J. Smith, Alex Hubbard, Nuoa Lei, Md~Abu~Bakar Siddik, Billie Holecek, Jonathan Koomey, Eric Masanet, and Dale Sartor.
\newblock 2024 united states data center energy usage report.
\newblock Technical report, Lawrence Berkeley National Laboratory, Berkeley, CA, December 2024.
\newblock Report prepared for the U.S. Department of Energy.

\bibitem{tucson_datacenter_water}
{The Associated Press}.
\newblock Proposed data center prompts tucson to regulate large water users, require conservation.
\newblock 2025.
\newblock Accessed: 2025-09-23.

\bibitem{loudoun2025datacenter}
Mike Turner.
\newblock Loudoun county, virginia: Data center capital of the world ``a strategy for a changing paradigm''.
\newblock Technical report, Loudoun County Board of Supervisors, Loudoun County, VA, August 2025.
\newblock Original Edition: July 1, 2024.

\bibitem{census_acs_5year}
{U.S. Census Bureau}.
\newblock American community survey 5-year data (2009-2023), 2023.
\newblock Accessed: 2025.

\bibitem{us_census_acs_2024}
{U.S. Census Bureau}.
\newblock American community survey ({ACS}).
\newblock \url{https://www.census.gov/programs-surveys/acs.html}, 2024.
\newblock Ongoing survey conducted since 2005.

\bibitem{us_census_category}
{U.S. Census Bureau}.
\newblock American community survey data, 2024.
\newblock Accessed: 2024.

\bibitem{eia_state_emissions_2025}
{U.S. Energy Information Administration}.
\newblock State carbon dioxide emissions data.
\newblock \url{https://www.eia.gov/environment/emissions/state/}, 2025.
\newblock Accessed: September 10, 2025.

\bibitem{virginia_deq_data_centers_2024}
{Virginia Department of Environmental Quality}.
\newblock Issued air permits for data centers.
\newblock \url{https://www.deq.virginia.gov/permits/air/issued-air-permits-for-data-centers}, 2024.
\newblock Accessed: December 2024.

\bibitem{Virginia_AirPermitsDataCenter_Report_JLARC_2024}
{Virginia Joint Legislative Audit and Review Commission}.
\newblock Report to the {Governor and the General Assembly of Virginia}: Data centers in {Virginia} ({JLARC} report 158), December 2024.

\bibitem{Wade2025electricity_grid}
Cameron Wade, Mike Blackhurst, Joe DeCarolis, Anderson de~Queiroz, Jeremiah Johnson, and Paulina Jaramillo.
\newblock Electricity grid impacts of rising demand from data centers and cryptocurrency mining operations.
\newblock Technical report, Carnegie Mellon University, Scott Institute for Energy Innovation and North Carolina State University, June 2025.
\newblock Open Energy Outlook Initiative.

\bibitem{policy_pulse}
Maggie Wang, Ella Colby, Jennifer Okwara, Varun Nagaraj~Rao, Yuhan Liu, and Andrés Monroy-Hernández.
\newblock Policypulse: Llm-synthesis tool for policy researchers.
\newblock In {\em Proceedings of the Extended Abstracts of the CHI Conference on Human Factors in Computing Systems}, CHI EA ’25, page 1–17. ACM, April 2025.

\bibitem{cot}
Jason Wei, Xuezhi Wang, Dale Schuurmans, Maarten Bosma, Brian Ichter, Fei Xia, Ed~Chi, Quoc Le, and Denny Zhou.
\newblock Chain-of-thought prompting elicits reasoning in large language models, 2023.

\bibitem{Carbon_SustainbleAI_CaroleWu_MLSys_2022_wu2022sustainable}
Carole-Jean Wu, Ramya Raghavendra, Udit Gupta, Bilge Acun, Newsha Ardalani, Kiwan Maeng, Gloria Chang, Fiona Aga, Jinshi Huang, Charles Bai, et~al.
\newblock Sustainable {AI}: Environmental implications, challenges and opportunities.
\newblock In {\em Proceedings of Machine Learning and Systems}, volume~4, pages 795--813, 2022.

\bibitem{heatmap_poll}
Matthew Zeitlin.
\newblock Heatmap poll: Only 44\% of americans would welcome a data center nearby, 2025.
\newblock Accessed: 2025-09-21.

\bibitem{election_simulation}
Xinnong Zhang, Jiayu Lin, Libo Sun, Weihong Qi, Yihang Yang, Yue Chen, Hanjia Lyu, Xinyi Mou, Siming Chen, Jiebo Luo, Xuanjing Huang, Shiping Tang, and Zhongyu Wei.
\newblock Electionsim: Massive population election simulation powered by large language model driven agents, 2024.

\end{thebibliography}
\end{document}